\documentclass[tradiabstract]{aa} 
\usepackage{natbib}
\usepackage{graphicx}
\usepackage{placeins}
\usepackage{amsmath}
\usepackage{arydshln}

\newcommand\mancha{\textsc{Mancha3D}}

\usepackage{xcolor}

\begin{document} 

\title{Two-fluid simulations of Rayleigh-Taylor instability in a magnetized solar prominence thread. I. Effects of prominence magnetization and mass loading}
\author{B. Popescu Braileanu
          \inst{1,2}
                \thanks{ \email{beatriceannemone.popescubraileanu@kuleuven.be}}
          \and
          V. S. Lukin\inst{3}
          \and
          E. Khomenko\inst{1,2}
          \and
          \'A. de Vicente\inst{1,2}} 
          
\titlerunning{The Rayleigh-Taylor instability}
\authorrunning{Popescu Braileanu et al.}

\institute{Instituto de Astrof\'{\i}sica de Canarias, 38205 La Laguna, Tenerife, Spain
\and Departamento de Astrof\'{\i}sica, Universidad de La Laguna, 38205, La Laguna, Tenerife, Spain
\and National Science Foundation\thanks{Any opinion, findings, and conclusions or recommendations expressed in this material are those of the authors and do not necessarily reflect the views of the National Science Foundation.}, Alexandria, VA, 22306, USA}

\date{Received 2020; Accepted XXXX}
 
\abstract{Solar prominences are formed by partially ionized plasma with inter-particle collision frequencies, which generally warrant magnetohydrodynamic treatment.  In this work, we explore the dynamical impacts and observable signatures of two-fluid effects in the parameter regimes when ion-neutral collisions do not fully couple the neutral and charged fluids. We performed 2.5D two-fluid (charge - neutrals) simulations of the Rayleigh-Taylor instability (RTI) at a smoothly changing interface between a solar prominence thread and the corona. The purpose of this study is to deepen our understanding of the RTI and the effects of partial ionization on the development of the RTI using non-linear two-fluid numerical simulations. Our two-fluid model takes into account viscosity, thermal conductivity, and collisional interaction between neutrals and charge: ionization or recombination, energy and momentum transfer, and frictional heating. In this paper, we explore the sensitivity of the RTI dynamics to the prominence equilibrium configuration, including the impact of the magnetic field strength and shear supporting the prominence thread, and the amount of prominence mass-loading. We show that at small scales, a realistically smooth prominence-corona interface leads to qualitatively different linear RTI evolution than that which is expected for a discontinuous interface, while magnetic field shear has the stabilizing effect of reducing the growth rate or eliminating the instability. In the non-linear phase, we observe that in the presence of field shear the development of the instability leads to formation of coherent and interacting 2.5D magnetic structures, which, in turn, can lead to substantial plasma flow across magnetic field lines and associated decoupling of the fluid velocities of charged particles  and neutrals.}


\maketitle
%

\section{Introduction}

The Rayleigh-Taylor Instability (RTI) is present in many astrophysical systems, and it is frequently  observed in solar prominences \citep{Berger2008, Berger_2017}.  RTI is also present in laboratory plasmas and can play an important role in inertial and magneto-inertial confinement fusion experiments \citep[see review by][and references therein]{ZHOU1,ZHOU2}.

On the Sun, prominences are defined as clouds of chromospheric material sustained in the solar corona by magnetic forces. They are colder and denser than the surrounding corona.
As a prominence sits above the lighter corona, a small perturbation at the interface between the prominence and the corona may grow without bound, giving rise to the RTI. Small-scale upflows, also called plumes, have been observed at the top border of prominence bubbles by \cite{Berger_2017}.  The plumes are highly dynamic columns of plasma, which rise from below the prominence and propagate upwards with speeds of $\approx$ 15 km/s. It is believed that they appear due to the RTI \citep{2010Berger,2014Ozorco,Berger_2017}.

The cold prominences contain a large fraction of neutral particles that get ionized when they enter the hotter corona during the development of the RTI.  Some observational studies show hints in support of the decoupling in the velocity of ions and neutrals in prominences \citep{2007Gilbert,2016Khomenko,2017Anan,Wiehr2019}, however, these observations are at the edge of the observational capabilities of the Sun. In Earth's ionosphere, plasma is also partially ionized, and many phenomena related to the uncoupled behavior of ions and neutrals are similar in the solar atmosphere and in the ionosphere \citep[see e.g., the review in][]{Leake2014}. In this work, we apply a two-fluid model to study the effect of the neutrals on the development of the RTI at the interface between the corona and a solar prominence thread, with a smooth transition between them, in order to resolve the effects of charge-neutral interactions.

One of the important parameters of the RTI, both in partially and in fully ionized plasma, is its growth rate. The growth of the RTI can be separated into two phases: linear and non-linear. During the initial linear evolution, different modes grow independently without interaction; however, mode-coupling effects become important with growing mode amplitudes and the instability evolves into the nonlinear phase when the bubbles (structures that rise) and spikes (structures that fall) form.

Most of the analytical studies of the linear phase of the RTI have been carried out based on the assumption of a discontinuous density profile. In the absence of the magnetic field (the hydrodynamic case), when a system composed of a heavier fluid with mass density, $\rho_2$, which is initially on top of a lighter fluid with mass density, $\rho_1$, is perturbed, an instability develops for all the wavelengths of the initial perturbation \citep[see][]{Ch1961}. With homogeneous densities, under the incompressible and inviscid assumption, and when the transition between the two fluids has negligible width (discontinuous density profile), the growth rate in the linear approximation is a monotonically increasing function of the wave number $k$:
\begin{equation}\label{eq:rti_li1} 
-\omega^2 = A g k\,,
\end{equation} 
where,  
\begin{equation} \label{eq:atwood}
A=(\rho_2-\rho_1)/(\rho_2+\rho_1)\,,
\end{equation} 
is the Atwood number \citep{Ch1961}.

Under the same assumptions, a component of the magnetic field in the perturbation direction, considered to be the $x$ direction (see Figure~\ref{fig:Bfield}, which illustrates the setup used in our simulations and uses the same conventions), stabilizes small scale perturbations with wavelengths smaller than the critical wavelength \citep{Ch1961}.
The influence of magnetic field shear at the discontinuous density interface was recently studied by \citet{2014Rude,2018Rude}. 

In order to capture the two-fluid effects, we have to resolve scales comparable to the collisional mean free path between ions and neutrals.  These very small scales are typically below the density gradient scale length for any solar structure.  Thus, the density profile cannot be approximated as discontinuous in two-fluid calculations.

The general case of RTI with arbitrary non-uniform equilibrium density profile with a continuous transition in the vertical direction does not have a known analytical solution. However, in the absence of the magnetic field and under the assumption of incompressibility, for an exponential density profile  $\propto \text{exp} (\beta z)$ with uniform $\beta$, \cite{Ch1961} finds the following expression for  the growth rate:
\begin{equation} \label{eq:gr_an_exp}
-\omega^2 = \frac{g \beta k^2 d^2}{k^2 d^2 + \frac{1}{4} \beta^2 d^2 + m^2 \pi^2}\,, 
\end{equation}
where it is assumed that the fluid is confined between the heights $z=0$ and $z=d,$ and $m$ is the vertical mode number. The stratification is unstable when $\beta >0$, with the maximum growth rate for $m=1$.  We note that the growth rate obtained in Eq.~(\ref{eq:gr_an_exp}) is a monotonically increasing function of $k$ bounded by $-\omega^2 = g \beta$. Thus, the linear growth rate of the RTI,  when the transition region between the heavy and light fluids if of a finite width, exhibits a different $k$-dependence from that calculated with a discontinuous profile discussed above, when the growth rate increases linearly with the wave number as shown in Eq.~(\ref{eq:rti_li1}).

The non-linear effects which occur from interaction of the modes cannot be accounted for analytically in the general case. There are analytical approaches which give qualitative results about the nonlinear (and linear) phase of the RTI in a 3D geometry \citep{Hillier2016} and statistical results obtained in the late nonlinear (turbulent) phase using a self-similar ansatz. However, the work of \citet{Hillier2016} does not consider partial ionization effects and the hypothesis used in their analytical derivations are restrictive. In the general case, only numerical solutions allow for a substantial tracking of the evolution of the RTI into the non-linear regime.

Magnetic field strength in prominence threads can be inferred from observations of density contrast of the RTI bubbles in an approximate way.
The shear flow at the bubble boundary gives rise to subsequent plumes as a result of the Kelvin-Helmholtz instability 
\citep[KHI,][]{Berger_2017}.
The growth rate of the associated KHI depends on the density profile and flow velocities above and below the interface and the magnetic field. By comparing the measured growth rate of the instability to analytic expressions, the magnetic field can be inferred.   \cite{Berger_2017} found a magnetic flux density across the bubble boundary of $\approx$ 10 G  at an angle of $\approx$ 70$^\circ$ to the prominence plane.
The review in \cite{Hillier2018} shows more examples of the determination  of the field strengths from the growth rates in several contexts, along with the uncertainties associated with these methods.

The degree of collisional coupling between charged particles 
and neutrals can influence the linear growth rate and the non-linear course of the instability, as well as the appearance of bubbles and spikes. However, these aspects have been investigated only scarcely. Most of the analytical studies reported in the literature neglect the partial ionization effects on the development of the RTI in solar prominences.  An analytical calculation of the linear growth rate of the RTI in the two-fluid model has been done for the discontinuous density profile in a 3D geometry by \citet{Diaz2012}.  In the regime studied in that work, the collisions between neutrals and charge were found to decrease the growth rate at all scales.  Viscosity is also known to act in a similar way \citep{Ch1961,visc_incomp}.  In a later work, \cite{DiazKh2013} introduced interaction between neutrals and charge through the ambipolar term in the generalized induction equation. The authors concluded that the neutrals decouple from the charge at small scales, below the collision scale, eliminating the stabilizing effect of the magnetic field on the neutrals, which removes the cut-off imposed by the magnetic field. 

Most of the numerical simulations of RTI have been done using the single-fluid magnetohydrodynamic (MHD) approach \citep{ANUCHINA}. Only a handful of numerical studies have included the partial ionization effects through the ambipolar term \citep{Arber_2007,2014bKh}. The latter studies were done in the context of magnetic flux emergence in the chromosphere and for the RTI developed at the prominence-corona transition region (PCTR).

Numerical studies of RTI using the two-fluid approach with application to solar prominences are extremely scarce \citep{Leake2014}. In the  work of \citet{Leake2014}, only one case was studied and only the initial evolution of the instability was considered, with no in-depth analysis of the results. 
Recently, \citet{hillier2019} studied Kelvin-Helmholtz instability numerically in the two-fluid approach, but this idealized study  cannot be applied to the prominence conditions and the type of the instability investigated by \citet{hillier2019} is different. 
Most of the studies, analytical and numerical, omit the effects of viscosity, thermal conductivity, and ionization-recombination.

The overarching purpose of this study is to explore the influence of and possible observables associated with incomplete collisional coupling between charge and neutrals in the partially ionized prominence plasma during the linear and non-linear phases of the RTI. To do so, we employ a 2.5D two-fluid model of the plasma and, for the first time, we perform a systematic study of the RTI including elastic collisions, ionization and recombination processes, viscosity, and thermal conductivity in the model.
In order to capture the two-fluid effects that occur on small spatial scales, we perform simulations at an adequate resolution and consider the equilibrium with a realistic transition between the prominence and the coronal part of the structure, where the properties of the plasma change in a continuous way.  { There have been a limited number of analytical and numerical studies examining properties of RTI in smoothly varying equilibria in the presence of magnetic field, and thus the novelty of this work includes incompressible MHD linear analysis of RTI in such equilibria.  This single-fluid analysis is necessary to begin to understand the more complex two-fluid RTI evolution in non-linear solar prominence simulations.}

This study starts with the same background model as in \cite{Leake2014}. We study how the development of RTI in the two-fluid model depends on the parameters of the background atmosphere, such as the density contrast and the magnetic field strength, because these parameters can be, in principle, extracted from observations.  From the simulations, we extract  the linear growth rates, study the development of bubbles, spikes and the associated magnetic structures in the non-linear phase, as well as of the flow decoupling between the neutral and charged fluids.

We describe the configuration and plasma parameters of the simulations in Section~\ref{sec:setup}, followed by an overview of the simulation dynamics and evolution of different scales in Section~\ref{sec:rti_results}. { We explore the analytical solution limits under the single-fluid incompressible MHD approximation for smoothly varying density and magnetic field profiles in Section~\ref{sec:an_gr_1f}.  We then study the linear behavior of RTI using semi-analytical calculations for the assumed background atmosphere and in the two-fluid simulations in Section~\ref{sec:gr}. In Section~\ref{sec:decoupling}, we study the decoupling in the velocity fields and in Section \ref{sec:power}, the magnetic structure formation.}  We discuss our results and present our conclusions in Section~\ref{sec:conclusions}.


\section{Description of the problem}\label{sec:setup}

\begin{figure}
 \includegraphics[width=8.5cm]{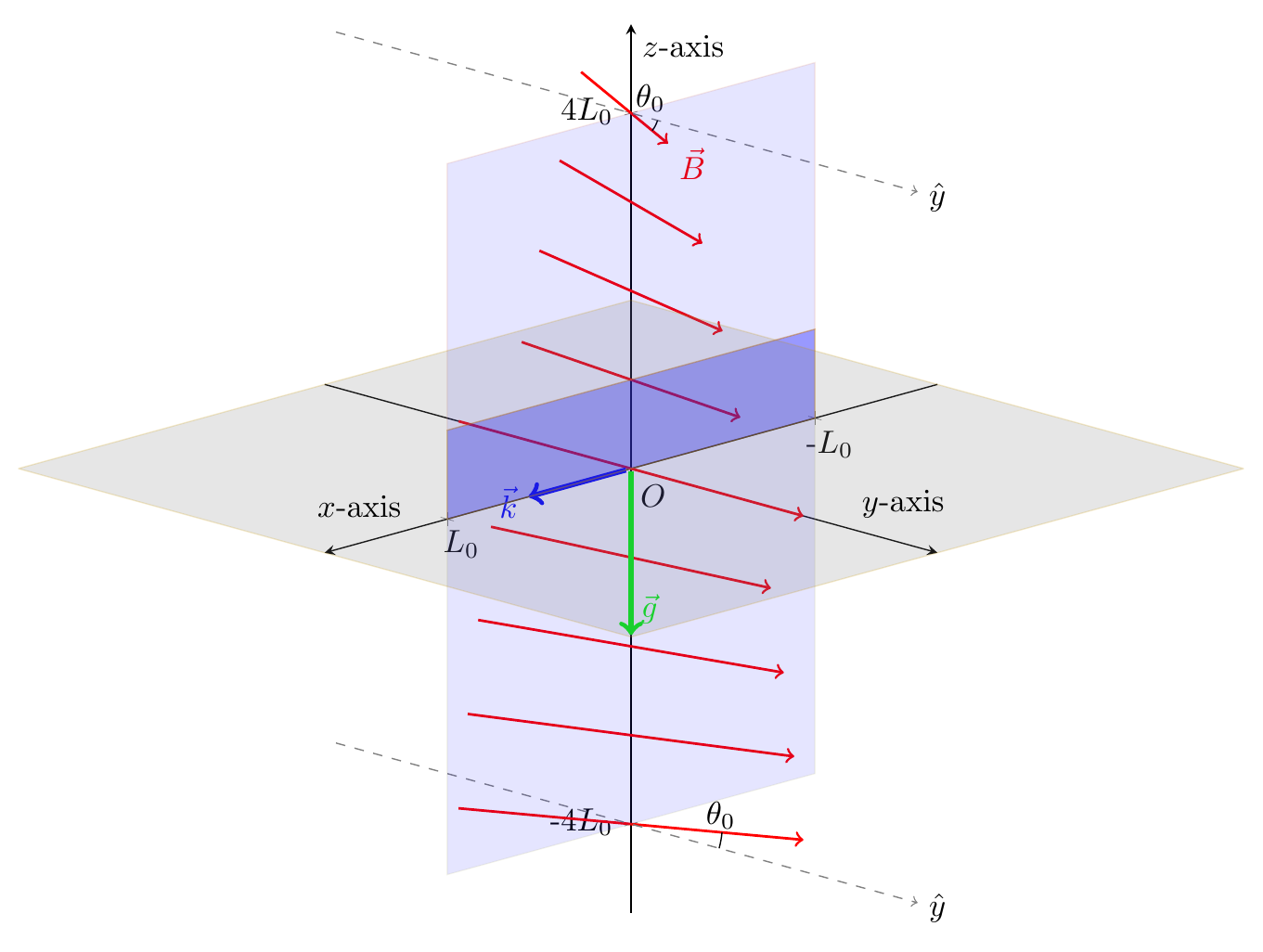}
\caption{Sketch of the initial configuration. The simulation domain (filled with blue color) is contained in the XOZ plane, where $x$ varies between $-L_0$ and $L_0$ and $z$ varies between $-4L_0$ and $4L_0$. The magnetic field, $\mathbf{B}$ (red lines), is contained in the  XOY plane (filled with gray color).  At the height $z=0,$ the magnetic field only has the component, $B_y$, and it is slowly rotated above and below $z=0$ by an angle which depends on height, keeping its modulus constant. The value of the angle is anti-symmetric with respect to $z=0$. The maximum angle obtained at the top and the bottom of the atmosphere is indicated by $\theta_0$.  When the field is perpendicular, the direction of $\mathbf{B}$ is along the $y$-axis for all the heights, $\theta_0=0^\circ$. For the sheared magnetic field configurations $\theta_0=1^\circ$.
The gravity, $\mathbf{g,}$ points in the negative $z$ direction and is indicated by a green  line.
The direction of the perturbation, $\mathbf{k,}$ is shown by a blue line.}
\label{fig:Bfield}
\end{figure}

We model the RTI at the border of a thin prominence thread. To approximate this situation, we adopt a configuration similar to that described by \cite{Leake2014}, shown in Figure~\ref{fig:Bfield}.  We use 2.5D geometry where the prominence is contained in the XOZ plane (filled with blue), above the height $z=0$, and sheared magnetic field, contained in the XOY plane (red lines). This configuration is described below by Eqs.~(\ref{rti_shear_setup}). The other configuration used for the simulations presented in this paper is the magnetic field, which is unidirectional along the $y$-axis. For this configuration, $\theta_0=0^\circ,$ as in the sketch in Figure~\ref{fig:Bfield}.

\subsection{Background atmosphere}
\begin{figure*}[!htb]
 \centering
 \includegraphics[width=8cm]{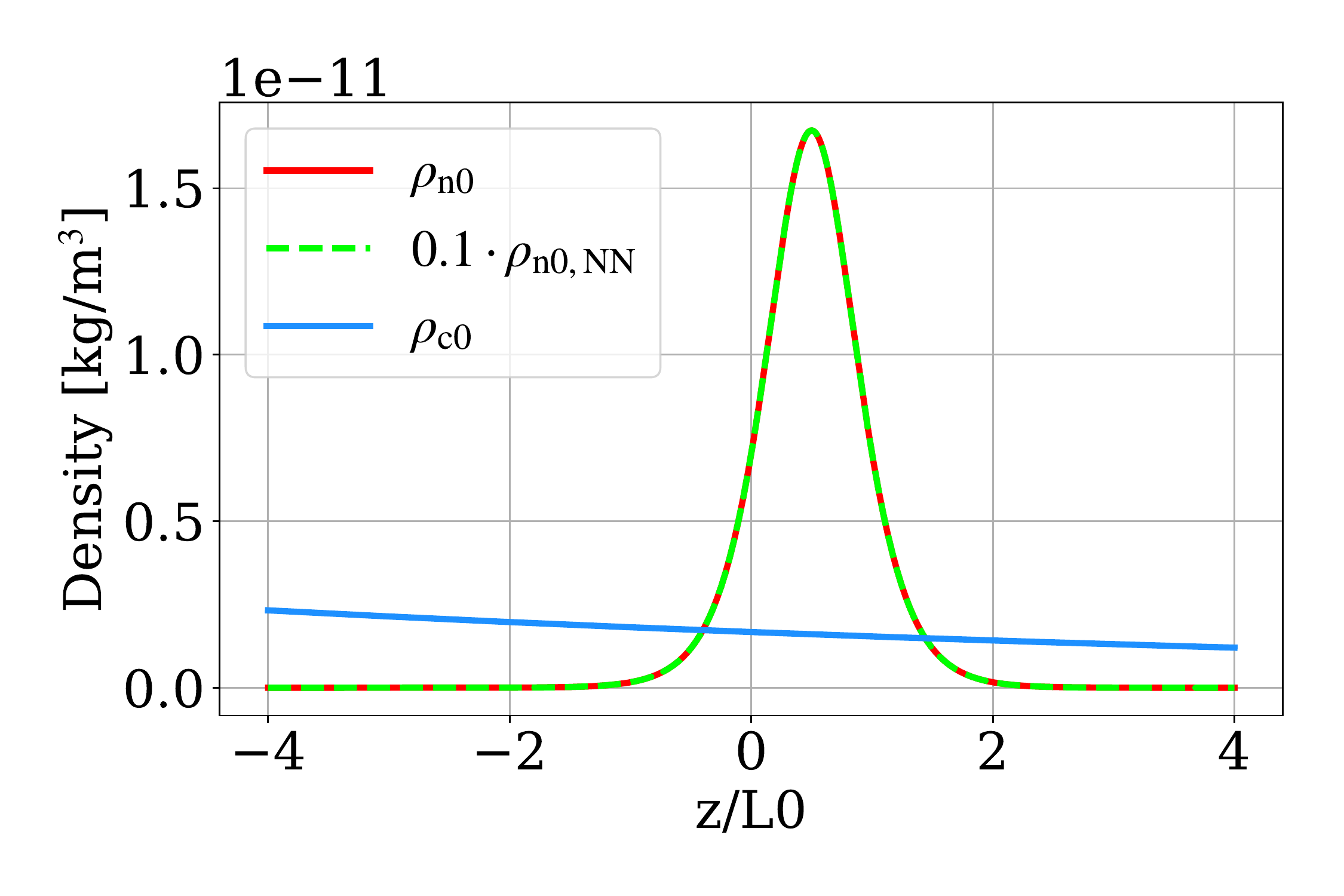}
 \includegraphics[width=8cm]{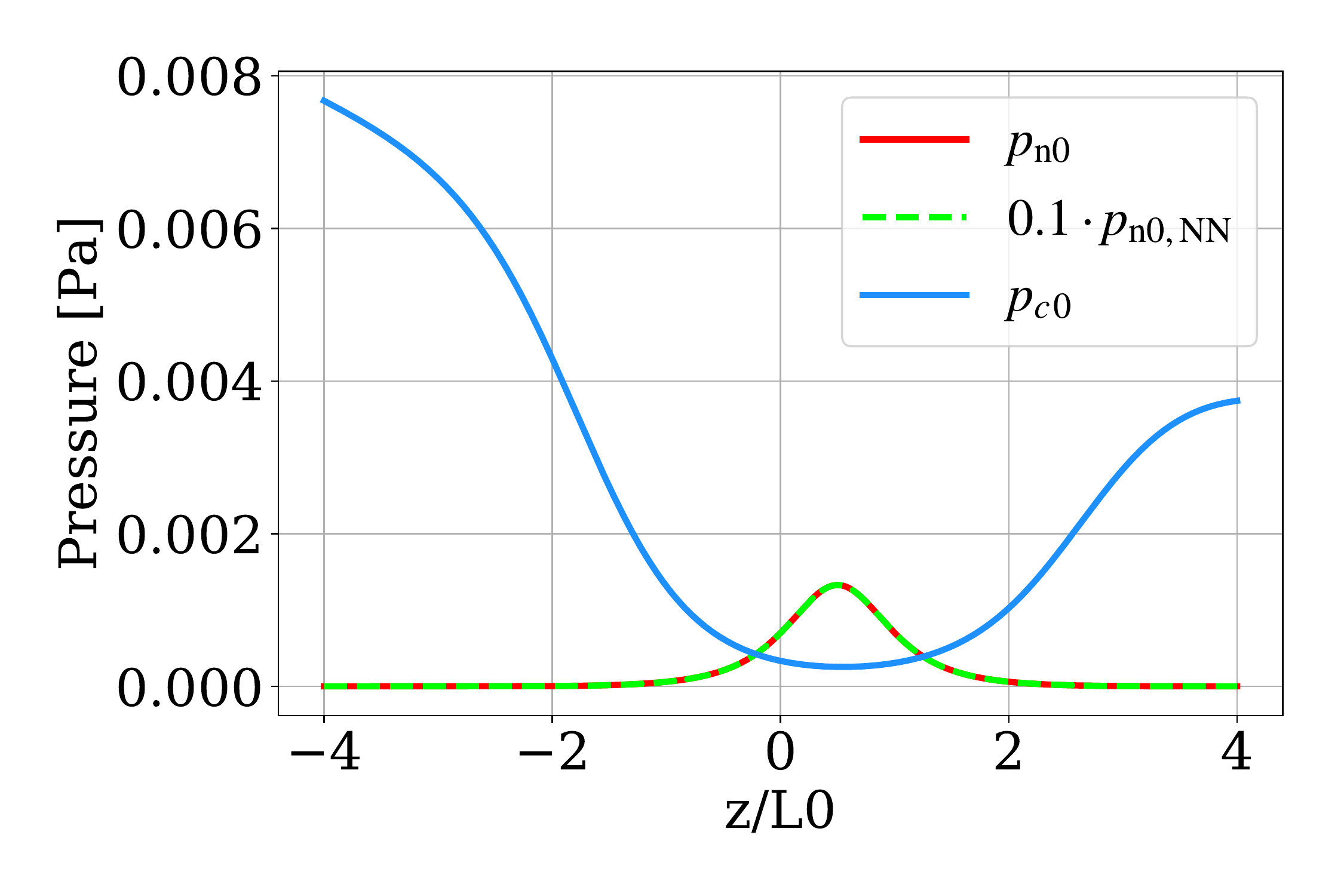}

 \includegraphics[width=8cm]{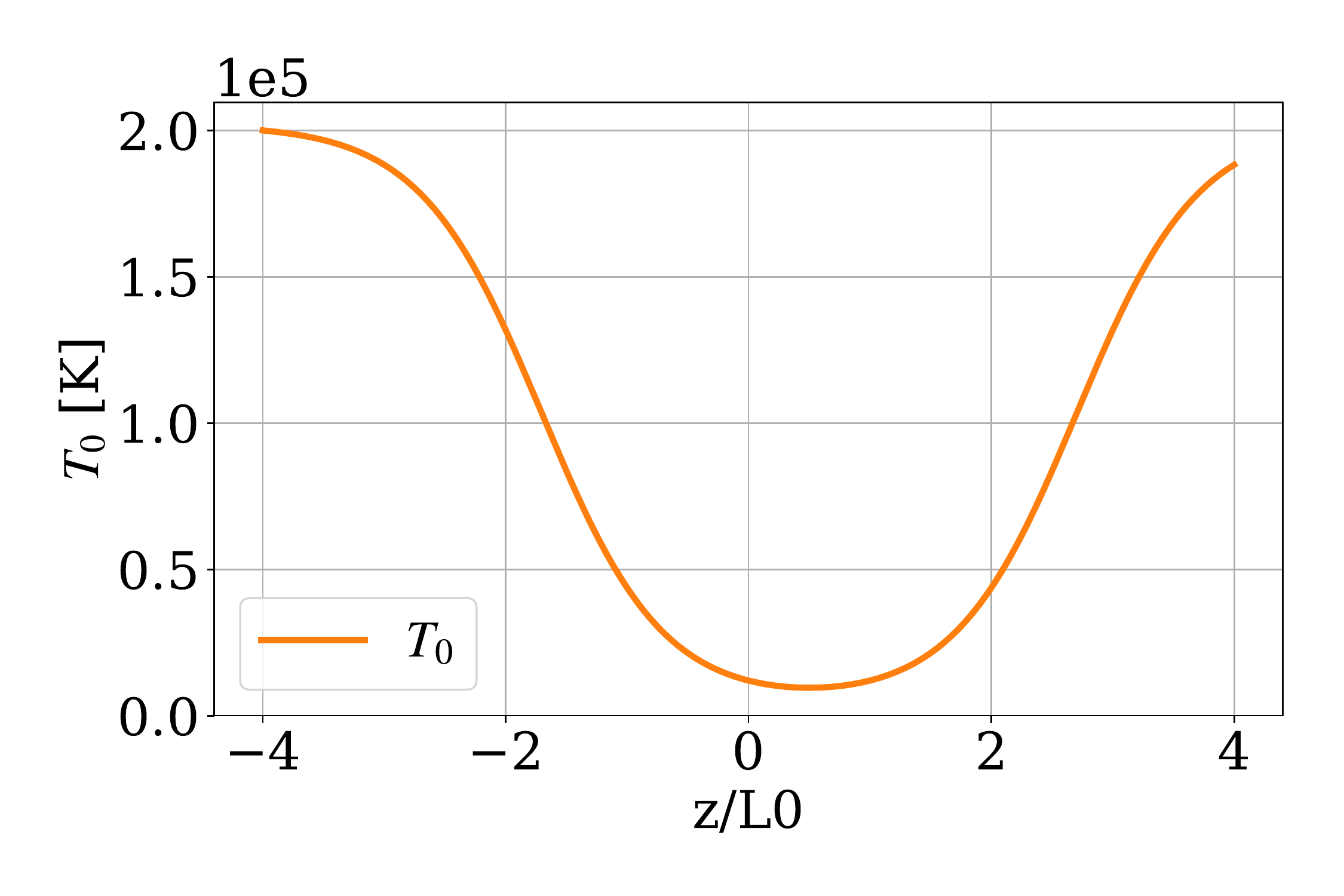}
 \includegraphics[width=8cm]{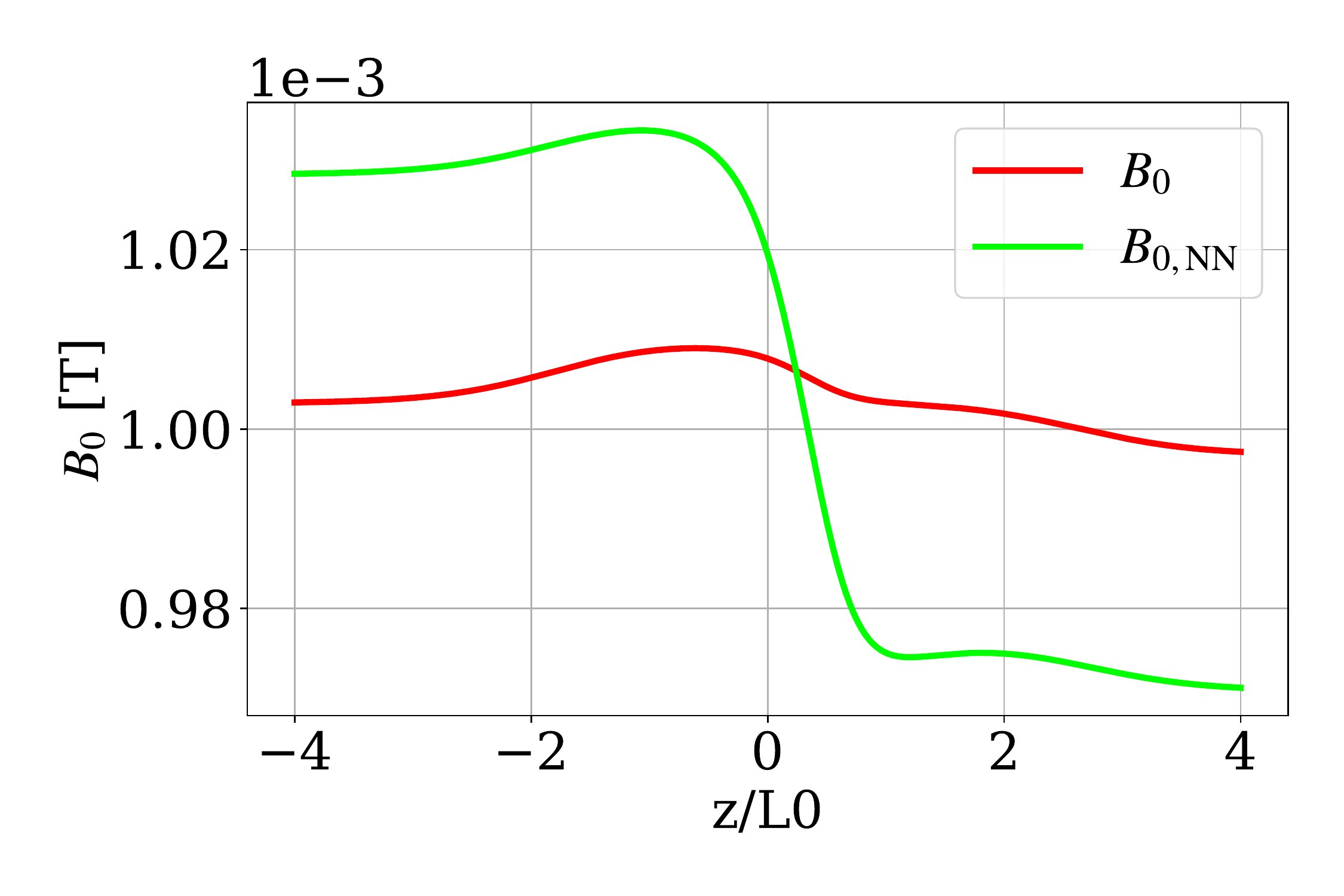}
\caption{Parameters of the initial model atmosphere as a function of the vertical coordinate, $z$. Top left: Density of neutrals in the models with the original (red) and enhanced (dashed green) density contrasts, and density of charged particles  (blue). Top right: Same for the pressure of neutrals and charged particles. Bottom left: Temperature of both the charged particles and neutrals. Bottom right: Modulus of the magnetic field in the models with the original (red) and enhanced (green) density contrasts.}
\label{fig_equi}
\end{figure*}

The equilibrium atmosphere is uniform in the $x$-direction. The equations that describe the number density of the ions and neutrals, the background temperature, the magnitude of the magnetic field, and the pressures in equilibrium configurations are \citep{Leake2014} the following:
\begin{align} 
n_{\rm i0} &= n_0 \text{ exp} \Big( -\frac{z}{H_c} \Big), \label{eqs:rti_setup_ni} \\
n_{\rm n0} &= n_{\rm n00} \text{sech}^2 \Big(2 \frac{z}{L_0} - 1 \Big) + n_{\rm nb}, \label{eqs:rti_setup_nn} \\
T_0 &= T_b f(z), \label{eqs:rti_setup_T}  \\ 
B_0 &= B_{00}  \Bigg\{1+ \beta_p  \bigg[\frac{n_{\rm i0}}{n_0}  \Bigg(1-f(z)\Bigg) -  \frac{1}{2}  
\frac{n_{\rm n0}}{n_0}  f(z) \nonumber \\    
 & -  \frac{1}{H_c  n_0}   \Bigg(\frac{1}{2}  L_0  n_{\rm n00}  \text{tanh}
\Big(2 \frac{z}{L_0}-1\Big) + n_{\rm nb}  z \Bigg)  \bigg]\Bigg\}^{0.5},  \label{eqs:rti_setup_B} \\
p_{\rm n0} &= n_{\rm n0} k_{\rm B} T_0\,, \nonumber  \\
p_{\rm c0} &= 2 n_{\rm i0} k_{\rm B} T_0\,,\label{eqs:rti_setup_pe}
\end{align}
where the ion number density at $z=0$ is $n_0$ = $10^{15}$ m$^{-3}$; the peak neutral number density,
reached at $z=L_0/2$ is $n_{\rm n00}$ = $10^{16}$ m$^{-3}$; 
the background temperature of the corona $T_b$ = 2.02 $\times$ 10$^5$ K; the neutral number density corresponding to the corona temperature ($T_b$) is $n_{\rm nb}$ = 3.5 $\times$ 10$^9$ m$^{-3}$; $B_{00}=10^{-3}$ T is the  gravitational 
scale height of the charged particles  of $H_{\rm c} = 2 {k_{\rm B} T_b}/{(m_{\rm H} g)}$. 
The characteristic length scale is $L_0$=1~Mm.
The plasma $\beta_p = {2 n_0 k_{\rm B} T_b}/({B_{00}^2}/2\mu_0)$ is calculated at $z=0$ using  $B_{00}$ as the value of the magnetic field 
and has the value of  $\beta_p \approx$  1.4 $\times$ 10$^{-2}$. 
The function $f$ is defined as follows: 
\begin{equation}
f(z) = \frac{\text{cosh}^2\Big(\frac{z}{L_0}-0.5\Big)}{\Big(\frac{z}{L_0}-0.5\Big)^2 + L_{\rm t}}\,.\label{eq:f_z}
\end{equation}
{ The temperature profile, shown in  Eq.~(\ref{eqs:rti_setup_T}), is described  using the function $f$.
 The value of $L_{\rm t}$ = 20, which appears in Eq.~(\ref{eq:f_z}), has been chosen so that the temperature is closest to the values observed in the Sun and the ionization fraction 
$\xi_{\rm i} = {\rho_{\rm c}}/{(\rho_{\rm c} + \rho_{\rm n})}$ = 0.091  remains small \citep[see][]{Leake2014}.}
Figure~\ref{fig_equi} shows the background profiles for the density, pressure, temperature and the module of the magnetic field. The neutrals and charged species have the same temperature profile, and are assumed coupled enough by collisions, so that the prominence, composed essentially by neutrals, is supported by the gradient in the magnetic pressure. 
The equilibrium for the density of neutrals represents a density enhancement at $z=L_0/2$, where it attains the maximum, $n_{\rm n00}$, as from Eq.~(\ref{eqs:rti_setup_nn}).
The minimum of the temperature is also attained at $z=L_0/2$, and the maximum temperature is considered to be the coronal temperature ($T_b$).

The equilibrium density of the charged particles  is gravitationally stratified.  Using these density and temperature profiles, the pressure profiles are obtained from the ideal gas law, Eq.~\ref{eqs:rti_setup_pe}. 
The magnetic field profile is obtained by integrating the magneto-hydrostatic equation where the total density and pressure are used. The integration constants permits the scaling of the magnetic field around some chosen value, $B_{00}$ in Eq.~(\ref{eqs:rti_setup_B}) shown above.
%

\subsection{Perturbation}

To initiate the instability, we use a perturbation in the density of neutrals, located at height $z=0$:
\begin{equation} 
\frac{\rho_{\rm n1}}{\rho_{\rm n0}} = \delta \times r_{\rm x} \times \text{exp}\Bigg(  -4 \Big( \frac{z}{L_0} \Big)^2 \Bigg),
\label{eq_rho_n1}
\end{equation}
where $\delta$ is the perturbation amplitude and $r_{\rm x}$ is the form of the a perturbation which depends on the $x$ coordinate. The equilibrium atmosphere is homogeneous in the $x$ direction and the dependence of the perturbation on the $x$ coordinate gives rise to the choice of the ansatz of separable solutions $\propto f(z) \text{exp}(i \omega t) \text{exp}(-i k_{\rm x} x)$. Thus, the perturbation is propagated in the $x$ direction, indicated by the vector $\mathbf{k}=(k_{\rm x},0,0)$ in Figure~\ref{fig:Bfield}.

While the initial perturbation should not have any influence on the linear growth rates of individual modes, the 
evolution in the nonlinear phase may depend on it. In most of the simulations reported in this paper, we use a white noise perturbation, defining $r_{\rm x}$ in Eq.~(\ref{eq_rho_n1}) as a random number in the interval \mbox{[-1,1]} for all the points. In addition, we compare the linear growth rate of modes with the white noise ("WN") perturbation with a simulation where the perturbation in the $x$-direction is generated as a multi-mode perturbation ("MM") specified by:
\begin{equation}
r_{\rm x} = \frac{1}{5}\sum_{n=1}^{m_{\rm x}} \text{sin} \Bigg( \frac{2 \pi n x}{L_{\rm x}} + \phi_{\rm n} \Bigg)\,,
\label{eq_rx_mm}
\end{equation}
 where $m_x$ is the number of points in the $x$ direction and 
 $\phi_{\rm n}$ is a random phase.

\subsection{Two-fluid evolution equations}

We solve the two-fluid set of equations, described in section 2.1 of  \cite{Popescu+etal2018}, referred to in this subsection as EqP.  To do so  we use the two-fluid version of the \mancha\ code, \mancha-2F. The single-fluid version of this code is described in \cite{Felipe2010, Pedro2018}, and the two-fluid version preserves several numerical features of \mancha, such as hyper-diffusive algorithms, filtering, and PML layer, see \cite{Popescu+etal2018}. 

The continuity equations of neutrals and charged species  (Eqs.~(1) in EqP) are coupled by the inelastic collisional term (Eq.~(2) in EqP) which models ionization and recombination processes.  The momentum (Eqs.~(3) in EqP) and energy (Eqs.~(9) in EqP) equations are coupled by  collisional terms (Eqs.~(6) and (10) in EqP, respectively), which include all the elastic and inelastic contributions.  The charge-exchange reactions are introduced via the elastic collisional parameter  (Eq.~(A.8) in EqP). We include the viscosities in the pressure tensor (Eq.~(8) in EqP). The thermal conductivity is considered only in the neutral energy equation. The expressions for the viscosity and thermal conductivity coefficients are given in Eqs.~(A.9) from Appendix A.2 in EqP. We use the ideal Ohm's law  (right hand side of Eq.~(14) in EqP is zero).

\subsection{Parameters of the simulations}
\begin{table*}
\caption{List of the simulations used in the paper. The columns represent, from left to right:
the abbreviated name of the simulation used in the text, the figures they are referenced in, and the parameters of the simulation.}
\begin{tabular}{llp{11cm}}
\hline                  
Name & Figures & Parameters \\
\hline  
L1-WN  & \ref{fig:time_snaps2}, \ref{fig_group_growing2}, \ref{fig_group_growing_dec2}, \ref{fig:gr_dec_sn}, \ref{fig_magf_twist2} & Sheared magnetic field described by Eq.~(\ref{rti_shear_setup}) with  $L_s=L_0$.  Values of $n_{\rm n00}=10^{16}$ m$^{-3}$ in Eq.~(\ref{eqs:rti_setup_nn}) and $B_{00}=10^{-3}$ T in Eq.~(\ref{eqs:rti_setup_B}).  White noise perturbation  with $\delta=10^{-2}$ in Eq.~(\ref{eq_rho_n1}). \\
L1-WN-NN  & \ref{fig:time_snaps2}, \ref{fig_group_growing2}, \ref{fig_group_growing_dec2}, \ref{fig:gr_dec_sn}, \ref{fig_magf_twist2} & Sheared magnetic field described by Eq.~(\ref{rti_shear_setup}) with $L_s=L_0$.  Values of $n_{\rm n00}=10^{17}$ m$^{-3}$ in Eq.~(\ref{eqs:rti_setup_nn}), and $B_{00}=10^{-3}$ T in Eq.~(\ref{eqs:rti_setup_B}).  White noise perturbation  with $\delta=10^{-2}$ in Eq.~(\ref{eq_rho_n1}). \\
L1-WN-NN-B & \ref{fig:time_snaps2}, \ref{fig_group_growing2}, \ref{fig_group_growing_dec2}, \ref{fig:gr_dec_sn}, \ref{fig_magf_twist2} & Sheared magnetic field described by Eq.~(\ref{rti_shear_setup}) with $L_s=L_0$.  Values of $n_{\rm n00}=10^{17}$ m$^{-3}$ in Eq.~(\ref{eqs:rti_setup_nn}), and of $B_{00}=3 \times 10^{-3}$ T in Eq.~(\ref{eqs:rti_setup_B}).  White noise perturbation with $\delta=10^{-2}$ in Eq.~(\ref{eq_rho_n1}). \\
P-WN-NN & \ref{fig:time_snaps2}, \ref{fig_group_growing2}, \ref{fig_group_growing_dec2}, \ref{fig:gr_dec_sn}, \ref{fig_magf_twist2} & Perpendicular magnetic field  described by Eq.~(\ref{rti_perp_setup}).  Values of $n_{\rm n00}=10^{17}$ m$^{-3}$ in Eq.~(\ref{eqs:rti_setup_nn}), and $B_{00}=10^{-3}$ T in Eq.~\ref{eqs:rti_setup_B}.  White noise perturbation  with $\delta=10^{-2}$ in Eq.~(\ref{eq_rho_n1}). \\
\hdashline \\
P-MM-S & \ref{fig_gr_comp} & Perpendicular magnetic field  described by Eq.~(\ref{rti_perp_setup}).  Values of $n_{\rm n00}=10^{16}$ m$^{-3}$ in Eq.~(\ref{eqs:rti_setup_nn}), and $B_{00}=10^{-3}$ T in Eq.~(\ref{eqs:rti_setup_B}).  Multimode perturbation with $\delta=10^{-4}$ in Eq.~(\ref{eq_rho_n1}). \\
P-WN-S & \ref{fig_gr_comp}, \ref{fig:gr_k_all} & Perpendicular magnetic field  described by Eq.~(\ref{rti_perp_setup}).  Values of $n_{\rm n00}=10^{16}$ m$^{-3}$ in Eq.~(\ref{eqs:rti_setup_nn}), $B_{00}=10^{-3}$ T in Eq.~(\ref{eqs:rti_setup_B}).  White noise perturbation with $\delta=10^{-4}$ in Eq.~(\ref{eq_rho_n1}). \\
L1-MM-S & \ref{fig_gr_comp} & Sheared magnetic field described by Eq.~\ref{rti_shear_setup} with $L_{\rm s}=L_0$.  Values of $n_{\rm n00}=10^{16}$ m$^{-3}$ in Eq.~(\ref{eqs:rti_setup_nn}), $B_{00}=10^{-3}$ T in Eq.~(\ref{eqs:rti_setup_B}).  Multimode perturbation with $\delta=10^{-4}$ in Eq.~(\ref{eq_rho_n1}). \\
L1-WN-S & \ref{fig_gr_comp}, \ref{fig:gr_k_all} & Sheared magnetic field described by Eq.~(\ref{rti_shear_setup}) with $L_s=L_0$.  Values of $n_{\rm n00}=10^{16}$ m$^{-3}$ in Eq.~(\ref{eqs:rti_setup_nn}), and $B_{00}=10^{-3}$ T in Eq.~(\ref{eqs:rti_setup_B}).
White noise perturbation  with $\delta=10^{-4}$ in Eq.~(\ref{eq_rho_n1}). \\
L1-WN-NN-S & \ref{fig:gr_k_all} & Sheared magnetic field described by Eq.~(\ref{rti_shear_setup}) with $L_s=L_0$.  Values of $n_{\rm n00}=10^{17}$ m$^{-3}$ in Eq.~(\ref{eqs:rti_setup_nn}), and $B_{00}=10^{-3}$ T in Eq.~(\ref{eqs:rti_setup_B}).  White noise perturbation with $\delta=10^{-4}$ in Eq.~(\ref{eq_rho_n1}). \\
L1-WN-NN-B-S & \ref{fig:gr_k_all} & Sheared magnetic field described by Eq.~(\ref{rti_shear_setup}) with $L_s=L_0$.  Values of $n_{\rm n00}=10^{17}$ m$^{-3}$ in Eq.~(\ref{eqs:rti_setup_nn}), and $B_{00}=3 \times 10^{-3}$ T in Eq.~(\ref{eqs:rti_setup_B}).
White noise perturbation  with $\delta=10^{-4}$ in Eq.~(\ref{eq_rho_n1}). \\
P-WN-NN-S  & \ref{fig:gr_k_all} & Perpendicular magnetic field  described by Eq.~(\ref{rti_perp_setup}).  Values of $n_{\rm n00}=10^{17}$ m$^{-3}$ in Eq.~(\ref{eqs:rti_setup_nn}), and $B_{00}=10^{-3}$ T in Eq.~(\ref{eqs:rti_setup_B}).  White noise perturbation with $\delta=10^{-4}$ in Eq.~(\ref{eq_rho_n1}). \\
\hline  
\end{tabular}
\label{tab:param_tests}
\end{table*}

Table~\ref{tab:param_tests} describes all the simulations used in this work and includes the abbreviated names of the simulations, the figures where they appear, and the summary of their parameters. There are four input parameters that are changed throughout the simulations:

 {\it Magnetic field inclination.}
The background magnetic field is contained in the XOY plane.
We use two types of configurations, namely, the magnetic field  perpendicular to the plane defined by the direction of the perturbation
and the gravity XOZ, and a slightly sheared  magnetic field, that is,

the perpendicular magnetic field
\begin{equation}\label{rti_perp_setup}
B_{\rm y0} = B_0\,,\qquad B_{\rm x0} = 0\,.
\end{equation}
        These simulations are referred in the table with the keyword ``P;''

the sheared magnetic field
\begin{align} \label{rti_shear_setup}
&B_{\rm x0} = B_0 \text{sin} (\theta)\,\,\,, B_{\rm y0} = B_0 \text{cos} (\theta)  \\
&\theta(z) = \left\{
\begin{array}{ll}
- \theta_0 \pi/180 & \text{, for } z < 0 \nonumber \\ 
\theta_0 \pi/180 & \text{, for } z \ge 0,
\end{array} 
\right.  \,\,\,\text{ if }L_{\rm s}=0,
\nonumber \\
&\theta(z) = \text{tanh}\left(\frac{z}{L_{\rm s}}\right) \theta_0 \times \pi/180, \,\,\, \text{ otherwise.} \nonumber 
\end{align}
We  use a fixed  value of $\theta_0$ = 1$^\circ$ and $L_{\rm s}=L_0$ in Eq.~(\ref{rti_shear_setup}) throughout the simulations. These simulations are marked in the table by  ``L1.''

 {\it Density contrast.}
The value of the peak density of neutrals is changed in some of the simulations by increasing the value of $n_{\rm n00}$ in Eq.~(\ref{eqs:rti_setup_nn}) by a factor of 10. These simulations appear marked with the ``NN'' keyword in their name.
The Atwood number values calculated using the total density at the heights $-L_0/2$ and $L_0/2$, that is, $\rho_2 = \rho_0(z=L_0/2)$ and $\rho_1 = \rho_0(z=-L_0/2)$, are 
$A = 0.72$ for the original density case and $A=0.85$ for the enhanced density case.
 {\it 

Magnetic field  magnitude.} 
The value of $B_{00}$ in Eq.~(\ref{eqs:rti_setup_B}), which describes the magnetic field configuration, is increased by a factor of 3 (the ``B'' keyword). 
{\it 

Amplitude of the perturbation.}
The amplitude of the perturbation is varied through the parameter $\delta$ in Eq.~(\ref{eq_rho_n1}) above.  We use the value $\delta=10^{-2}$ in most of the simulations. But for the study of the linear growth rate, we use a smaller amplitude by a factor of 100, $\delta=10^{-4}$, in order to avoid the nonlinear effects. The names of these simulations in the table have an additional suffix ``-S.''

Figure~\ref{fig_equi} shows the background profiles for both the original and enhanced value of the density contrast.  We note that when the density peak is increased, larger magnetic pressure gradient is needed to support the prominence. Therefore, the magnetic field profile also changes, as shown in Figure~\ref{fig_equi} in the bottom right plot.

We choose the domain size  $L_{\rm x} = 2 L_0$, $L_{\rm z} = 8 L_0$.
We use a uniform grid for the two dimensions and a resolution of $512 \times 2048$ points; this gives the values of the grid cell sizes: $d_{\rm x}=d_{\rm z}=3.9$ km.  We choose the point (0,0) in the middle of the plane XOZ, so that $x$ varies from  $-L_0$ to $L_0$ , and $z$ varies from $-4 L_0$ to $4 L_0$.

In all our simulations, we avoid using hyper-diffusion to stabilize the simulations. In order to remove small scale noise, we use the filtering method described in \citet{2007ManchaPa}.  We use periodic boundary conditions in the $x$ direction, reflecting boundary conditions for the vertical velocities of neutrals and charged particles (velocities are zero at the boundaries) and open boundary conditions for the rest of the variables (space derivatives in the vertical direction are zero at the boundaries) in the $z$ direction. Similarly to \cite{Soler-rti1}, the reflecting boundary condition for the velocities at the bottom of the atmosphere plays an important role in supporting the prominence against gravity.

\subsection{Collisional dissipation coefficients and scales} \label{ssec:coll_effects}

\begin{figure*}[!htb]
 \includegraphics[width=8cm]{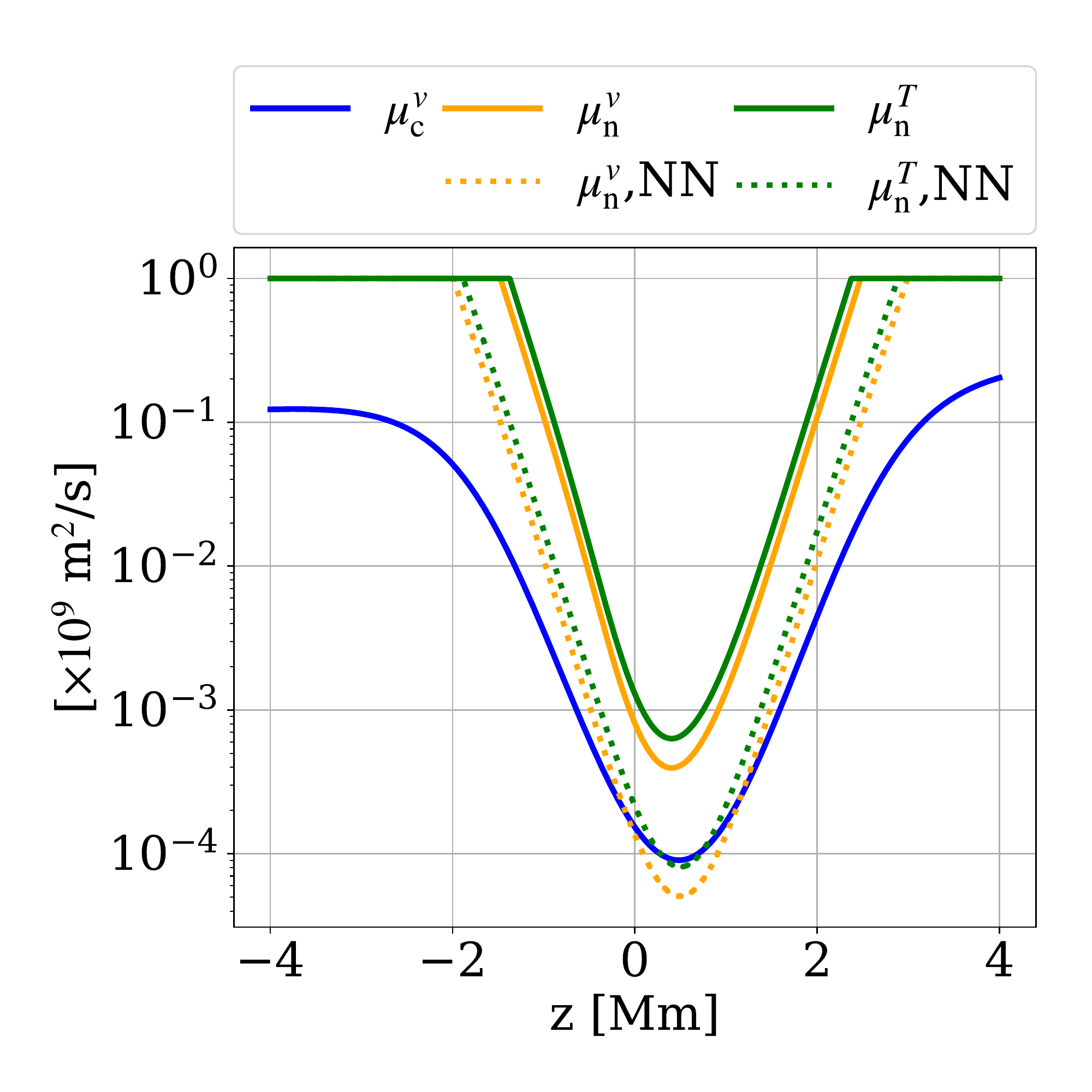}
 \includegraphics[width=8cm]{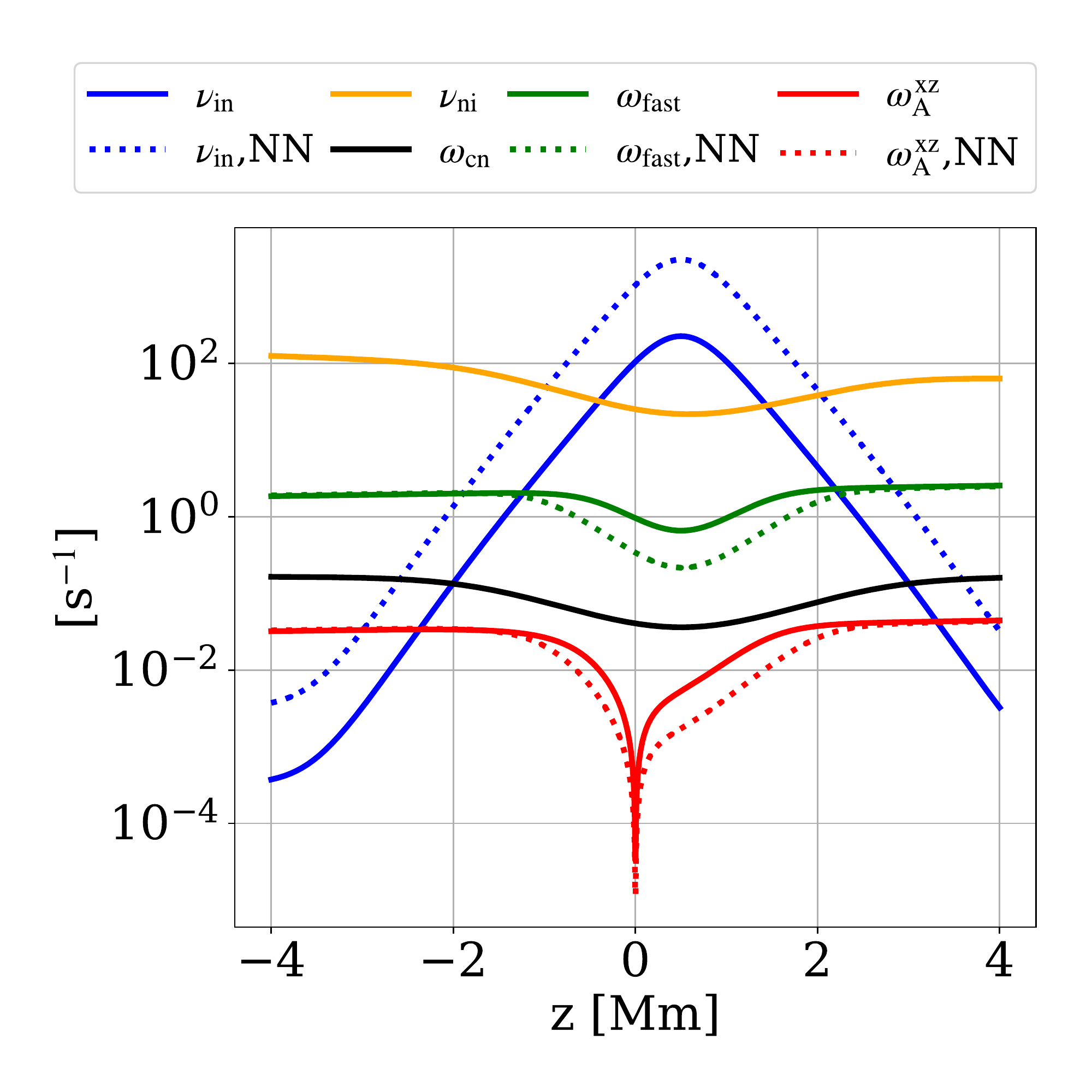}
\caption{Left: Coefficients of the viscosity of charged particles ($\mu^v_{\rm c}$, blue line) and neutrals ($\mu^v_{\rm n}$, orange lines), and the thermal conductivity of neutrals ($\mu^T_{\rm n}$, green lines), in units of m$^2$/s, as from Eq.~(\ref{eq_diff_coef}). The neutral viscosity and thermal conductivity coefficients are limited to $10^9$ m$^2$/s. Right: Ion-neutral ($\nu_{\rm in}$, blue lines),  neutral-ion ($\nu_{\rm ni}$,  orange line) collision frequencies, and frequencies associated with waves: fast magneto-acoustic  ($\omega_{\rm fast}$, green lines), in-plane Alfv\'en for ``L1'' shear ($\omega_{\rm A}^{\rm xz}$, red lines), acoustic in the neutral fluid ($\omega_{\rm cn}$, black line) as functions of height. Values for the model with the original density contrast model are shown with solid lines, and for the enhanced density contrast with dotted lines.}
\label{diff_coef1}
\end{figure*}

Our simulations are set to have sufficient resolution to capture the scales impacted by diffusion due to neutral viscosity or neutral-charge collisions, or both.  We evaluated these scales using the parameters of the background model atmospheres by computing the coefficients of the viscosity of charged particles  ($\mu_{\rm c}^{\rm v}$) and neutrals ($\mu_{\rm n}^{\rm v}$), and the neutral thermal conductivity ($\mu_{\rm n}^{\rm T}$), as well as the neutral-ion ($\nu_{\rm ni}$) and ion-neutral ($\nu_{\rm in}$) collision frequencies.  The latter two are also compared to the time-scales associated with the fast magneto-acoustic wave ($\omega_{\rm fast}$), the in-plane Alfv\'en wave ($\omega_{\rm A}^{\rm xz}$), and the neutrals sound wave ($\omega_{\rm cn}$) within the background atmospheres as characteristic of dynamical time-scales in the system.  These are calculated as follows, 

\begin{eqnarray} \label{eq_diff_coef}
& \mu^{\rm v}_{\rm n} = \xi_{\rm n}/\rho_{\rm n}\,;\quad 
\mu^{\rm v}_{\rm c} = \xi_{\rm c}/\rho_c\,; \nonumber \\
& \mu^{\rm T}_{\rm n} = K_{\rm n} m_{\rm H} (\gamma-1)/( \rho_{\rm n} k_{\rm B})\,;
\nonumber \\ 
& \nu_{\rm in} = \alpha \rho_{\rm n}\,;\quad \nu_{\rm ni} = \alpha \rho_{\rm c}\,;
\nonumber \\
& \omega_{\rm fast} = v_{\rm fast} (2 \pi)/L_{\rm x}\,;\quad
 \omega_{\rm A}^{\rm xz} = v_{\rm A}^{\rm xz} (2 \pi)/L_{\rm x}\,;\nonumber \\
& v_{\rm fast} = \sqrt{\gamma ( p_n + p_c)/\rho_{\rm tot} + B^2/(\mu_0 \rho_{\rm tot})}\,;\nonumber \\
& v_{\rm A}^{\rm xz} = \sqrt{(B_x^2 + B_z^2)/(\mu_0 \rho_{\rm tot})},
\end{eqnarray}
where $\gamma$ is the adiabatic index, and  $\rho_{\rm tot} =\rho_n + \rho_c $ is the total density. The collisional parameter $\alpha$ has been defined in Eq.~(A.8)  and $\xi_\alpha$ and $K_\alpha$ in Eqs.~(A.9) in \cite{Popescu+etal2018}. The density of the neutrals is very low in the corona, and the temperature is high, therefore, the coefficients $\mu^v_{\rm n}$ and $\mu^T_{\rm n}$ are very high. To overcome the very small time-step imposed by such large coefficients of neutral viscosity and thermal conductivity, the values of these coefficients have been limited to 10$^9$ m$^2$/s in order to speed up the computation. This limit does not significantly impact the regions of the domain where the instability develops and therefore does not make qualitative difference for the outcome of the computations. An improved implementation of the viscosity and thermal conductivity operators for low collisionality systems will be considered in future work.

The left panel of Figure~\ref{diff_coef1} shows the values of the neutral viscosity and thermal conductivity coefficients, limited as described above, and the viscosity coefficient of the charged particles  for the two background atmosphere neutral density profiles.  We observe that the neutral viscosity and thermal conductivity coefficients have almost the same profile and the charge viscosity coefficient is an order of magnitude smaller than the neutral viscosity coefficient.

Right panel of Figure~\ref{diff_coef1} shows the ion-neutral, $\nu_{in}$, and the neutral-ion, $\nu_{ni}$, collision frequency profiles as functions of height for the background atmospheres.  To compare these to ideal MHD timescales of interest, we also plot the frequencies associated with the largest scale wave modes in the $x$-direction for the fast magneto-acoustic wave, $\omega_{\rm fast}$, the in-plane component of the Alfv\'en wave for ``L1'' magnetic field shear, $\omega_{\rm A}^{\rm xz}$, and the neutrals sound wave, $\omega_{\rm cn}$.

We observe that in the immediate neighborhood of the prominence thread between $z \approx -1$ Mm and $z \approx 2$ Mm, collision frequencies between the charged and neutral fluids are higher than the large-scale ideal MHD frequencies, and, thus, large-scale flows are expected to be well coupled.  However, it is also apparent that the frequency separation is not so large within the thread, and disappears altogether outside the core of the thread, so that smaller scale and higher frequency ideal MHD dynamics is likely to be strongly impacted by the two-fluid effects. 

\section{Global overview of simulation dynamics}\label{sec:rti_results}

\begin{figure*}
 \centering
 \includegraphics[width=8cm,height=11cm]{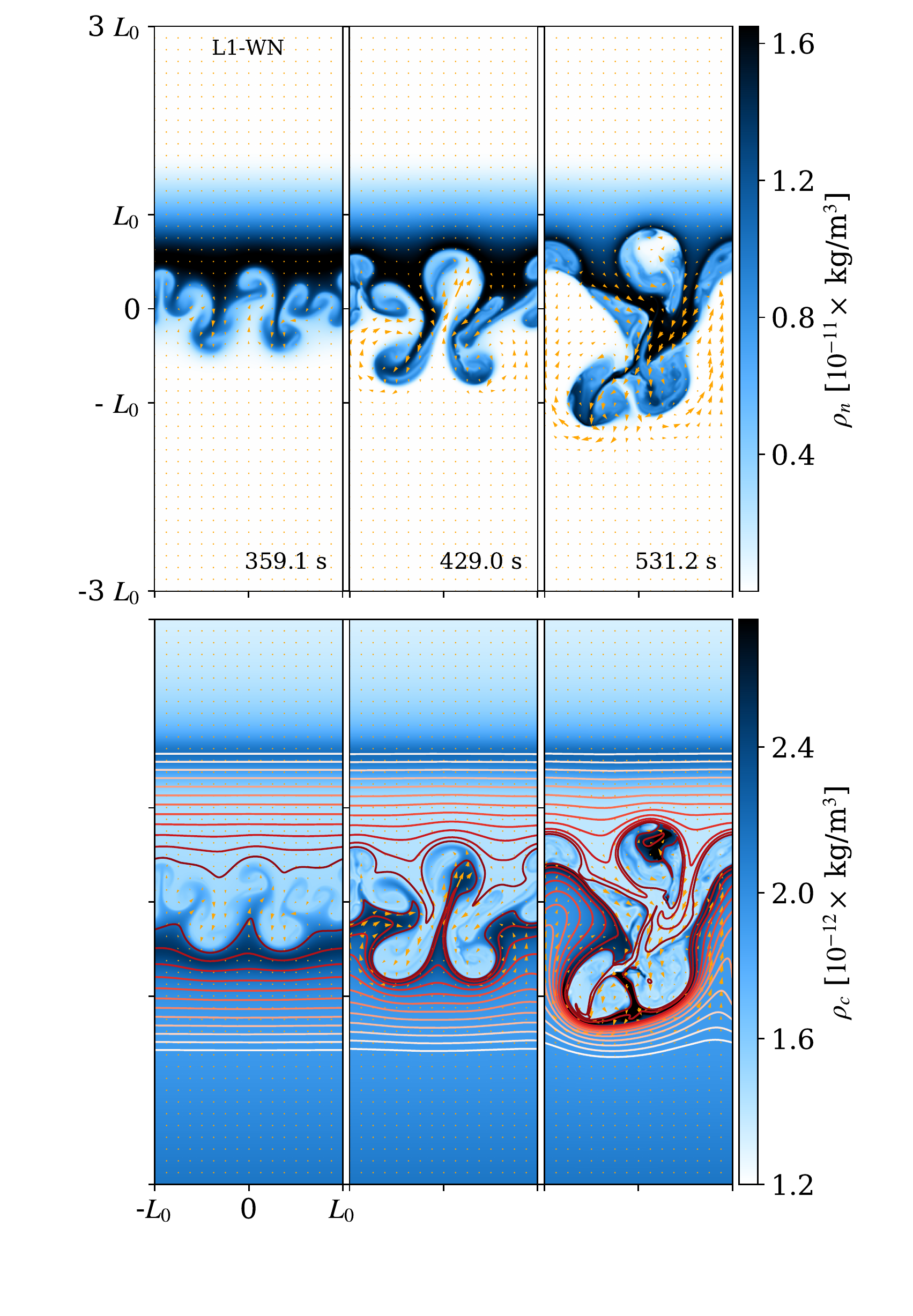}
 \includegraphics[width=8cm,height=11cm]{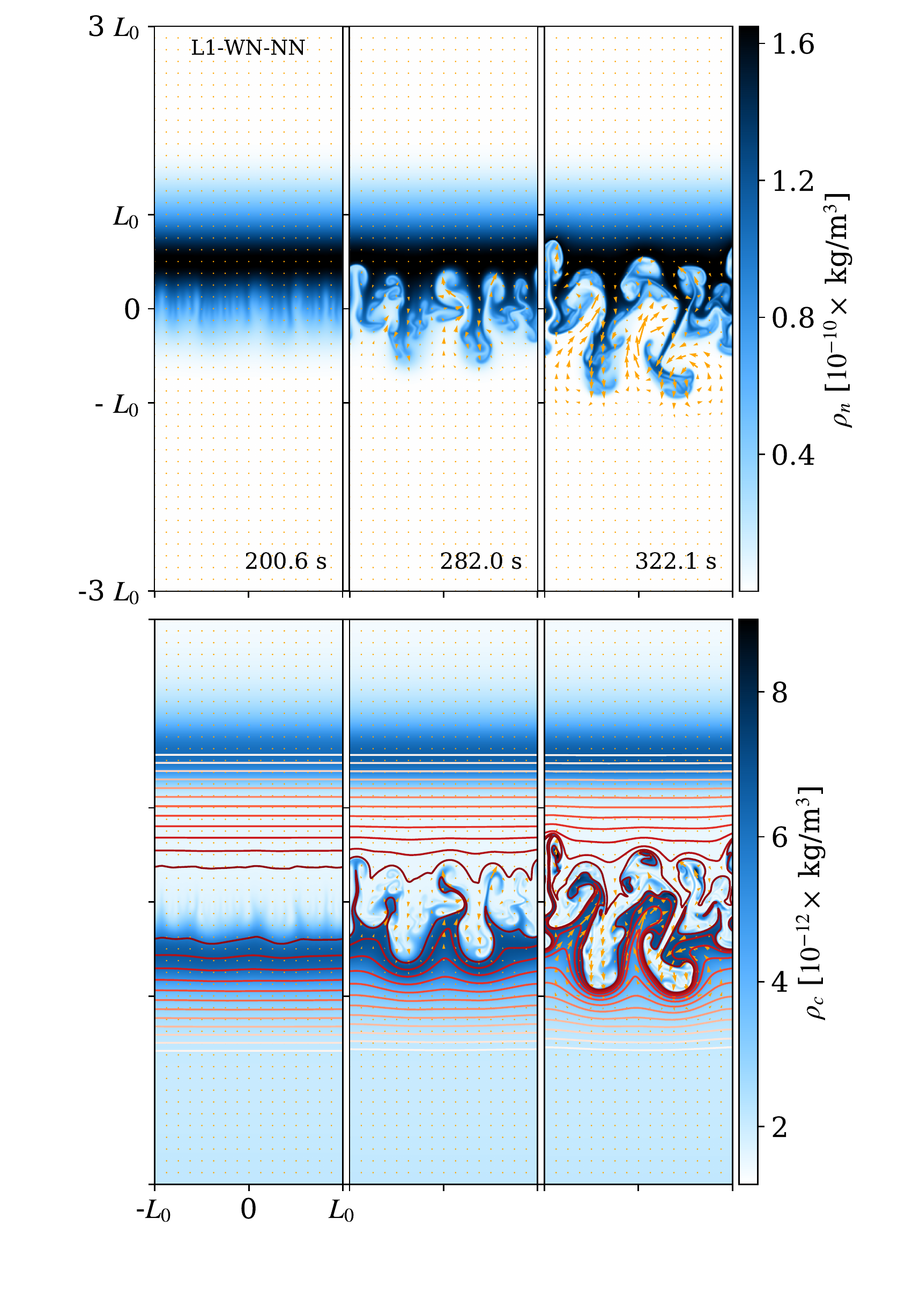}

 \includegraphics[width=8cm,height=11cm]{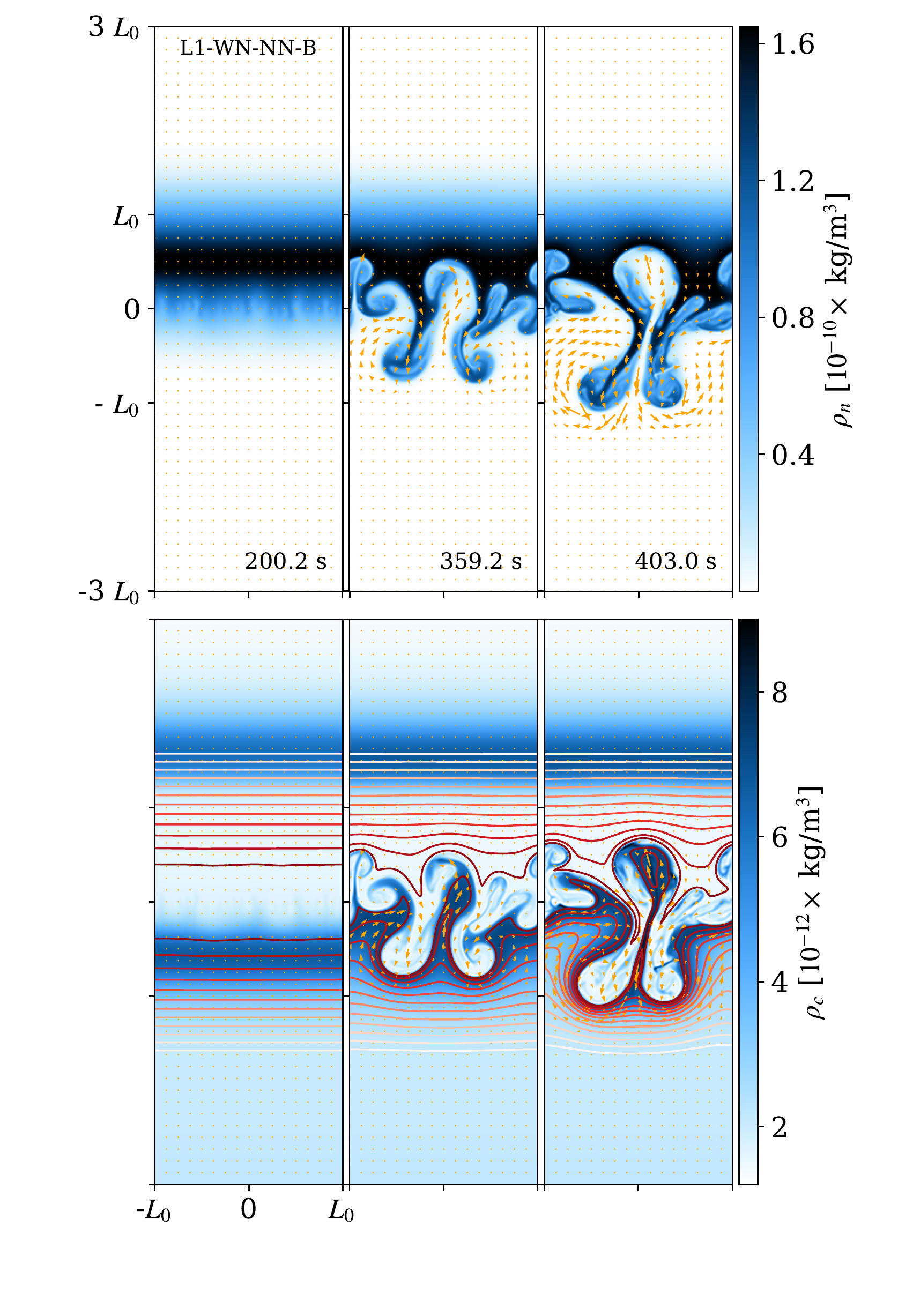}
 \includegraphics[width=8cm,height=11cm]{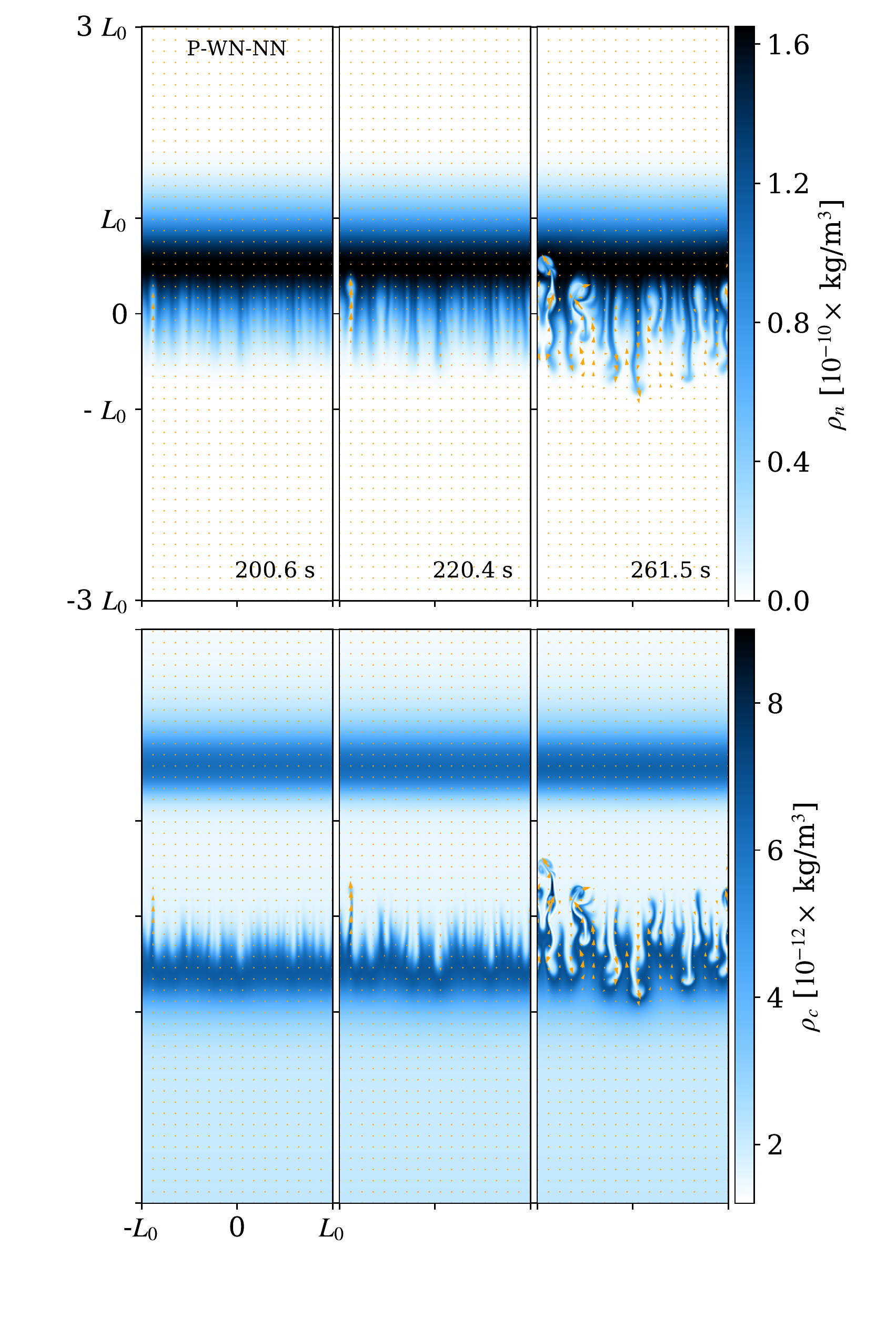}
\caption{Time series of the snapshots of neutral and charged densities for the simulations L1-WN (top left group of panels), L1-WN-NN (top right group of panels), L1-WN-NN-B (bottom left group of panels), P-WN-NN (bottom right group of panels). The in plane ($x$ and $z$ components) velocities of neutrals and charged particles  are represented by orange arrows in the plots which show the densities of neutrals and charged particles, respectively. For the simulations with sheared magnetic field (L1-WN, L1-WN-NN, L1-WN-NN-B), the iso-contours of the absolute value of the $y$-component of the magnetic field potential, $|A_{\rm y}|$, are shown at the bottom panels. The iso-contours are equally spaced between 0.6$A_{\rm y}^{\rm max}$ and $A_{\rm y}^{\rm max}$, where $A_{\rm y}^{\rm max}$ is the maximum value of $|A_{\rm y}|$.}
\label{fig:time_snaps2}
\end{figure*}

\begin{figure*}[]
 \centering
 \includegraphics[width=16cm]{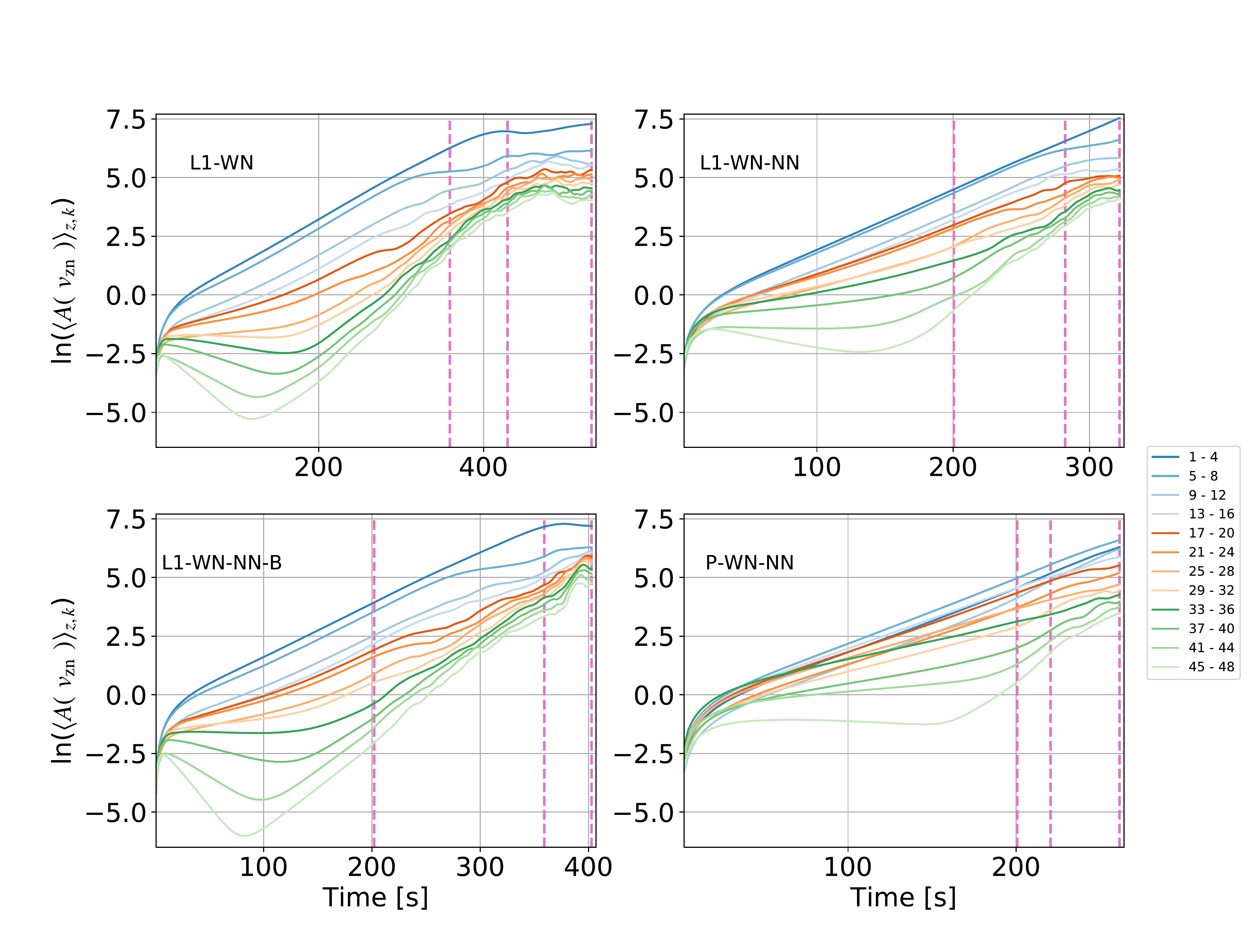}
\caption{Temporal evolution of individual harmonics, computed from the vertical velocity of neutrals, as a function of time. Different color lines show the Fourier amplitudes of the harmonics, grouped in sets of four, from $n=1$ to $n=48$. The amplitudes are averaged between heights -$L_0$ and $L_0$. Different panels are for different simulations: L1-WN (top left); L1-WN-NN (top right); L1-WM-NN-B (bottom left); P-WN-NN (bottom right).The vertical dashed pink lines mark the times of the snapshots shown in Figure~\ref{fig:time_snaps2}.}
\label{fig_group_growing2}
\end{figure*}

Figure~\ref{fig:time_snaps2} shows time series of snapshots of the four simulations, L1-WN, L1-WN-NN, L1-WN-NN-B, and P-WN-NN. This figure illustrates the development of the instability in cases with different density contrast, magnetic field strength and orientation. The moments where the snapshots are shown to correspond to times indicated in the following Figure~\ref{fig_group_growing2} and taken at approximately similar evolution stages of the RTI in each case.  The three simulations with a sheared magnetic field (L1-WN, L1-WN-NN, L1-WN-NN-B) have been initialized with exactly the same perturbation, so that it is easy to compare them. The velocities of neutrals and charged particles are indicated by arrows in the panels of the corresponding densities. For the simulations with the sheared field, the bottom panels for each simulation show the in-plane magnetic field lines plotted as iso-contours of the absolute value of the  $y$-component of the magnetic potential  obtained by numerical integration.
The iso-contour lines span uniformly the range between 0.6$A_{\rm y}^{\rm max}$ and $A_{\rm y}^{\rm max}$, where $A_{\rm y}^{\rm max}$ is the maximum value of $|A_{\rm y}|$ and they are darker for larger values.

By comparing the snapshots of L1-WN to the snapshots of L1-WN-NN, we can observe that small-scale structures are more abundant when the density contrast is increased. Increasing the magnitude of the magnetic field has the opposite effect. As the density of the neutrals increases with height across the interface, and that of charged species  decreases, the spikes contain mostly neutral fluid and the bubbles contain mainly charged particles. We observe this effect in the density contrast between charged particles and neutrals in the snapshots for all the simulations in Figure~\ref{fig:time_snaps2}. The collisional coupling is strong and the neutrals drag the charged particles  during the development of the instability.  

In order to see how the changes in the equilibrium parameters influence the development of different scales in the system, we measure the time evolution of the horizontal modes averaged over the vertical extent of the unstable region.  We compute the amplitudes of the modes as the amplitudes of the Fourier coefficients obtained by performing a FFT of a given quantity (for example vertical velocity of neutrals or charged particles ), in the $x$ direction, at each time moment. We then average the amplitudes in the vertical direction between $z=-L_0$ and $z=L_0$, and over four consecutive modes.  Figure~\ref{fig_group_growing2} shows the natural logarithm of the Fourier amplitudes for the first 48 modes. The average over four modes and the limit of 48 modes are chosen for clarity of presentation.

Noting that the time extent of the horizontal axes in all the panels of Figure~\ref{fig_group_growing2} is different, by comparing top-left (L1-WN) and top-right panels (L1-WN-NN) we observe that the increase in the density contrast increases the growth rate of all modes. 
By comparing top-right (L1-WN-NN) and bottom-right (P-WN-NN) panels, we observe that the magnetic field shear has a stabilizing effect; and by comparing top-right (L1-WN-NN) and bottom-left panel (L1-WN-NN-B), we observe that, in the presence of shear, the increase in the magnetic field magnitude similarly decreases the growth rate of all modes.

In all cases shown in Figure~\ref{fig_group_growing2}, the modes corresponding to the largest scales (modes 1--8) have a clear linear phase. Small scales (modes 45--48) do not have an initial linear growth for any of the four simulations. As discussed further in Section \ref{ssec:diss_modes} below, these are the modes affected by the viscosity and ion-neutral collisions.
The largest mode number $n$ that grows in the linear phase in the enhanced density case, L1-WN-NN (top right), is $n=33$, while for the original density case, L1-WN (top left), it drops to $n=25$. Increasing the magnetic field causes an opposite effect to increasing the density contrast, the largest $n$ growing in the linear phase becomes $n=29$. We note that in all cases, at some point in the simulation, the smallest scales begin to grow. This reversal marks the end of the linear phase and the beginning of the non-linear phase. 
The small scales involved, which are suppressed during the linear phase by the dissipation effects, are then driven by the energy cascade from low $n$ to high $n$ in the nonlinear phase.

\section{Analytical approach} \label{sec:an_gr_1f}



We now explore the expected linear properties of the RTI for the given equilibrium analytically in order to better understand and interpret the simulations, as well as to identify the impacts of the collisional effects on the development of the instability.  The following derivations use a single-fluid ideal incompressible MHD approximation of the plasma.


We note that while the derivation below does not take into account the effects of compressibility, it has been shown that compressibility can have a destabilizing effect \citep{1961Vander, 1974Bhatia, 1983Newcomb, 2005Ribeyre,2010Xue}.
The analytical calculation in the hydrodynamic case for our density profile (not presented in this study) also shows that the compressibility has a destabilizing effect, which is larger for smaller Atwood numbers, similarly to the conclusion of \cite{2005Ribeyre}.


Under the assumption of incompressibility, substituting the zero-divergence condition on velocity for the energy equation, the equations become:

\begin{eqnarray} \label{eq:eqs_rti}
\frac{\partial \rho}{\partial t} + \mathbf{\nabla}\cdot \left(\rho\mathbf{u}\right) =  0, \nonumber  \\ 
\mathbf{\nabla}\cdot\mathbf{u} = 0, \nonumber  \\
\frac{\partial (\rho\mathbf{u})}{\partial t} + \mathbf{\nabla}\cdot (\rho\mathbf{u} \mathbf{u} +p)  = \mathbf{J}\times\mathbf{B} + \rho\mathbf{g},  \nonumber \\
\frac{\partial\mathbf{B}}{\partial t}  -  \mathbf{\nabla}\times (\mathbf{u}\times\mathbf{B}) = 0.
\end{eqnarray}

{ We consider a 2.5D case, where there are no variations in the $y$-direction, the equilibrium atmosphere is homogeneous in the $x$ direction but not uniform in the $z$ direction, and $B_{z0}=0$.  We use the ansatz of separable solutions with Fourier decomposition of the dependent variables in time and in the $x$ spatial direction, looking for solutions of the form:

\begin{eqnarray} \label{eq:sol_ch6}
\{ \tilde \rho_1, \tilde p_1, \tilde v_{\rm x}, \tilde v_{\rm y}, \tilde  v_{\rm z}, 
\tilde  B_{\rm x1}, \tilde  B_{\rm y1}, \tilde  B_{\rm z1}  \} =  
\{  \rho_1(z),  p_1(z),  \nonumber \\ v_{\rm x}(z),  v_{\rm y}(z),   v_{\rm z}(z), 
 B_{\rm x1}(z),   B_{\rm y1}(z),  B_{\rm z1}(z)  \}  \times \nonumber \\
{\rm exp}\big(i\omega t - i k_{\rm x} x \big)\,,
\end{eqnarray}
with the general solution in the linear regime being a superposition of such single-mode solutions.
Linearizing Eqs.~(\ref{eq:eqs_rti}) and substituting  perturbations of the form defined in Eq.~(\ref{eq:sol_ch6}), we have:
%
\begin{align} \label{eq:eqs_rti_p2}
& \rho_1 = -\frac{1}{i \omega} \frac{\text{d} \rho_0}{\text{d} z} v_{\rm z} , \nonumber  \\
& -i k_{\rm x} v_{\rm x} + \frac{\text{d} v_{\rm z}}{\text{d} z} = 0 , \nonumber  \\
& i \omega \rho_0 v_{\rm x} = i k_{\rm x} p_1 + \frac{1}{\mu_0} \left( i k_{\rm x} B_{\rm y1} B_{\rm y0} +  
  \frac{\text{d} B_{\rm x0}}{\text{d} z} B_{\rm z1} \right) , \nonumber  \\
& i \omega \rho_0 v_{\rm y} = \frac{1}{\mu_0} \left( - i k_{\rm x} B_{\rm y1} B_{\rm x0} +  
  \frac{\text{d} B_{\rm y0}}{\text{d} z} B_{\rm z1} \right) , \nonumber  \\
& i \omega \rho_0 v_{\rm z} = -\rho_1  g -  \frac{\text{d} p_1}{\text{d} z} -  \frac{1}{\mu_0} \left[ \frac{\text{d} (B_{\rm x1} B_{\rm x0})}{\text{d} z} + \right.\nonumber \\
&\quad \quad  \left. + \frac{\text{d} ( B_{\rm y1} B_{\rm y0})}{\text{d} z} + i k_{\rm x} B_{\rm z1} B_{\rm x0} \right]  , \nonumber  \\
& B_{\rm x1} = - \frac{1}{i \omega} \frac{\text{d} (v_{\rm z} B_{\rm x0})}{\text{d} z}  , \nonumber  \\
&B_{\rm y1} = - \frac{1}{i \omega}  \left(i k_{\rm x} v_{\rm y} B_{\rm x0} + v_{\rm z}  \frac{\text{d} B_{\rm y0}}{\text{d} z} \right) , \nonumber  \\
& B_{\rm z1} = -\frac{1}{\omega} k_{\rm x} v_{\rm z} B_{\rm x0}  \,. \quad
\end{align}
After manipulating the above we obtain a second order differential equation for $v_{\rm z}$:
\begin{equation} \label{eq:rti_an_gen}
\frac{\text{d}}{\text{d} z}\left(a \frac{\text{d} v_{\rm z}}{\text{d} z}\right) -
k_{\rm x}^2 \left(a - g \frac{\text{d} \rho_0}{\text{d} z} \right) v_{\rm z} = 0,
\end{equation}
where $a$ is defined as:
\begin{equation} \label{eq:rti_an_gen_a}
a =  \frac{1}{\mu_0} \left( k_{\rm x} B_{\rm x0} \right)^2 - \omega^2 \rho_0\,,
\end{equation}
}
{
\noindent where $\rho_0$ is the total density, calculated from our two-fluid model as:
\begin{equation} \label{eq:density}
\rho_0 = (n_{\rm i0} + n_{\rm n0}) m_H\,,
\end{equation}
where $n_{\rm i0}$ and $n_{\rm n0}$ have been defined in Eqs.~(\ref{eqs:rti_setup_ni}) and (\ref{eqs:rti_setup_nn}), respectively.
}
\noindent When $B_{\rm x0}=B_{\rm y0}=0$, we recover Eq.~42 from \cite{Ch1961}.
In the hydrodynamic case, neglecting the spatial derivatives in Eq.~\ref{eq:rti_an_gen} recovers the Brunt-V\"ais\"al\"a frequency in the incompressible limit.

Equation~(\ref{eq:rti_an_gen}) does not have a general known analytical solution for generic continuous density and magnetic field profiles.
{ In order to obtain an approximate solution,  Eq.~(\ref{eq:rti_an_gen}) is numerically solved as a generalized eigenvalue problem with fixed boundary conditions,  $v_{\rm z}(-4 L_0)=v_{\rm z}(4 L_0)=0$.  In doing so, we look for the fastest growing mode for a given $k_x$.}
We refer to this method as semi-analytical and use it for exploring the linear properties of RTI in our equilibrium in Section~\ref{sec:gr}.


\subsection{Analytical expressions for the growth rate} \label{ssec:an_gr}

The solution for the growth rate above is obtained semi-analytically by solving a general eigenvalue problem. However, this method does not give an analytical expression for the growth rate. We can estimate the growth rate by making further assumptions in some limiting cases.

\subsubsection{Discontinuous density profile} \label{sssec:an_gr_discont}

\begin{figure}[!htb]
 \includegraphics[width=8.5cm]{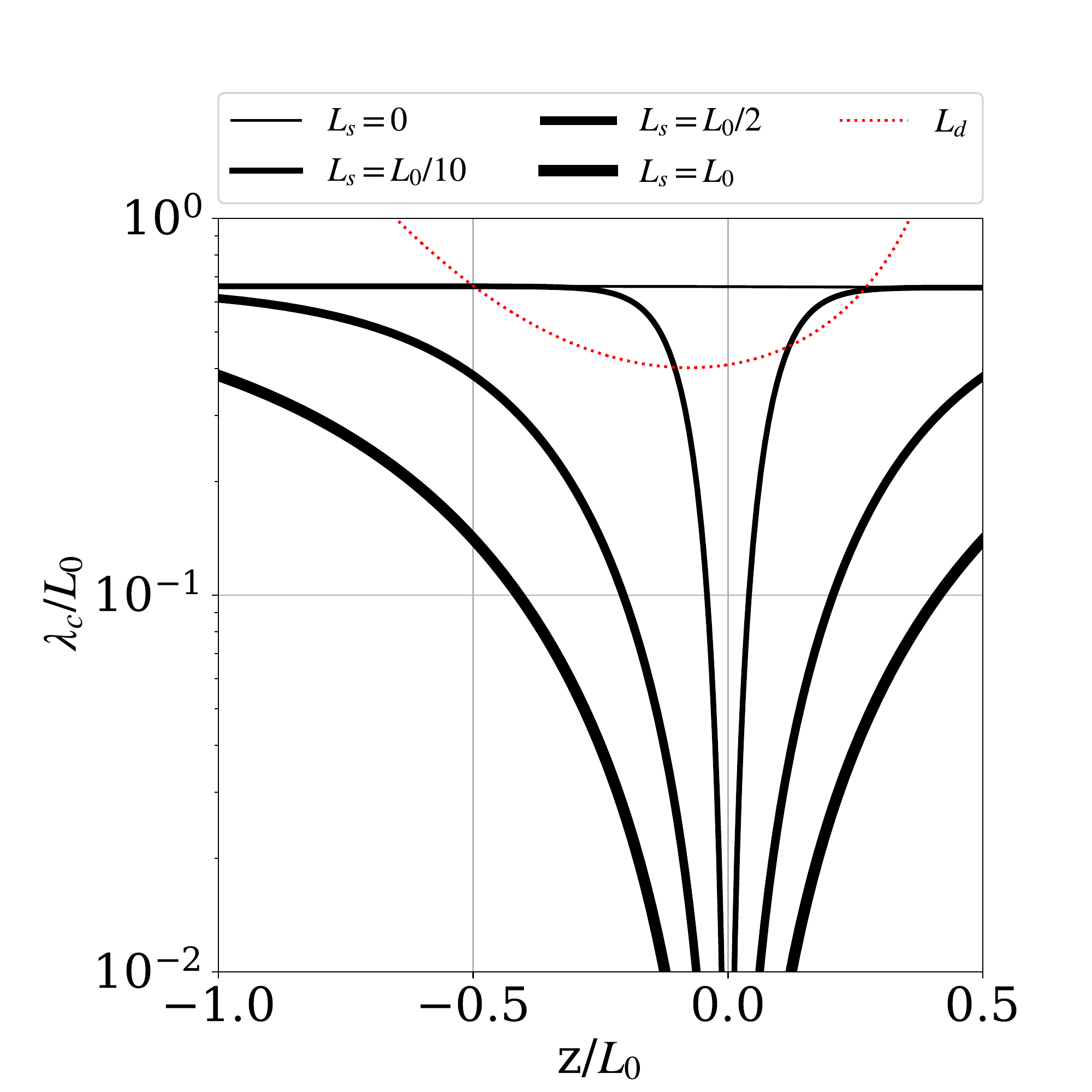}
\caption{
Cutoff wavelengths (black lines) due to the in-plane component of the magnetic field as a function of height, computed from Eq.~(\ref{eq:lc})  for the equilibrium profile  with the values of $n_{\rm n00}=10^{16}$ m$^{-3}$ and $B_{00}=10^{-3}$ T in Eqs.~(\ref{eqs:rti_setup_nn}) and (\ref{eqs:rti_setup_B}). 
Cutoff wavelengths for the following four shear lengths: $L_s=[0, L_0/10, L_0/2, L_0]$ are shown with increasing width of the line for increasing $L_s$. The red dotted line corresponds to the density gradient scale length $L_{\rm d}$. 
}
\label{fig_equi_lc}
\end{figure}

In our case, the transition between the prominence and the corona is smooth, and the minimum of the density gradient scale length, $L_{\rm d}=\rho_0 \big/\frac{\partial \rho_0}{\partial z}$ is non-zero.  However, { for modes with $k_x L_{\rm d}/2\pi \lesssim 1$,} we can approximate our vertical profile as a discontinuous profile, using values of $\rho_2 = \rho_0(z=L_0/2) \approx 1.83 \times 10^{-11}$ kg/m$^3$, and $\rho_1 = \rho_0(z=-L_0/2) \approx 2.93 \times 10^{-12}$ kg/m$^3$, with $\rho_1$ and $\rho_2$ as in Eq.~(\ref{eq:atwood}). 

It can be shown \citep[see][]{Ch1961} that in a magnetized plasma with discontinuous density profile and a{ uniform} magnetic field component parallel to the direction of perturbation, considered here to be the $x$ direction (see Figure~\ref{fig:Bfield}), the growth rate becomes:
\begin{equation}
-\omega^2 = A g k_x - \frac{2 B_{\rm x0}^2 k_x^2}{\mu_0   \Big(\rho_2+\rho_1\Big)}\,.
\label{eq:disc_growth_rate}
\end{equation}
The condition for stability $-\omega^2 <0$ gives the expression for the critical wavelength $\lambda_c$, so that the scales with wavelength { $\lambda  = 2\pi/k_x < \lambda_c$ do not grow:
\begin{equation}
\lambda_c = \frac{2 \pi (v^x_A)^2}{A g}\,,
\label{eq:lc}
\end{equation}
 where $v^x_A \equiv B_{\rm x0}/\sqrt{\mu_0(\rho_2+\rho_1)/2}$ is the characteristic in-plane  Alfv\'{e}n speed at the interface. Thus, Eq.~(\ref{eq:disc_growth_rate}) can be used to estimate the expected RTI growth rate for modes with wavelength $\lambda > {\rm max}(L_{\rm d},\lambda_c)$.}

Figure~\ref{fig_equi_lc} shows the cutoff wavelengths $\lambda_c$, defined in Eq.~(\ref{eq:lc}), as a function of height calculated using the values of $\rho_1$ and $\rho_2$ specified above for different magnetic field shear length scales,  $L_{\rm s}=[L_0/10, L_0/2, L_0]$. { These values provide a visual representation of the local magnetic field stabilization effect.} The density gradient scale length (red dotted line), attains its minimum of $L_{\rm d} \approx$ 409 km at z $\approx$ 0, and corresponds to the mode number $n_{\rm d} =  L_{\rm x}/(2\pi L_{\rm d}) \approx$ 1. The density gradient becomes negative in the region $z>L_0/2$. We observe that for RTI modes with $\lambda > L_{\rm d}$ that  can be treated using the approximation of a discontinuous interface, the horizontal magnetic field may be expected to begin to stabilize such modes for shear scales of $L_{\rm s} \approx L_0/10$ and smaller.



\subsubsection{Exponential density profile} \label{sssec:an_gr_exp}

\begin{figure}
 \includegraphics[width=8.5cm]{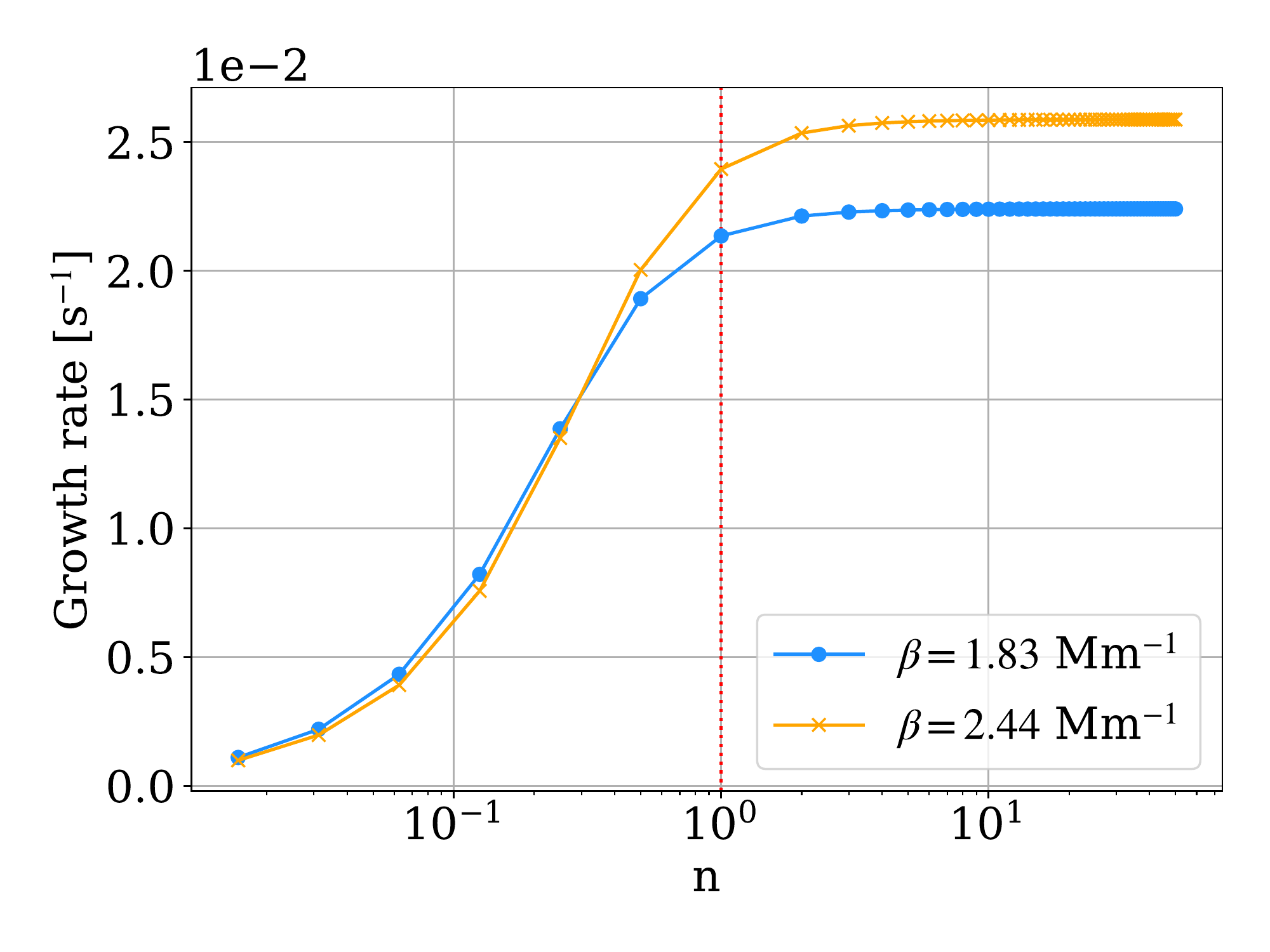}
\caption{Growth rate obtained from the exact solution for the exponential density profile, Eq.~(\ref{eq:gr_an_exp}) with uniform $\beta = 1.83 \times 10^{-6}$ m$^{-1}$ (blue curve marked with ``o'') and $\beta = 2.44 \times 10^{-6}$ m$^{-1}$ (orange curve marked with ``x''). The values of n are between $n=1/64$ and $n=50$.}
\label{fig:gr_an_cha}
\end{figure}


Another limiting case is the case of an exponential density profile, $\rho_0 \propto \text{exp} (\beta z),$ with a constant, $\beta > 0$, { extended over height, $d$. In this case, the growth rate can be calculated exactly \citep{Ch1961} with $\omega(k)$ given by Eq.~(\ref{eq:gr_an_exp}) and the vertical extent of the mode $v_z(z)$ given by:
\begin{equation} \label{eq:eig_ch}
v_z(z) \propto \exp\left(-\frac{1}{2} \beta z \right)\text{sin}\left(\frac{m \pi z}{d} \right).
\end{equation}}
Using the same heavy and light fluid density values as in the sub-section above, $\rho_2=1.83\times 10^{-11}$ kg/m$^3$ and $\rho_1=2.93 \times 10^{-12}$ kg/m$^3$; we match them to an exponential density profile $\rho_0 \propto \text{exp}(\beta z)$, for $z$ between -$L_0/2$ and $L_0/2$, with constant $\beta$.  We obtain $\beta = \ln{(\rho_2/\rho_1)}/L_0 = 1.83 \times 10^{-6}$ m$^{-1}$. { Alternatively, we can calculate  $\beta$ from the local density gradient scale length at $z=0$ and we obtain a value of $\beta = 2.44 \times 10^{-6}$ m$^{-1}$.} Figure~\ref{fig:gr_an_cha} shows the dispersion relations obtained by substituting the two values of $\beta$ and $d=8 L_0$, corresponding to the vertical extent of the computational domain, into Eq.~(\ref{eq:gr_an_exp}).  We note that for long wavelength modes with $\lambda\beta \gtrsim 1,$ the growth rate begins to decrease with decreasing $n$ due to the nature of an exponentially stratified density profile and the finite vertical extent of the modes.  However, for modes with wavelength much smaller than both the density gradient scale length and the vertical extent of the domain, the growth rate becomes independent of $n$.

The growth rate calculations with an exponential density profile give a rather satisfactory result compared to the semi-analytical calculations presented below in Section~\ref{sec:gr}.  On small scales, for $\beta = 2.44 \times 10^{-6}$ m$^{-1}$, the value of the growth rate in the high $n$ limit is the Brunt-V\"ais\"al\"a frequency $\approx 2.58 \times 10^{-2} s^{-1}$.

While the analytical approximation of the dispersion relation given by Eq.~(\ref{eq:gr_an_exp}) appears to be a good match to that obtained for the equilibrium configuration considered here, that does not imply that the RTI eigenmodes given by Eq.~(\ref{eq:eig_ch}) are also the eigenmodes for an equilibrium with a vertically limited extent of the RTI-unstable density profile.  In particular, we note that the functional form of Eq.~(\ref{eq:eig_ch}) for the fastest growing { $m=1$} mode specifies $v_z(z)$ with non-negligible magnitude over the full extent of the domain and shows no dependence on $k_x$. At the same time, it is easy to show from Eq.~(\ref{eq:rti_an_gen}) that for an equilibrium with $B_{x0}=0$:
\begin{eqnarray} \label{eq:eig_limited}
    &&\omega^2\left[\int_{-4L_0}^{4L_0} \rho_0\left(k_x^2 v_z^2 + \left(\frac{d v_z}{dz}\right)^2\right)dz \right.\\ \nonumber 
    &&\left. - \frac{1}{2}\rho_0\frac{d(v_z^2)}{dz}\biggr\vert_{-4L_0}^{4L_0} \right] = -g k_x^2 \int_{-4L_0}^{4L_0} \frac{d\rho_0}{dz} v_z^2 dz
\end{eqnarray}
and the most unstable eigenmodes $v_z(z)$ will be those with $v_z \approx 0$ in the regions of stable stratification, $(d\rho_0/dz) \leq 0$, corresponding to a maximal value of the right-hand-side integral.  It follows that for $\rho_0$ as given by Eq.~\ref{eq:density}, we can expect the fastest growing RTI modes to have $v_z \approx 0$ for $|z| \gtrsim L_0/2$, and for the functional form of $v_z(z)$ to depend on { $k_x$}.

\subsubsection{Stabilization by sheared magnetic field} \label{sssec:an_gr_Bshear}

As shown in Sec.~\ref{sssec:an_gr_discont} above, for a RTI with $\vec{k}\cdot\vec{B} \neq 0$, magnetic field acts to reduce the growth rate or entirely eliminate the instability due to the magnetic field-line tension counteracting the gravity force driving the instability.  Thus, in magnetized 3D systems with unidirectional B-field, the fastest growing modes will always be those with $\vec{k}\cdot\vec{B} = 0$.  

For more realistic systems where the magnetic field may be sheared, such as in some of the background equilibria considered here, the situation is more complex.  To explore the impact of a sheared field on the growth rate for most unstable RTI modes in a 2D system with $\vec{k}=k_x\hat{x}$, we note that the vertical extent of an RTI eigenmode in a magnetized plasma may be limited by the magnetic field tension away from the height where $B_{x0}=0$.

We now consider spatially varying horizontal magnetic field $B_{x0}(z)$ such that:
\begin{equation} \label{eq:appr1}
B_{\rm x0}(z) = b_0 \frac{z}{L_s}\,,
\end{equation}
where $b_0$ is constant and $L_s$ is the characteristic length of magnetic field shear.  We assume the density profile to be exponentially stratified, $\rho_0 = \rho_{00}\exp(\beta z)$, as in Section~\ref{sssec:an_gr_exp} above.

For sufficiently large $b_0$, the vertical extent of an RTI eigenmode for a given $k_x$ can then be estimated by balancing the gravitational force driving the instability against the magnetic tension in the $z$-momentum equation { (Eqs. \ref{eq:eqs_rti_p2})} and looking for $z=z_d$ where:
\begin{eqnarray} \label{eq:appr2}
&&-\rho_1 g = i k_x \frac{1}{\mu_0} B_{\rm z1} B_{\rm x0} \implies \nonumber\\
&&g \beta \rho_0|_{z=z_d} = k_x^2 \frac{B_{\rm x0}^2|_{z=z_d}}{\mu_0} = k_x^2 \frac{b_0^2}{\mu_0 L_s^2} z_d^2.
\end{eqnarray}
It follows that
\begin{equation} \label{eq:appr3}
\frac{\exp(\beta z_d)}{(\beta z_d)^2} = 
\frac{k_x^2}{g \beta} \frac{b_0^2}{\mu_0 \rho_{00}}\frac{1}{\beta^2 L_s^2},
\end{equation}
and it is easy to show that positive solutions of Eq.~(\ref{eq:appr3}) for $z_d$ only exist when 
\begin{equation} \label{eq:appr4}
\frac{k_x^2}{g \beta} \frac{b_0^2}{\mu_0 \rho_{00}}\frac{1}{\beta^2 L_s^2} > e^2/4,
\end{equation}
implying that the RTI mode is not vertically confined by sheared magnetic field in the long wavelength limit such that Eq.~(\ref{eq:appr4}) is not satisfied.  Equation~(\ref{eq:appr4}) can be approximately evaluated for our equilibrium parameters showing that, for example, for field shear of $L_s = L_0$ modes with $k_x L_0 < 16.3$ are not expected to be confined by the sheared magnetic field. 

In the opposite limit of $(k_x/\beta) \gg 1$, Eq.~(\ref{eq:appr3}) has two roots for $z_d$ with that of $(\beta z_d) \ll 1$ as the one of interest.  In this limit, it follows from Eq.~(\ref{eq:appr3}) that:
\begin{equation} \label{eq:appr5}
z_d^2 \approx \frac{g \beta}{k_x^2} \frac{L_s^2}{b_0^2/(\mu_0 \rho_{00})}
,\end{equation}
implying that the vertical extent of unstable RTI modes in the presence of sheared magnetic field is inversely proportional to $k_x$ for $(k_x/\beta) \gg 1$.  Equation~(\ref{eq:appr5}) can be evaluated for our equilibrium parameters and field shear of $L_s = L_0$ to give $k_x z_d \approx 4.91$, with proportionally smaller values for smaller $L_s$.

We now substitute the expression for $z_d^2$ from Eq.~(\ref{eq:appr5}) in place of $d^2$ in Eq.~(\ref{eq:gr_an_exp}), as that would be the estimated extent of the RTI eigenmode as stabilized by the sheared magnetic field, and take the $(k_x/\beta) \gg 1$ limit, resulting in:
\begin{equation} \label{eq:appr6}
-\omega^2 = g \beta \left(1 + \frac{m^2\pi^2}{g \beta L_s^2 } \frac{b_0^2}{\mu_0\rho_{00}}\right)^{-1},
\end{equation}
with $m=1$ for the fastest growing mode.  

The estimate of sheared field stabilization provided by Eqs.~(\ref{eq:appr2})-(\ref{eq:appr6}) can be further refined by again applying the $\int v_z*[...] dz$ operator to Eq.~(\ref{eq:rti_an_gen}), which is now in the presence of a sheared magnetic field.  In observing that for a mode that is vertically confined by magnetic field tension (or stable density stratification outside of the instability layer), the boundary integral terms can be neglected, we get the following relationship:
\begin{eqnarray} \label{eq:eig_Blimited}
    &&\omega^2\int_{-4L_0}^{4L_0} \rho_0\left[v_z^2 + \frac{1}{k_x^2}\left(\frac{d v_z}{dz}\right)^2\right]dz = \\ \nonumber 
    && \int_{-4L_0}^{4L_0} \left[\left(k_x^2 \frac{B_{x0}^2}{\mu_0} - g \frac{d\rho_0}{dz} \right) v_z^2 + \frac{B_{x0}^2}{\mu_0}\left(\frac{d v_z}{dz}\right)^2\right] dz,
\end{eqnarray}
where the first two integrant terms on the right hand side represent the balance between stabilizing magnetic field tension and destabilizing gravitational stratification discussed above, and the third term on the right hand side represents an additional stabilizing effect of compression of in-plane magnetic field within a vertically confined mode. { We note that this magnetic field compression effect is present even under the assumption of incompressible flow used to derive Eq.~(\ref{eq:rti_an_gen}), as incompressible vortex flow across the field can close via flow along the in-plane component of magnetic field thus locally compressing the field without compressing the fluid.  We also note that this effect is independent of $k_x$ and is purely stabilizing.}  

An estimate of the importance of the magnetic field compression effect for magnetic field configurations with different shear scales on a mode with wave number $n = (k_x L_0)/\pi$ can be obtained by assuming mode's vertical extent to be $z_d$ given by Eq.~(\ref{eq:appr5}) and approximating $|d v_z/dz| \approx v_z / z_d$ while assuming $v_z(z) \approx \tilde{v}_{z}$ of some arbitrary constant magnitude for $z\in (-z_d,z_d)$, and negligible otherwise.  Further approximating $(d \rho_z/dz) \approx \beta \rho_{00}$ for $z\in (-z_d,z_d)$ and substituting for $B_{x0}$ from Eq.~\ref{eq:appr1}, under these assumptions, the right hand side of Eq.~\ref{eq:eig_Blimited} can be approximated as:
\begin{eqnarray} \label{eq:eig_Blimited2}
      && \int_{-4L_0}^{4L_0} \left[\left(k_x^2 \frac{B_{x0}^2}{\mu_0} - g \frac{d\rho_0}{dz} \right) v_z^2 + \frac{B_{x0}^2}{\mu_0}\left(\frac{d v_z}{dz}\right)^2\right] dz \nonumber \\
      && \approx 2z_d \tilde{v}_{z}^2\left[\frac{k_x^2 z_d^2}{3 L_s^2}\frac{b_{0}^2}{\mu_0} - g \beta \rho_{00} + \frac{1}{3 L_s^2}\frac{b_{0}^2}{\mu_0}\right] \nonumber \\
      && = 2z_d \tilde{v}_{z}^2\left[-\frac{2}{3} g \beta \rho_{00} + \frac{1}{3 L_s^2}\frac{b_{0}^2}{\mu_0}\right]
,\end{eqnarray}
indicating that such a mode is stable for any field shear length $L_s < \sqrt{[b_0^2/(\mu_0 \rho_{00})] / (2 g \beta)}$, with the stability criterion independent of $k_x$ as long as $(k_x/\beta) \gg 1,$ as assumed above.  For the equilibrium configuration used in this work, this stability criterion is $L_s < 0.14 L_0$.

The above derivation shows that presence of sheared magnetic field with finite shear length can stabilize RTI for sufficiently small shear length, but the stabilization effect is largely independent of the wavelength of the mode as long as the magnetic field is strong enough to confine the mode within the RTI-unstable layer.  On the other hand, unlike the derivation in Sec.~\ref{sssec:an_gr_discont} for uniform $B_{x0}$, if the magnetic field shear length is sufficiently large, in the limit of $(k_x/\beta) \gg 1$, the RTI growth rate appears to be independent of $k_x$, with the asymptotic value reduced from the unmagnetized value of $\sqrt{g \beta}$ due to the presence of the magnetic field.

\begin{figure}[!htb]
 \includegraphics[width=8cm]{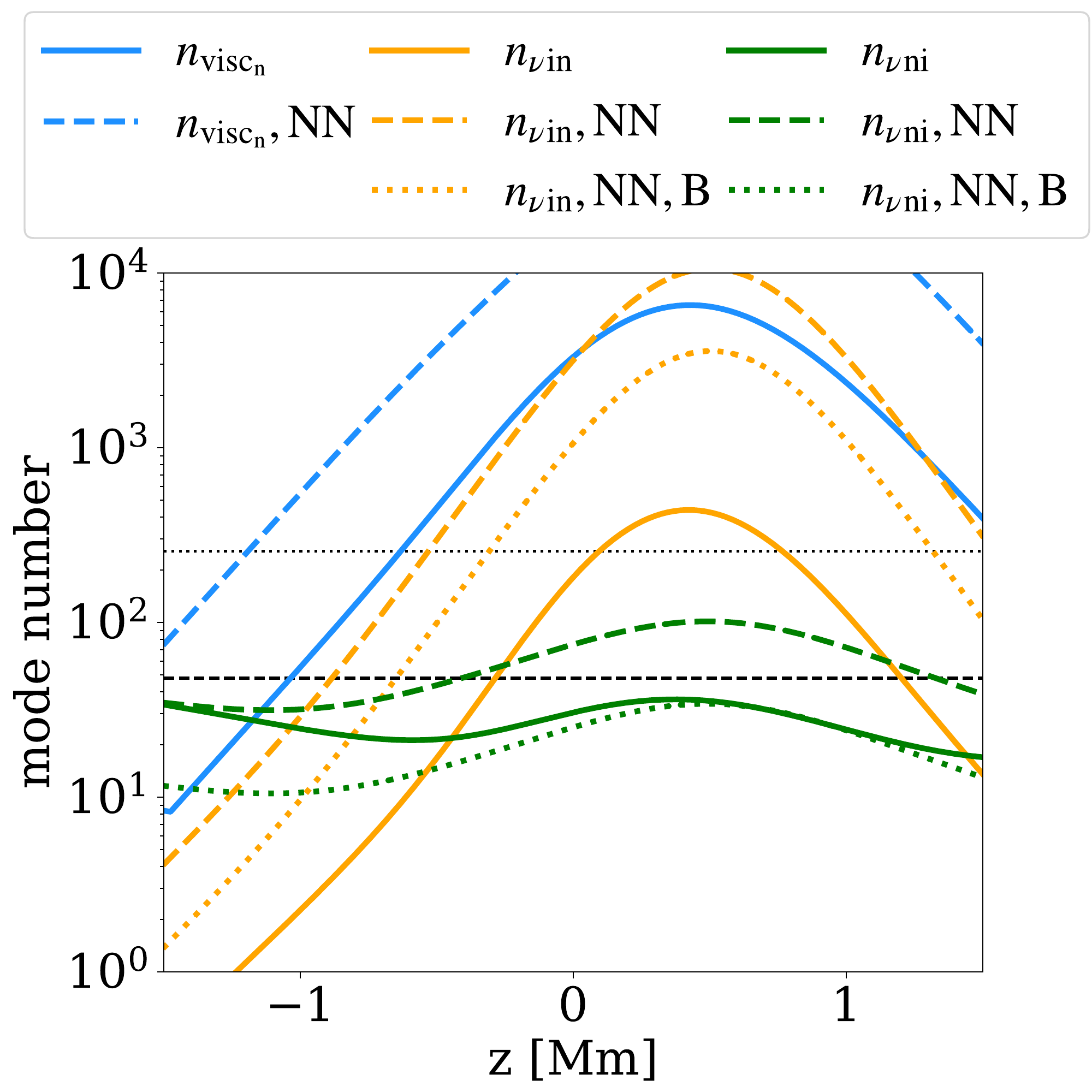}
\caption{Modes corresponding to the neutral viscosity (blue lines), ion-neutral collision scale (orange lines), and neutral-ion collision scale (green lines) as a function of height, computed for the original density profile (solid lines), for the enhanced density profile (dashed lines), and for the enhanced density  and magnetic field profiles (dotted lines),  from Eqs.~(\ref{eq:visc_lengthScales})--(\ref{eq:ni_lengthScales}). The horizontal dashed line marks the maximum mode number shown in Figure~\ref{fig_group_growing2}, $n=48$, and the dotted line represents the largest mode number that can be resolved in the domain, using two points, $n=256$.}
\label{fig:visc_and_dec_scale}
\end{figure}

\subsection{Modes affected by collisional dissipation} \label{ssec:diss_modes}


Dissipation terms affect the growth of linear modes on scales similar to their characteristic length scales. We can evaluate which modes will be affected by collisions by estimating the mode numbers at which the collisional dissipative terms in the corresponding equations become comparable to the dominant convective terms.

As discussed in Section~\ref{ssec:coll_effects}, neutral viscosity is the dominant viscous dissipation mechanism for the plasma parameters considered in this work. Taking the sound speed of the neutrals, $c_{\rm n}$, as the characteristic flow velocity of neutrals and comparing the convective derivative term to the viscous term in the momentum equation for neutrals, the length scale, and mode number associated with the neutral viscosity can be estimated as:
\begin{equation} \label{eq:visc_lengthScales}
L_{\rm visc_n} = \mu^v_{\rm n}/c_{\rm n0}\,, \quad n_{\rm visc_n}=L_{\rm x}/(2\pi L_{\rm visc_n}).
\end{equation}

The length scales associated with the effects of ion-neutral collisions can be similarly estimated in the limit of strong coupling by balancing the magnetic plus thermal pressure gradient force against the inertial terms and the ion-neutral friction term.  Doing so leads to:
\begin{align} \label{eq:ni_lengthScales}
& L_{\rm \nu_{in}} = v_{\rm fast}/\nu_{\rm in}\,,\quad n_{\rm \nu_{in}}=L_{\rm x}/(2\pi L_{\rm \nu_{in}})\,, \nonumber  \\
& L_{\rm \nu_{ni}} = v_{\rm fast}/\nu_{\rm ni}\,,\quad n_{\rm \nu_{ni}}=L_{\rm x}/(2\pi L_{\rm \nu_{ni}}),
\end{align}
where $\nu_{\rm in}$, $\nu_{\rm ni}$, and $v_{\rm fast}$ have been defined in Eq.~(\ref{eq_diff_coef}).

Figure~\ref{fig:visc_and_dec_scale} shows the mode numbers $n$ corresponding to the ion-neutral decoupling scale and to the neutral viscosity scale. 
The modes associated with the scales mentioned above are plotted for the original (solid lines), enhanced density contrast (dashed lines), and for the atmosphere where both the density contrast and the magnetic field are increased (dotted lines).  We limit the horizontal axis between $z=-1.5$~Mm and $z=1.5$~Mm, these values are considered to be the maximum extent of the structures that form during the instability development in our simulations.
The horizontal dashed and dotted lines are $n=48$ and $n=256$, and represent the maximum mode number which appear in Figure~\ref{fig_group_growing2} and the maximum mode number that can exist in the domain (sampled by two points), respectively.

We observe that both the viscosity and the ion-neutral collisions affect modes that can exist in our domain, that is, the modes below the horizontal dotted line.  However, the scales associated with neutral-ion collisions likely provide the strongest damping mechanism.  We note that higher neutral density contrast and, thus, higher peak neutral density, is expected to reduce the impact of collisional dissipation.  We also note that stronger background magnetic field is expected to lead to more damping due to collisions between ions and neutrals.  Both of these effects are particularly consistent with the expectation from damping due to ambipolar resistivity in a single-fluid approximation.



\section{Linear phase of the RTI}\label{sec:gr}

\begin{figure*}[!htb]
 \centering
  \includegraphics[width=8cm]{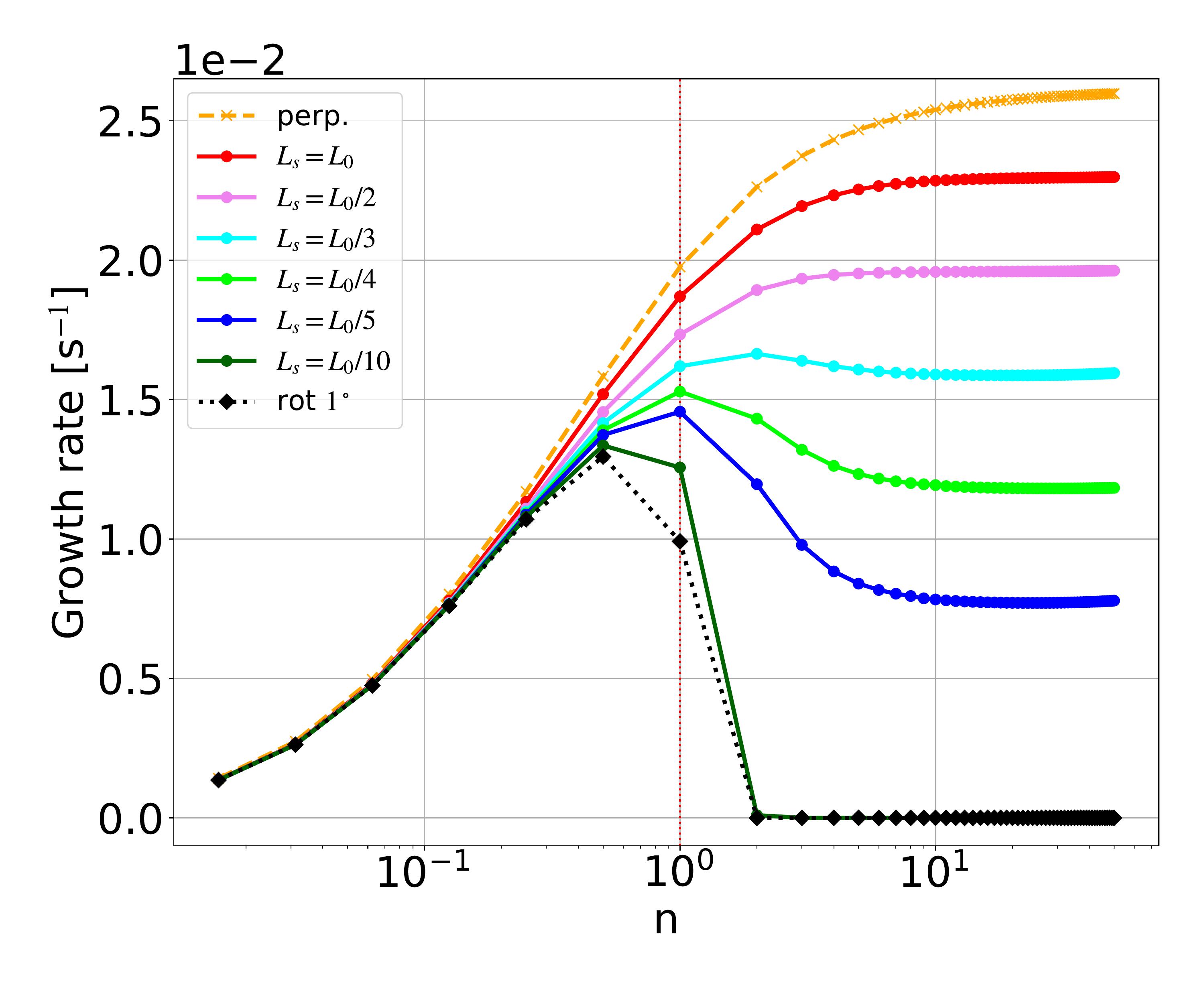}
 \includegraphics[width=8cm]{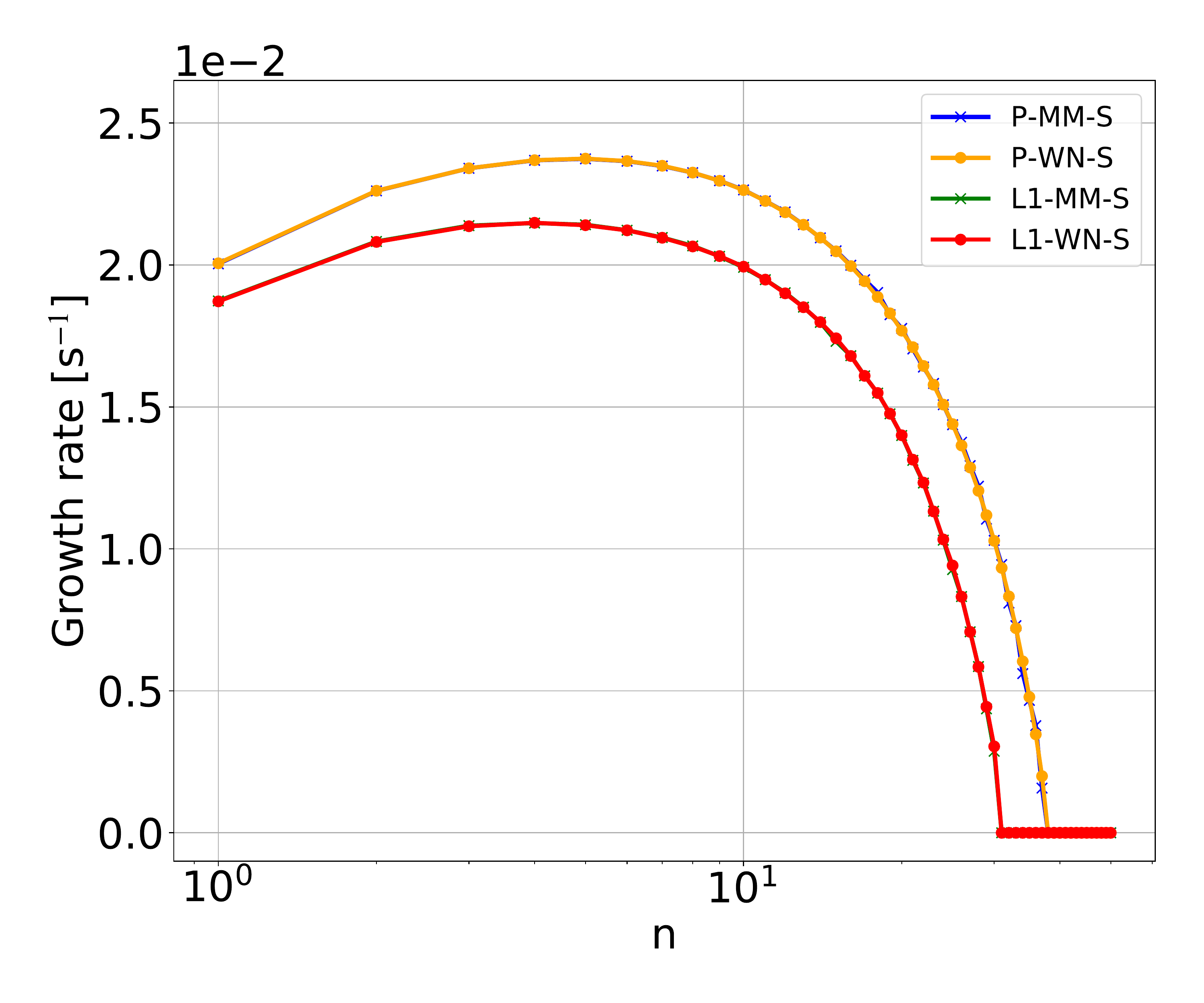}
\caption{Growth rates as a function of the mode number, $n,$ in logarithmic scale up to $n=50$. 
The  first value is $n=1/64$ for the left panel and $n=1$ for the right panel.
Left: Growth rate obtained semi-analytically for the perpendicular magnetic field (orange); sheared field with $L_{\rm s}$ equal to $L_0$ (red), $L_0/2$ (violet), $L_0/3$ (cyan), $L_0/4$ (lime), $L_0/5$ (blue), 
$L_0/10$ (dark green); and for uniformly rotated magnetic field by 1$^\circ$ (black). Right: Growth rate obtained from the simulations P-MM-S (blue), P-WN-S (orange), L1-MM-S (green), L1-WN-S (red).}
\label{fig_gr_comp}
\end{figure*}

\begin{figure*}[!htb]
 \includegraphics[width=8cm]{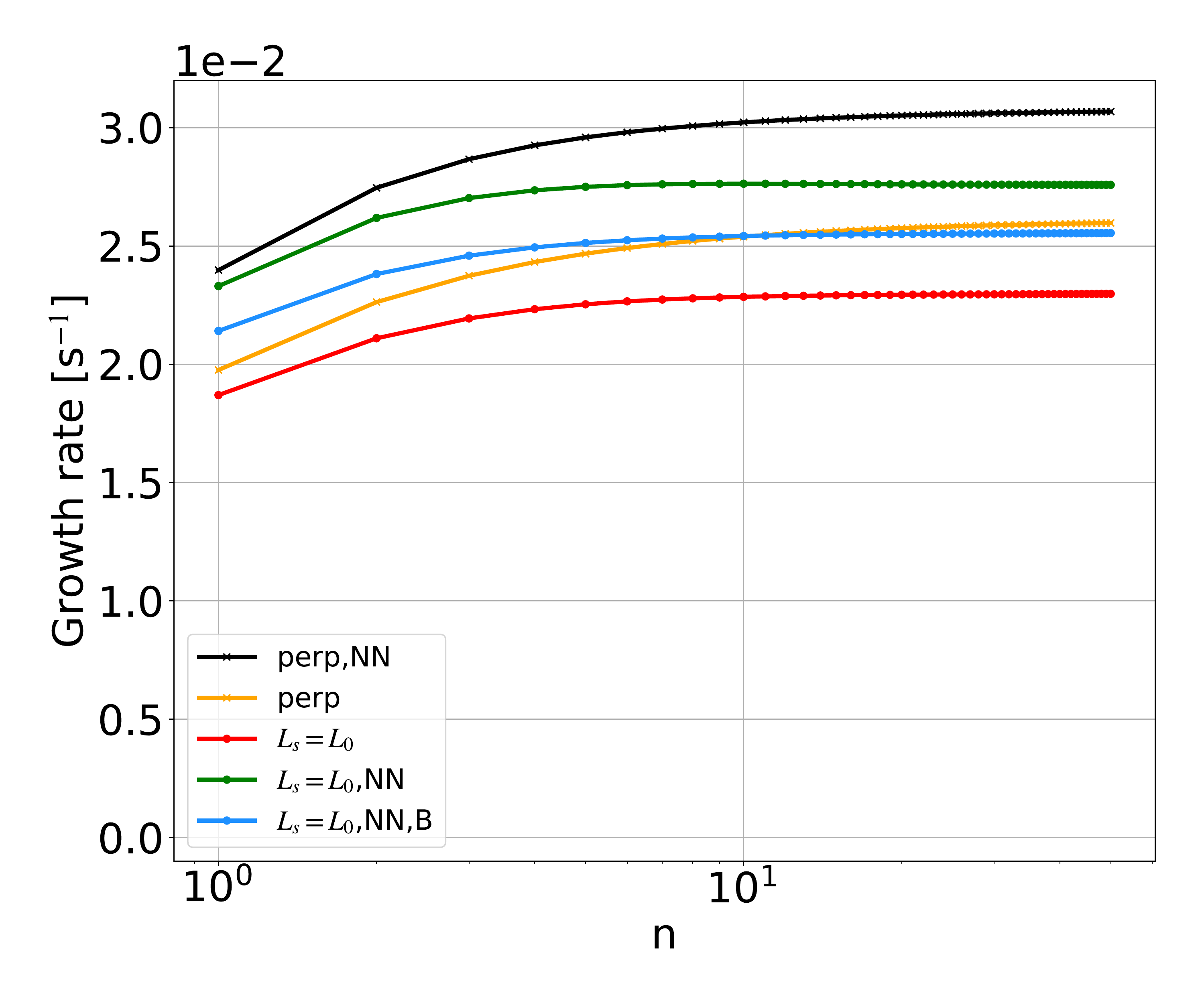}
 \includegraphics[width=8cm]{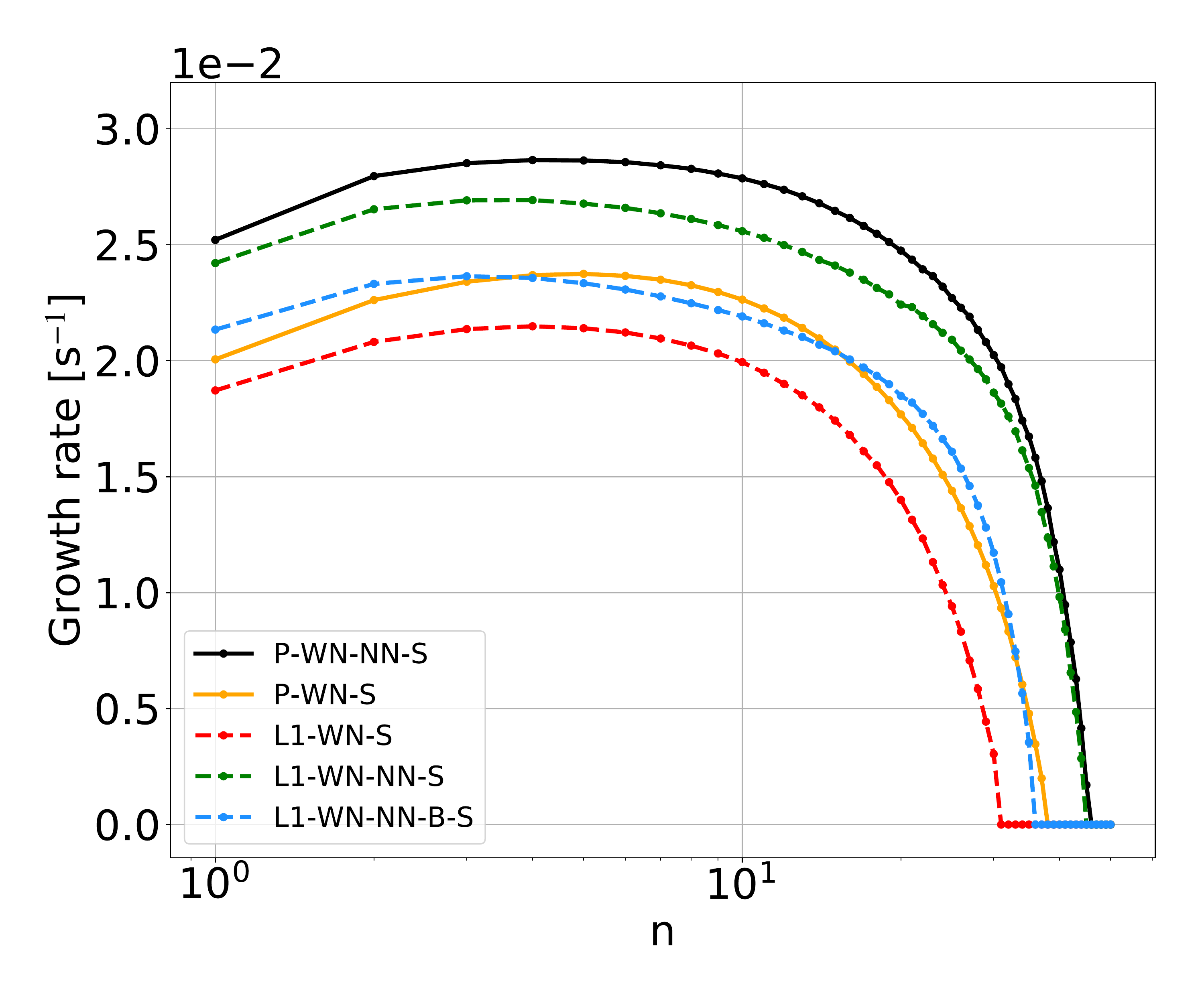}
\caption{Comparison between the growth rates obtained semi-analytically in the incompressible MHD approximation (left) and the growth rates obtained from simulations (right) and  as functions of the mode number, $n,$ in logarithmic scale, limited to $n=50$. In both panels, the perpendicular magnetic field cases are shown with solid lines and sheared case are shown with dashed lines. Orange lines: P-WN-S; black lines: P-WN-NN-S; red lines: L1-WN-S; green lines: L1-WN-NN-S, and blue lines for L1-WN-NN-B-S.}
\label{fig:gr_k_all}
\end{figure*}

\begin{figure*}[!htb]
 \centering
 \includegraphics[width=8cm,height=8cm]{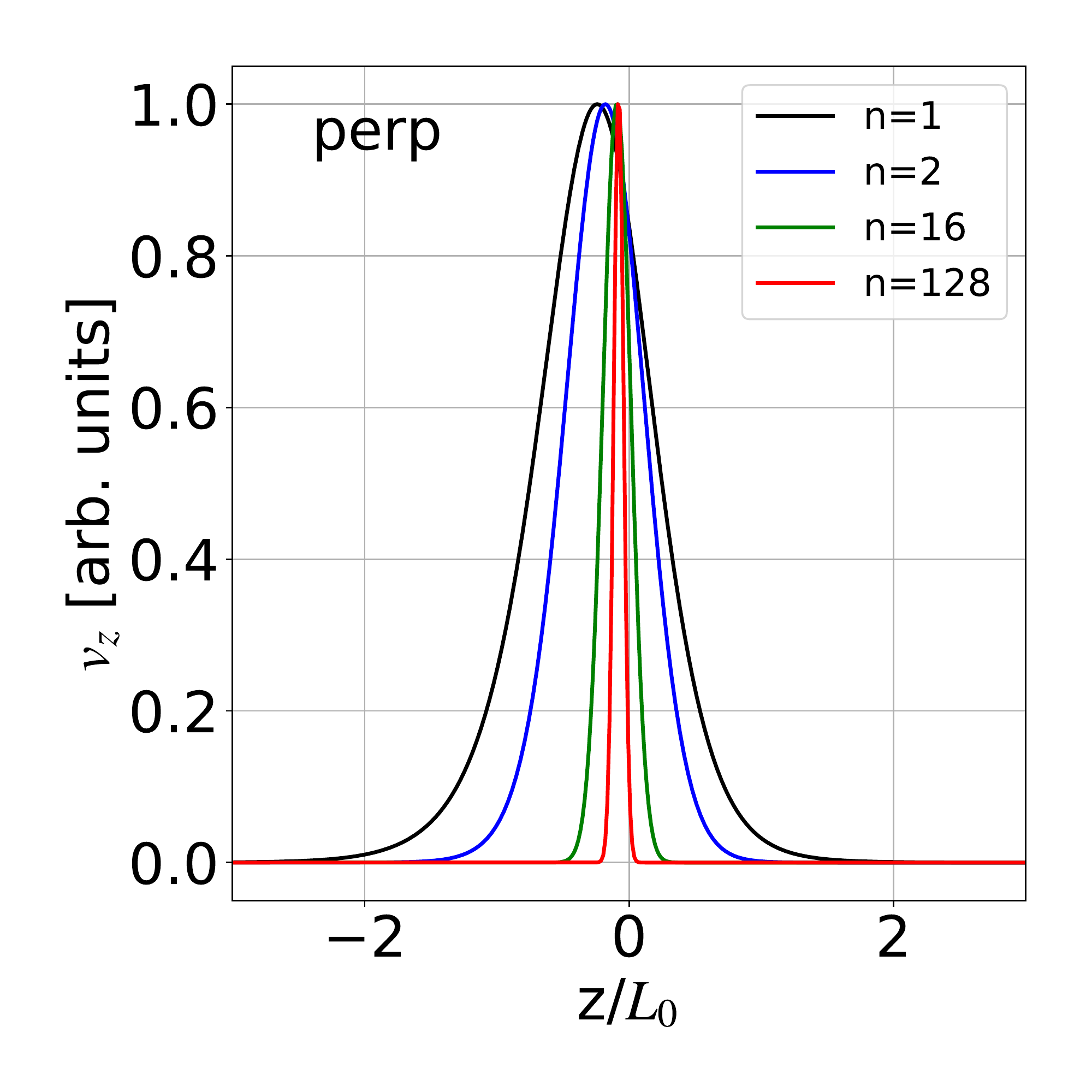}
 \includegraphics[width=8cm,height=8cm]{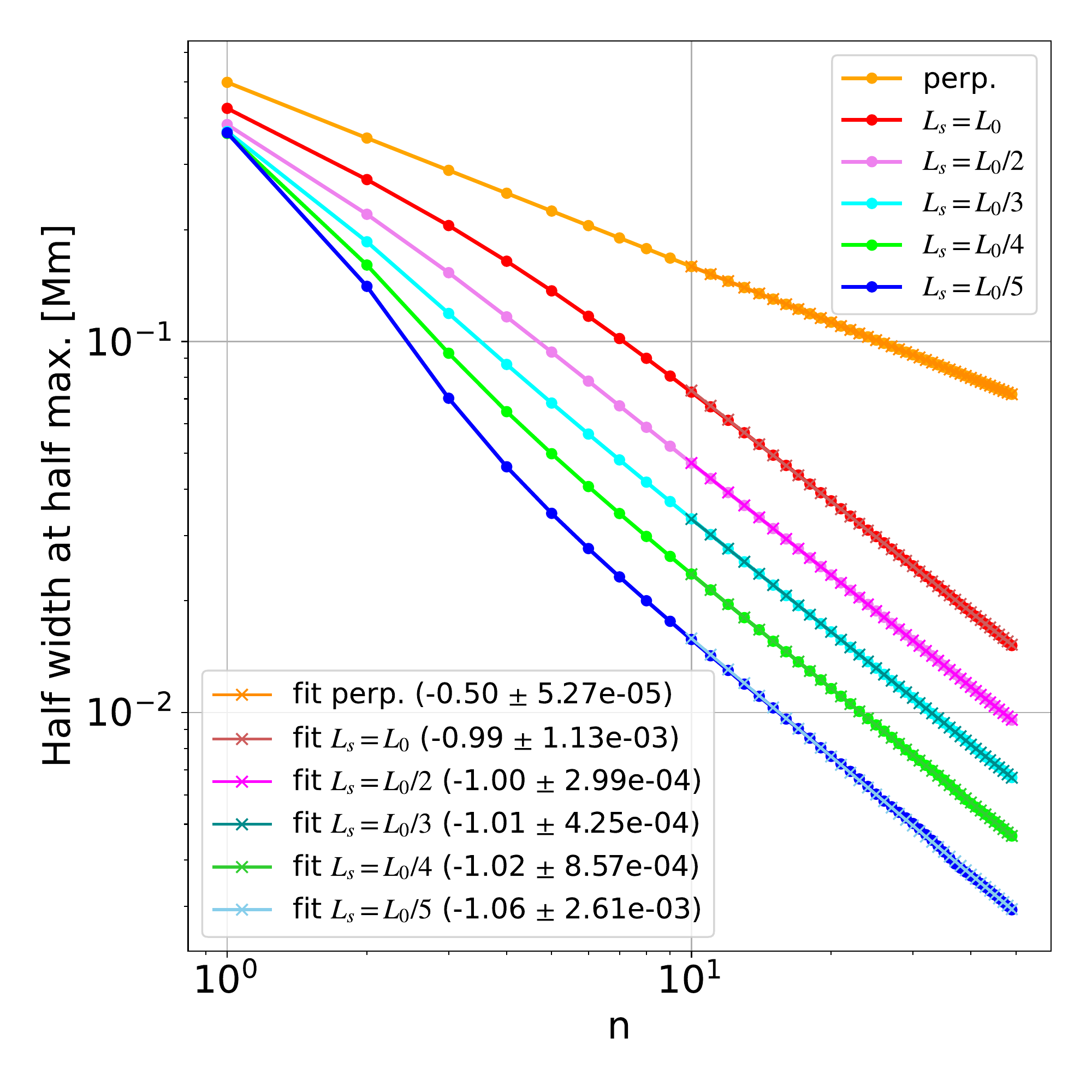}
\caption{Left: Eigenfunctions corresponding to the largest growth rates for the modes $n=1$ (black line), $n=2$ (blue line), $n=16$ (green line), and $n=128$ (red line), normalized to unity, as a function of height for the perpendicular magnetic field configuration. Right: Log-log plot of half width at half maximum, $\Delta$, of the eigenfunctions as a function of the mode number $n$ for the range $1\le n \le 50$ for different configurations of the magnetic field: perpendicular (orange), sheared with shear length equal to: $L_0$ (red), $L_0/2$ (violet), $L_0/3$ (cyan), $L_0/4$ (lime), $L_0/5$ (blue). The slopes of the curves calculated from a linear fit for the modes $10 \le n \le 50$ and the uncertainties of the fit are indicated in the legend.
}
\label{fig_gr_eigen}
\end{figure*}

In this section, we study the linear phase of the RTI. We compare the linear growth rates obtained semi-analytically in the single-fluid incompressible ideal MHD limit to the  { analytical estimates presented in Sec.~\ref{sec:an_gr_1f} and} linear growth rates obtained from the simulations.

The left panel of Figure~\ref{fig_gr_comp} shows growth rates as functions of the wave number obtained semi-analytically for the original density contrast and magnetic field strength, but different magnetic field shear scales, as well as for a uniform magnetic field rotated by 1$^\circ$ with respect to the $y$-axis.  To compare these to the analytical estimated in both long wavelength and short wavelength limits, we show asymptotic behavior for $n \ll 1$, as well as for $n \gg 1$. We observe that for long wavelength modes with $k_x L_0 = \pi n < 1$, RTI growth rates become insensitive to the presence of an in-plane magnetic field and converge towards $\omega = 0$ for $n \to 0$.  This is consistent with lack of magnetic field confinement of the modes for small $k_x L_0$ as shown in Eq.~\ref{eq:appr4}.
{We can compare the growth for the perpendicular case
 (the orange line labeled ``perp'') to the dispersion relations shown in Figure~\ref{fig:gr_an_cha}. We note that the growth rate curves match the analytical estimate well, based on the local density gradient scale $\beta = 2.44 \times 10^{-6}$ m $^{-1}$ giving a particularly good approximation.}

The cases where the equilibrium magnetic field has shear (and, thus, an in-plane component of the field in the domain) show the peak value of the growth rate decrease with decreasing shear scale, $L_{\rm s}$.  This is consistent with the analytical calculation of RTI stabilization by sheared magnetic field described in Sec.~\ref{sssec:an_gr_Bshear} and shown in Fig.~\ref{fig_equi_lc} for a discontinuous density profile.  For $k_x L_0 > 1$, as we consider equilibria with shorter and shorter shear scale, $L_s$, we observe that the growth rate is decreasing with $L_s$ but is independent of $k_x$ for a range of $L_s$ values, and is fully stabilized for $L_s$ between $0.2 L_0$ and $0.1 L_0$, becoming essentially equivalent to the $L_s\to 0$ limit with uniformly rotated magnetic field.  This result is again fully consistent with the analytical derivations presented in Sec.~\ref{sssec:an_gr_Bshear} where the stability criterion was estimated to be $L_s < 0.14 L_0$ due to the stabilization of RTI by a combination of magnetic field tension and compression effects.\

In order to calculate the growth rate from the simulations, we perform the following procedure. First, we re-run the simulations with smaller perturbation amplitudes to ensure a long linear phase, see Table 1. For each simulation, we identify a time interval of the linear phase, usually observed to be between 100~s and 200~s. We then compute the Fourier amplitude of $v_{\rm zn}(x)$ along the $x$ direction at every height, for every snapshot of the series. These Fourier amplitudes were averaged in the interval of heights $z_1 < z < z_2$, defined as the region where the normalized Fourier amplitude is larger than 0.3 by the end of the linear phase.
We thus obtain average amplitudes of the harmonics as a function of time. The growth rate is obtained as a linear fit to the natural logarithm of the amplitudes as a function of time.  

The resulting growth rates of the first 50 modes are plotted in the right panel of Figure~\ref{fig_gr_comp} for two of the equilibrium configurations: "P" with magnetic field perpendicular to the plane of the simulation and "L1" with sheared field with $L_{\rm s} = L_0$. 
To verify the numerical accuracy of the simulations, we use two different types of perturbations and show that the spectrum of linear growth rates of the RTI depends on the equilibrium but is independent of the form of the perturbation. The right panel of Figure~\ref{fig_gr_comp} compares simulations with exactly the same parameters other than the initial perturbations being the white noise "WN" (as used everywhere else in this paper) or the multi-mode perturbation "MM" described by Eq.~(\ref{eq_rx_mm}). The multi-mode and white noise perturbations give growth rate spectra that match almost perfectly for both magnetic field configurations, verifying our numerical procedure.

By comparing the linear growth rates obtained from the simulations to the semi-analytical calculations, we observe that there is a cutoff for the growth rates obtained from the simulations which is not present in the semi-analytical calculation.  However, we also note that the growth rate values match very well for the low $n$ mode numbers for both "P" and "L1" magnetic field configurations.  The difference between the idealized semi-analytical calculations and the simulations is due to dissipative effects such as viscosity, thermal conductivity, and collisions between neutrals and ions, which are included in the simulations and act on smaller scales.  The reduction in the growth rates around $n\approx 10$ and cutoff near $n\approx 30-40$ is consistent with the estimates provided in Sec.~\ref{ssec:coll_effects} and shown in Fig~\ref{fig:visc_and_dec_scale}.



Figure~\ref{fig:gr_k_all} compares the growth rates obtained semi-analytically (left) and numerically (right) for the cases with different density contrasts, magnetic field strength and shear.  We can observe that for smaller mode numbers, $n \leq 10$, there is good agreement between the simulated and the semi-analytical results, both in the magnitude of the growth rates obtained, and in the ordering of the curves for different cases. For higher $n$ we observe that, similarly to Figure~\ref{fig_gr_comp}, the growth rates decrease and a cut-off is produced for all simulated cases, unlike the semi-analytical cases.


We observe that the increase in density contrast leads to the increase in both the peak growth rate and the cut-off wave number, $n_{\rm cd}$. Increasing the magnetic field has an opposite effect, that is, the values of the growth rates and $n_{\rm cd}$ become smaller.  Collisions between neutrals and ions contribute to the creation of the cut-off observed in the simulations in Figure~\ref{fig:gr_k_all}.  As shown in Figure~\ref{fig:visc_and_dec_scale}, the lowest modes estimated to be impacted by neutral-ion and ion-neutral collisions, $n_{\rm \nu_{ni}}$ and $n_{\rm \nu_{in}}$, are below those for the viscosity scale, $n_{\rm visc_n}$. Therefore, it is expected that the observed cut-off is due to collisions between neutrals and ions. In fact, the ordering of the $n_{\rm cd}$ values for different atmosphere profiles in the simulations with the sheared magnetic field (L1-WN-S, L1-WN-NN-B-S, L1-WN-NN-S, from the smallest to the largest) is the same as the ordering of $n_{\rm \nu_{in}}$ in Figure~\ref{fig:visc_and_dec_scale} (yellow curves). 




We obtain further information about the RTI instability in a smoothly stratified atmosphere by studying the eigenfunctions of the instability. In particular, the eigenfunctions of the system provide information on the vertical spatial extent of the perturbations for different modes and can therefore be used to estimate which of the modes may be more affected by the collisional effects in the simulations. { The left panel of} Figure~\ref{fig_gr_eigen} shows the vertical eigenfunctions, $v_z$, obtained semi-analytically for the perpendicular field case { for several values of $n$} with the original values of the density contrast and the magnetic field. The eigenfunctions are normalized to unity. They correspond to the fastest growing modes for the wave numbers $n$ indicated in the legend, and we can observe that the high $n$ eigenmodes are more spatially localized in the vertical direction.  { The right panel of  Fig.~\ref{fig_gr_eigen} quantifies the vertical extent of the eigenmodes by plotting half width at half maximum, $\Delta$, of the fastest growing modes as a function of $n$ for different configurations of the magnetic field corresponding to the growth rates shown in the left panel of Fig.~\ref{fig_gr_comp}.  A linear fit of $\log(\Delta)$ vs. $\log(n)$ is made for high-$n$ modes for each of the magnetic field configurations.  

We observe from the right panel of Fig.~\ref{fig_gr_eigen} that there is a clear distinction in the $\Delta(n)$ functional form for the case of purely perpendicular magnetic field from those with sheared field.  The configuration with perpendicular magnetic field has $\Delta \propto n^{-1/2}$, while all of the RTI-unstable sheared configurations consistently show $\Delta \propto n^{-1}$.  The wavelength dependence of the vertical extent of the RTI modes in the case with no in-plane field and a vertically limited RTI-unstable region with $(d\rho_0/dz) > 0$ was discussed in Sec.~\ref{sssec:an_gr_exp} and the $\Delta \propto n^{-1/2}$ dependence is consistent with the conclusions derived from Eq.~\ref{eq:eig_limited}.  In turn, the $\Delta \propto n^{-1}$ dependence for the sheared field configurations is as estimated in Eq.~\ref{eq:appr5}, with the eigenfunctions becoming progressively more localized for smaller shear lengths.} This implies that high-$n$ modes are more prone to collisional dissipation effects; the effect should be substantially more pronounced for the sheared field configurations, which is consistent with the temporal behavior of different $n$ modes observed in Fig.~\ref{fig:time_snaps2}. 


\section{Decoupling} \label{sec:decoupling}

\begin{figure*}[!htb]
 \centering
 \includegraphics[width=16cm]{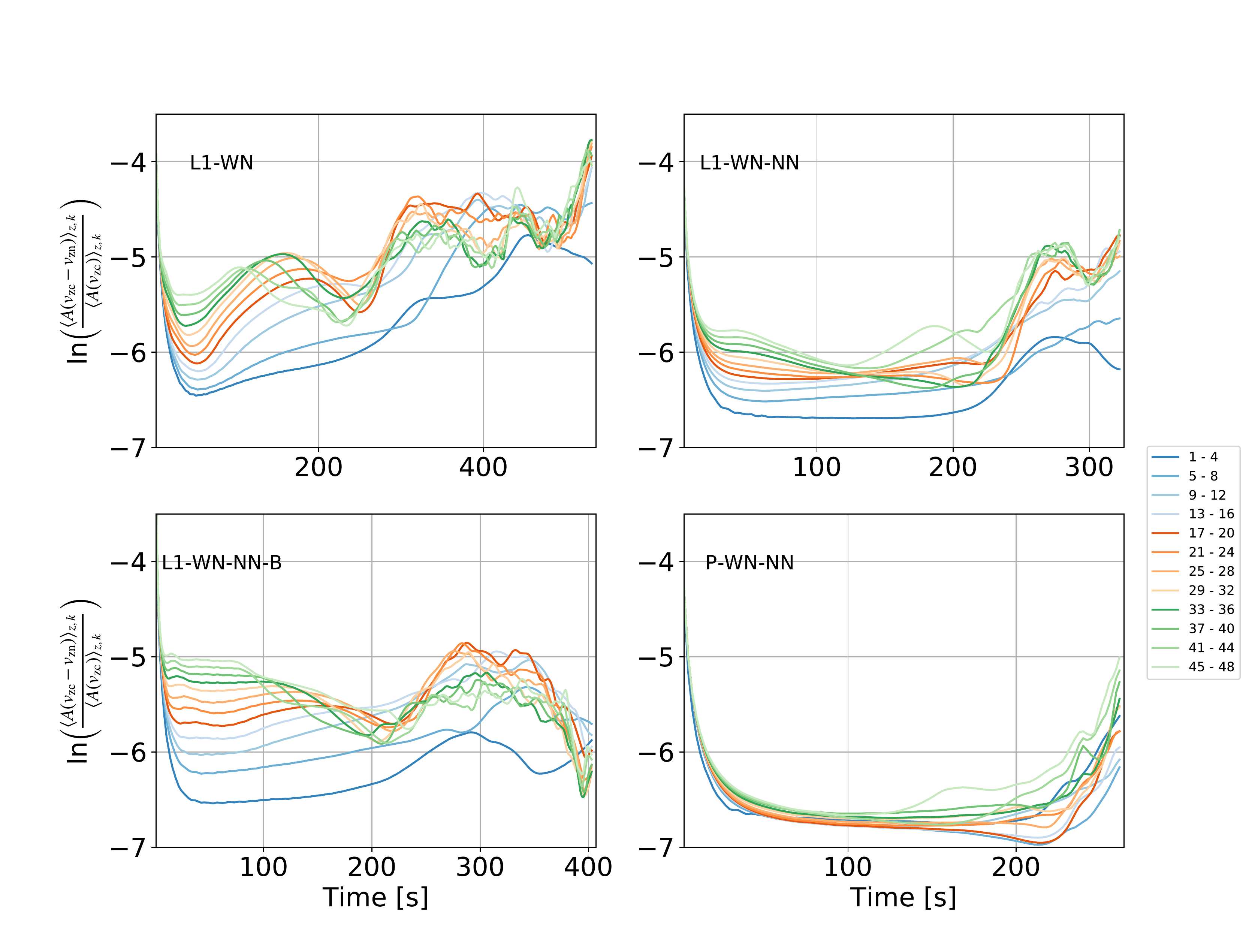}
\caption{Normalized vertical decoupling in vertical velocity for different harmonics  as a function of time. Different color lines correspond to different harmonics grouped in sets of four, from $n=1$ to $n=48$. Top left:  L1-WN simulation; Top, right: L1-WN-NN; Bottom left: L1-WN-NN-B; Bottom right: P-WN-NN.}
\label{fig_group_growing_dec2}
\end{figure*}

\begin{figure*}[!htb]
 \includegraphics[width=8cm]{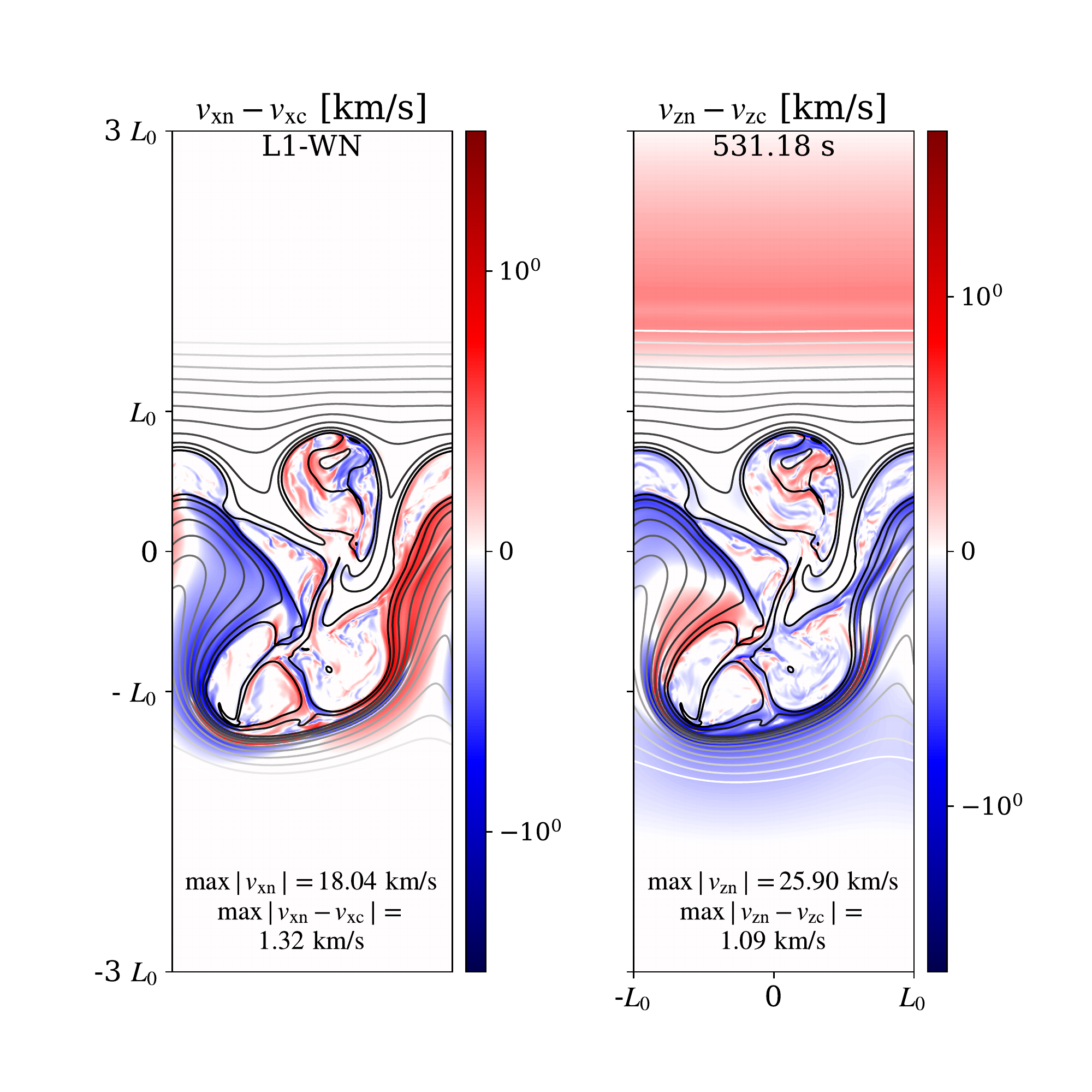}
 \includegraphics[width=8cm]{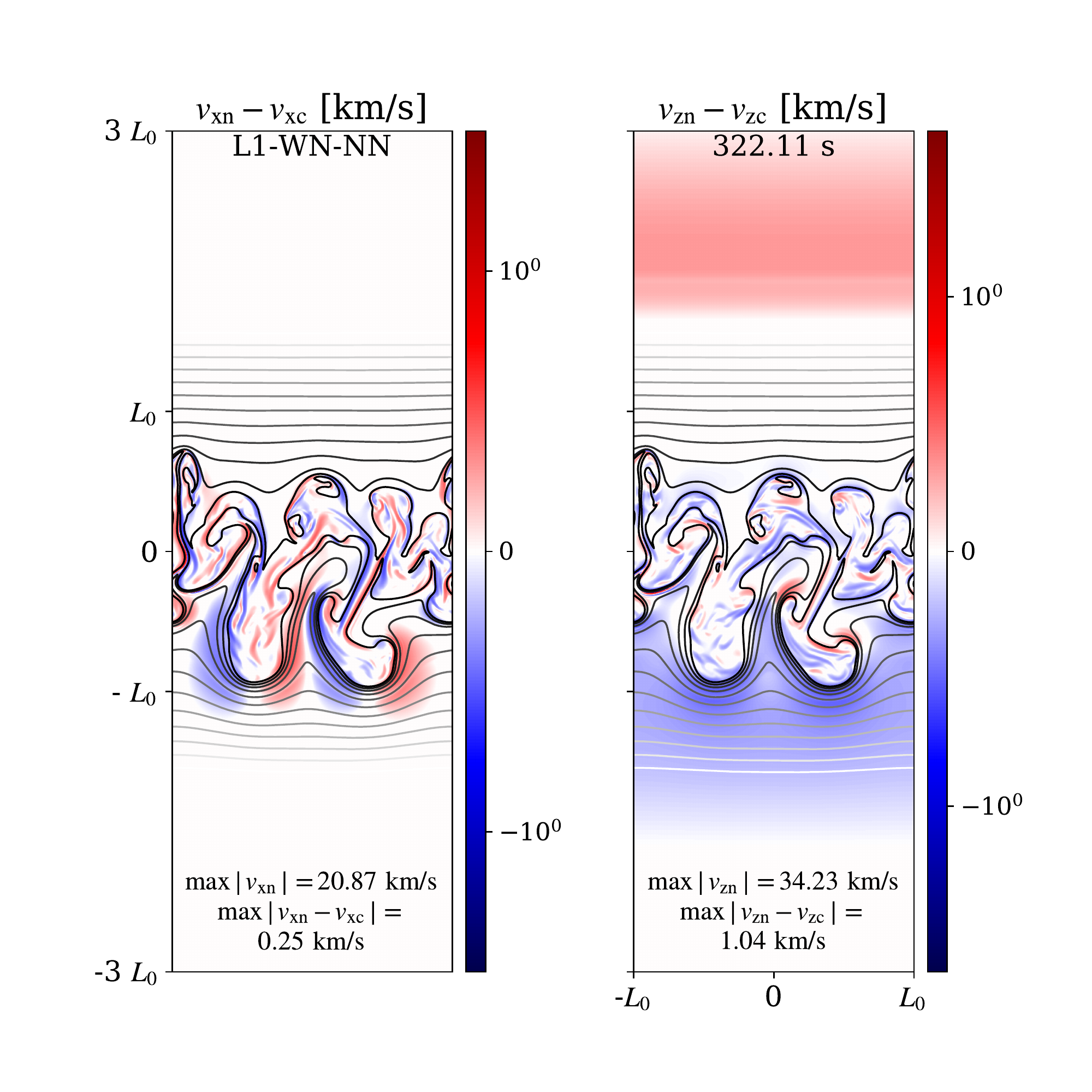}
 
 \includegraphics[width=8cm]{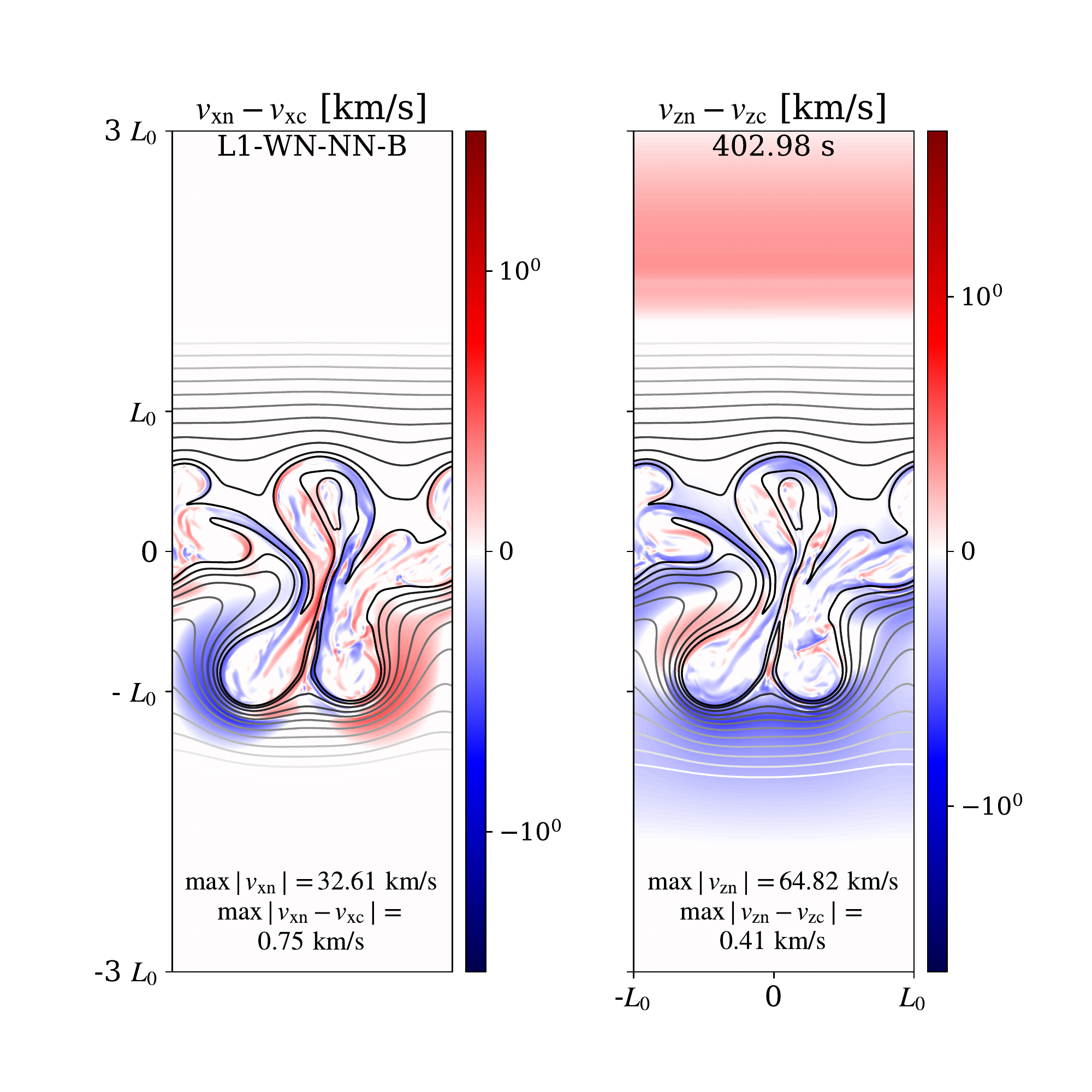}
 \includegraphics[width=8cm]{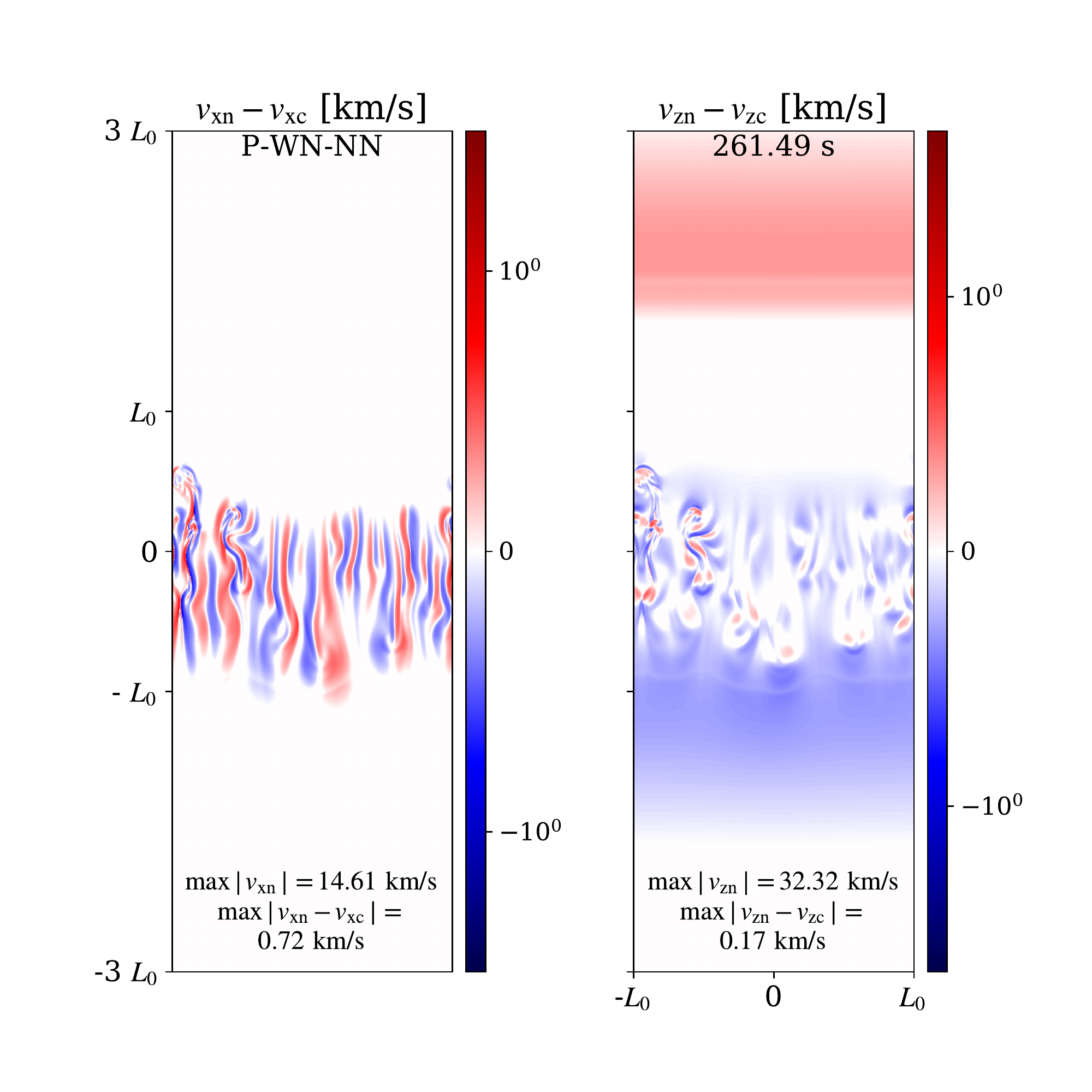}
 \caption{Pairs of snapshots of the horizontal and vertical decoupling in velocities for the simulations P-WN (top left), P-WN-NN (top right), L1-WN-NN-B (bottom left), and P-WN-NN (bottom right). The snapshots are taken at the time indicated at the right-hand-side image of each pair. The  maximum absolute values of the corresponding decoupling and velocity are indicated at the bottom of each image. The black lines represent the 
  iso-contours of the absolute value of the $y$-component of the magnetic field potential, $|A_{\rm y}|$. The iso-contours are equally spaced between 0.6$A_{\rm y}^{\rm max}$ and $A_{\rm y}^{\rm max}$, where $A_{\rm y}^{\rm max}$ is the maximum value of $|A_{\rm y}|$.
 }
\label{fig:gr_dec_sn}
\end{figure*}

Here, we consider the decoupling effects as observed in the simulations. To do so, we quantify the differences in the horizontal and vertical velocity components between the neutrals and charged particles.  We recall that the driver for the RTI instability is in the gravitational free energy of a prominence thread with a high density of neutrals; that is, the instability acts on the neutral fluid.  At the same time, electromagnetic forces act only on charged particles, so it can be expected that incomplete collisional coupling may result in some differences in the velocities of the charged and neutral species.  The viscous force is another important agent since in the plasmas under consideration, the viscosity of neutrals is much greater than viscosity of charged particles. The decoupling can occur when the differentiating viscous or electromagnetic forces (or both) become larger than the collisional coupling between neutrals and charged particles. 

The normalized decoupling in the vertical velocity (and equivalent for the horizontal velocity) is computed according to the following steps: (i) the Fourier amplitude of the difference between the vertical velocities of charged particles  and neutrals, $v_{zc} - v_{zn}$, is computed for all modes $n$ as a function of height; (ii) the Fourier amplitude of the vertical velocity of charged particles, $v_{zc}$, is computed for all modes $n$ as a function of height; (iii) the above quantities are averaged separately for the heights between $-L_0$ and $L_0$ and over four consecutive modes; (iv) the normalized decoupling is obtained as the natural logarithm of the ratio between the two quantities.

Figure~\ref{fig_group_growing_dec2} compares the vertical normalized decoupling between flows of neutrals and charged particles  for the simulations L1-WN (top left), L1-WN-NN (top right), L1-WN-NN-B (bottom left), and P-WN-NN (bottom right) for the first 48 modes. Similarly to Figure~\ref{fig_group_growing2}, we choose the averaging over four consecutive modes and the upper limit of $n=48$ for clarity of the presentation. 

In all four cases considered, the normalized decoupling for all $n$ has an initially high value that rapidly drops early in the simulations.  The initial magnitude of the decoupling is due to the instability being seeded by perturbations in the density of neutrals, without a corresponding perturbation in the density of charged particles.  This leads to the generation of neutral flows that are then rapidly followed by corresponding flows in the charged fluid.  We can observe that the amount of decoupling strongly varies with $n$ for the sheared magnetic field configurations, while for the perpendicular magnetic field case all modes behave nearly identically during the linear phase.  This is explained by the effect of magnetic tension on the development of RTI in the "L1" sheared magnetic field cases, which is absent in the "P" case.  Magnetic tension acts only on charged particles, thus seeding decoupling, and more so on the low $n$ modes that are less localized and therefore sample regions of stronger in-plane magnetic field.  Thus, the L1-WN-NN simulation (top right) shows larger decoupling than the P-WN-NN one (bottom right). In a similar vein, since the Lorentz force is larger for larger magnetic field, the decoupling in the L1-WN-NN-B case (bottom left) is generally greater than in the L1-WN-NN case (top right).

In the non-linear phase, in all cases, the decoupling increases with the mode number $n$. Such behavior is expected because the decoupling should occur for scales below the ion-neutral (or neutral-ion) collisional scales as shown in Figure~\ref{fig:visc_and_dec_scale}.  The increase of the density contrast produces lower decoupling (L1-WN vs L1-WN-NN case). This is also in agreement with the analysis of the ion-neutral collisional scales from Figure~\ref{fig:visc_and_dec_scale}, as $n_{\rm \nu_{in}}$ for the increased density case is much larger than for the original density case. Moreover, the viscosity is smaller for the increased density contrast case, therefore neutrals are less affected, and the decoupling can be expected to be less pronounced. 


The above conclusions can also be confirmed by a visual inspection of the snapshots of the decoupling shown in Figure~\ref{fig:gr_dec_sn}. The locations where the decoupling increases, both in horizontal and vertical velocity, locally coincide with locations where the magnetic field gets compressed (corresponding to higher density of the magnetic field potential contours). This greater decoupling is produced as a result of different movements on the part of neutrals and charged particles across the field lines. 
The decoupling in the vertical velocity is preferentially negative in a larger area right below the instability eddies. A negative value of the decoupling implies that the downward velocities of neutrals are larger than the downward velocities of charged particles. This behavior is consistent with the fact that neutrals drive the instability as they cannot be sustained by the magnetic forces other than through collisions with charged particles. Nevertheless, the absolute value of the normalized decoupling in a horizontal velocity is generally larger (except for the case of the enhanced density L1-WN-NN, top right). The simulation L1-WN shows the largest decoupling among the four cases, reaching values as high as 5-10\% of the corresponding velocity. When the density contrast is increased (top right), the values of the vertical decoupling become larger than the values of the horizontal decoupling. On the contrary, when the magnetic field strength is increased (bottom left), the horizontal decoupling relative to the horizontal velocity becomes larger than the vertical decoupling relative to the vertical velocity.

\section{Magnetic structures}\label{sec:power}

\begin{figure*}[!htb]
 \centering
 \includegraphics[width=16cm]{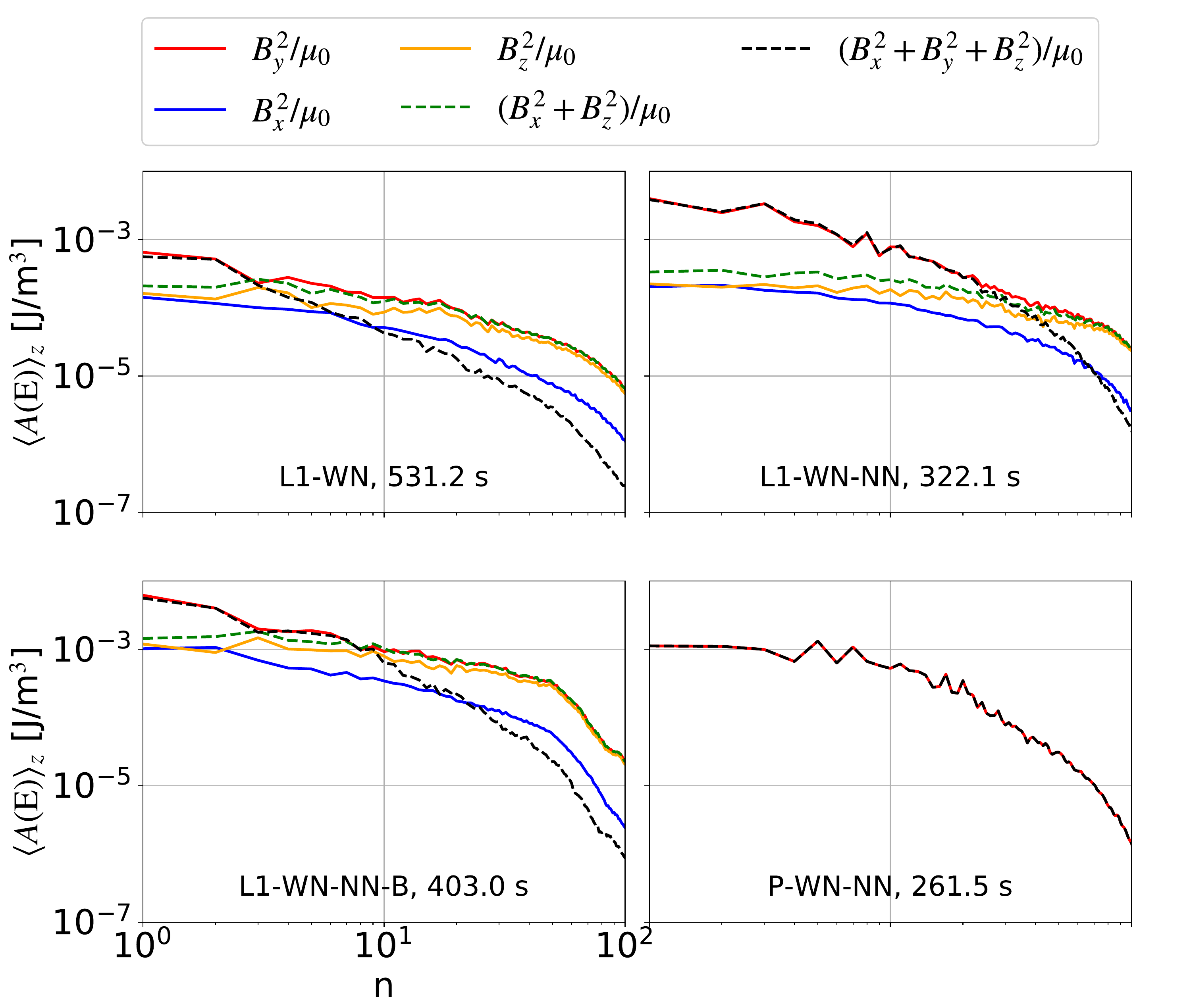}
\caption{Power spectra of magnetic energy for each of the  components of the magnetic field calculated
 at the final time for the simulations L1-WN (top left),  L1-WN-NN (top right), and L1-WN-NN-B (bottom left), and P-WN-NN (bottom right). We note that the simulation P-WN-NN has no magnetic energy in the $x$ and $z$ directions. Solid lines are individual components of the magnetic energy, $B_x^2/\mu_0$ (blue), $B_y^2/\mu_0$ (red), $B_z^2/\mu_0$ (orange). Dashed lines are for the in-plane component, $(B_x^2+B_z^2)/\mu_0$ (green), and the total energy, $(B_x^2+_y^2+B_z^2)/\mu_0$ (black).}
\label{fig_magf_twist2}
\end{figure*}

To get further insights on the formation of magnetic structures, Figure~\ref{fig_magf_twist2} shows the power spectra of magnetic energy for each of the components of the magnetic field calculated at the last available time for the four simulations L1-WN, L1-WN-NN, L1-WN-NN-B, and P-WN-NN.  The power spectra have been constructed as follows: (i) different components of the magnetic energy are computed as functions of $x$ and $z$; (ii) Fourier transform of these components in $x$ direction is done to obtain their Fourier amplitudes as a function of the vertical direction; (iii) the Fourier amplitudes are averaged at heights between $z=-L_0$ and $z=L_0$.

Figure~\ref{fig_magf_twist2} shows qualitatively similar behavior of the magnetic energies for the three $L1$ simulations.  In particular, in all three cases and for all modes $n$, the mode amplitude of total in-plane magnetic energy ($(B_x^2+B_z^2)/\mu_0$, dashed green lines) is greater than in each of the in-plane components of the magnetic energy ($B_x^2/\mu_0$ and $B_z^2/\mu_0$, blue and orange solid lines). This implies that the in-plane components of magnetic field energy vary approximately when they are in-phase with each other. 

On the other hand, the mode amplitudes of the total in-plane magnetic field energy and the energy in the out-of-plane $B$-field component ($B_y^2/\mu_0$, red solid lines) appear to be additive at low $n$, but not at higher $n$.  At low $n$, the mode amplitude of the total magnetic energy is approximately a sum of the corresponding amplitudes of its components. The value of $n$ where this behaviour breaks is different for the three $L1$ cases, with the largest shown for the enhanced density case (upper right panel). For the higher $n$ modes, the mode amplitudes of the in-plane and out-of-plane $B$-field energy components converge to nearly the same values and both greatly exceed the mode amplitudes of the total magnetic energy; that is, black dashed lines are below the green dashed and the red solid lines, which begin to overlap. This indicates that the high-$n$ modes of the in-plane and the out-of-plane components of magnetic field energy have the same amplitude and are nearly exactly out of phase.

The scale where the total in-plane, $(B_x^2+B_z^2)/\mu_0$, and out-of-plane, $B_y^2/\mu_0$, components transition from being additive to canceling each other seems to be related to the largest size of the in-plane plasmoids seen in Figure~\ref{fig:time_snaps2} for the same time moment. We can observe that the structures formed in the L1-WN case are the largest between three $L1$ simulations, followed by the sizes in the L1-WN-NN-B case, and then by the L1-WN-NN case. The ordering between these scales is the same as the ordering of the scales where the out-of-plane and the in-plane components of magnetic energy in Figure~\ref{fig_magf_twist2} become out of phase. 
{
Some authors found breaks in the power spectra at a certain scale, \citep{freed,hillier,leonardis} which might be related to the formation of the magnetic structures, which will be discussed in a forthcoming paper.}
The 3D configuration of the magnetic field, with a dominant component in the $y$-direction being out of phase with the in-plane component, may resemble a flux rope. 
 
\section{Discussion and conclusions} \label{sec:conclusions}

The main goal of our paper is to the study of the effects of partial ionization on the development of the RTI at a realistically smooth transition layer between a magnetized prominence thread and the solar corona.  We use a two-fluid model, which evolves neutral and charged species, coupled by collisional terms, where we consider both the elastic and inelastic processes.  
We explore the dependence of the linear and non-linear evolution of the instability on the magnetic field configuration and mass loading.  Our main conclusions can be summarized as follows:

\begin{itemize}
\item[(i)]
The consideration of a smooth RTI-unstable transition layer as opposed to the typically considered discontinuous interface is both necessary to self-consistently evaluate two-fluid effects on the RTI instability and to demonstrates a different set of RTI characteristics than previously observed in discontinuous interface studies.  
\item[(ii)] 
For sheared magnetic field equilibria, the in-plane component of the magnetic field suppresses RTI growth in the linear phase. Both in simulations and in analytical calculations using single-fluid ideal MHD treatment, the growth rate at all scales is smaller when the magnetic field is sheared compared to the perpendicular configuration. { The linear growth rate decreases with decreasing shear length, stabilizing the RTI for sufficiently small shear length for wavelength shorter than the density gradient scale; however, sheared equilibria that are unstable to wavelengths of order of the density gradient scale are also unstable with regard to modes of arbitrarily short wavelengths with the same growth rate.}
%
%
\item[(iii)] 
For a smooth transition layer, low wave-number RTI eigenfunctions have vertical extents that encompass the prominence thread, while higher $n$ modes become localized at the location of highest density gradient.  The presence of sheared magnetic field provides further and more rapid vertical localization of the eigenfunctions as a function of $n$, making high-$n$ modes even more prone to collisional damping effects. 
\item[(iv)]  When mass loading and the density contrast is increased, both semi-analytical ideal MHD calculations and simulations show that the linear growth rate is larger at all scales.  Additionally, greater mass loading leads to stronger ion-neutral collisions and lower viscosity. Thus, smaller scales develop in the linear stage of the high density contrast simulations, while the relative decoupling between charged particles  and neutrals observed during the non-linear phase is also smaller.
\item[(v)] 
 Ion-neutral collisions and, to a smaller extent, neutral viscosity effects appear to suppress growth of short wavelength modes in the linear phase of the RTI instability.  This is established by estimating the scales expected to be impacted by the collisional effects in Fig.~\ref{fig:visc_and_dec_scale} and then comparing the linear growth rates calculated from the simulations to the linear growth rates obtained semi-analytically under the single-fluid ideal MHD approximation.  These effects will be further explored in follow-up work. 
\item[(vi)]  The increase of the magnetic field magnitude decreases RTI growth rate. At the same time, in the non-linear phase, the flow decoupling between charged particles  and neutrals appears to be strongest near the locations of highest magnetic field stress.  We thus infer that much of the decoupling observed in the simulations is likely driven by the magnetic activity due to charged particles, but not neutrals, as the former are directly impacted by magnetic field forces.
\item[(vii)]  In simulations with the sheared magnetic field equilibria, the spatial spectra of the out-of-plane and in-plane components of the magnetic energy show strong out-of-phase coherence at small scales, but not at the largest scales, during the non-linear phase of the evolution. The transition scale is different in simulations with different mass loading and magnetic field strength, with the L1-WN simulation showing both the strongest magnetic activity and highest degree of such magnetic self-organization across the scales. These effect will be further explored in follow-on work.
\end{itemize}

It is known that the linear growth rate of RTI is related to the properties of the background atmosphere, such as the density contrast and the magnetic field. In prominences, the growth rate and the degree of mass loading can be measured from observations and such measurements have been used in the past to derive the magnetic field properties based on observations of the RTI. Here, we obtain that other measurable quantities, such as the decoupling in velocity of ions and neutrals,
are also related to the background variables. These latter quantities are much more difficult to measure because they require high resolution and high temporal cadence of the data. Such data could possibly be obtained using the large-aperture 4-m solar telescope facilities (DKIST\footnotemark \footnotetext{\url{https://dkist.nso.edu/}}, EST
\footnotemark \footnotetext{\url{http://www.est-east.eu/}}). 


In all the simulations performed, we found that ion-neutral interactions and, to a smaller extent, viscosity damp linear RTI development at smallest scales. The comparison of the numerical linear growth rate to the linear growth rates obtained semi-analytically in the single-fluid ideal MHD approximation, neglecting the  effect of the collisions, confirmed that the suppression of the small scales is indeed produced by dissipation effects.
 
In our simulations, the largest scales do not show effects of dissipation due to ion-neutral collisions and viscosity, with the global RTI growth rate remaining the same as in the single-fluid ideal MHD regime. Some of the previous numerical simulations have similarly shown that the characteristics of RTI flows are identical for the viscous and inviscid models \citep{rti_visc3}. On the other hand, \cite{visc1} and \cite{visc2} found an increased growth rate at all wave numbers in the inviscid case, compared to the case with viscosity.  The apparent contradiction between the two results may come from the different scales considered in the simulations since the dissipative effects occur only for the scales below a certain scale.

We  find that the linear growth rate increases with increasing the density contrast. This result is in agreement with the classical hydrodynamic model from \citet{Ch1961} for the discontinuous density profile, as well as for the density profile exponentially stratified with height \citep{visc_incomp}. It has been shown that compressibility causes a destabilizing effect, which becomes more important for smaller density contrasts \citep{visc_incomp,2005Ribeyre}. In the nonlinear phase, it has been shown that increasing the density contrast results in a higher growth rate, thinner fingers, and less roll-up at the tip of the fingers due to secondary KHI \citep{YOUNGS1984, 1988Gardner, 1995Stone}. The magnetic field affects the growth rate at small scales, but even in the magnetic RTI, the growth rate is still larger for a larger density contrast \citep{rti-atwood1}.

In our case, the viscosity coefficient is smaller in the model with enhanced density contrast, and the ion-neutral collisional frequency, is higher. This reduction of viscosity and the importance of ion-neutral interaction act in favor of increasing of the growth rate, together with the proper density contrast enhancement. 

We find that the viscous force acting dominantly on neutrals along with the electromagnetic force acting only on the charged particles  produce the observed decoupling of the respective flows in the non-linear phase of RTI development. This decoupling decreases with decreasing importance of these forces. For example, when the density contrast is enhanced, the magnitude of the Lorentz force relative to the gravitational force decreases and the decoupling also decreases. Similarly, when the density contrast is enhanced, the neutral viscosity becomes smaller, leading to lower values for the decoupling. Overall, we observe in our simulations that the decoupling can reach up to 5\% of the individual flow velocities. There is a correlation between the locations with larger decoupling and the locations of greatest magnetic field stress (i.e., the locations where the Lorentz force is increased). 

The magnetic field parallel to the perturbation plane has a stabilizing effect on the RTI.  Early analytical studies by \cite{Ch1961} have shown that a component of the magnetic field parallel to the direction of the perturbation suppresses the small-scale perturbations with wavelengths smaller than a cutoff wavelength, proportional to the square of the in-plane field strength, as seen from Eq.~(\ref{eq:lc}). In the nonlinear phase, \cite{Hillier2017} measured the growth of the magnetic RTI by means of numerical simulations in a 3D setup.  These authors found that the growth rate decreases with increasing magnetic field and they attributed it to  the effects of higher magnetic tension, requiring greater energy for the frozen-in plasma to move. In our simulations, we found that the linear growth rate is smaller when the field is sheared compared to the perpendicular field case. { The ideal MHD semi-analytical calculations for the sheared magnetic field configurations similarly showed that the peak growth rate decreases with decreasing shear length $L_s$; furthermore, for sufficiently small shear lengths, all modes with a wavelength shorter than the density gradient scale are stabilized due to a combination of magnetic tension and in-plane magnetic pressure effects.  However, both analytically estimated and semi-analytical calculations have also shown that for a range of shear lengths, in the absence of collisional effects, the RTI growth rate becomes wavelength-independent in the short wavelength limit.}

Using a single-fluid approximation with two-fluid effects introduced through the ambipolar term, \cite{DiazKh2013} found that collisions may remove the stabilizing effect of the magnetic field on the small scales. Thus, the linear growth rate at small scales would become larger when the collisions are taken into account. In our case, the small scales are suppressed when the collisions are taken into account through neutral-charge interaction and viscosity.  This is due, primarily, to our consideration of a smooth rather than a discontinuous prominence-corona interface with a gradually rotating magnetic field, such that the scales associated with magnetic field stabilization are much smaller than those stabilized by collisional dissipation, as shown in Figure~\ref{fig_gr_comp}.

At small scales, suppressed by dissipation effects in the linear phase, are introduced in the early non-linear stage by energy cascade.  In the non-linear phase, we observe a continued production of magnetic energy at intermediate and small scales due to the twisting of the field lines. The in-plane plasmoids observed in the snapshots, together with the dominant perpendicular component of the magnetic field, create 3D structures resembling flux ropes. This is confirmed by the Fourier analysis of the three components of the magnetic energy. The strong out-of-phase correlation between the in-plane and the out-of-plane magnetic energy for intermediate and high $n$ modes hints at magnetic self-organization processes taking place during non-linear RTI evolution.  The scales where the two components of the magnetic energy become coherent seem to be related to the largest scales of the in-plane magnetic structures.  Further study of this magnetic field dynamics will be presented in future work.

\begin{acknowledgements}
This work was supported by the Spanish Ministry of Science through the project AYA2014-55078-P and the US National Science Foundation. It contributes to the deliverable identified in FP7 European Research Council grant agreement ERC-2017-CoG771310-PI2FA for the project ``Partial Ionization: Two-fluid Approach''. The author(s) wish to acknowledge the contribution of Teide High-Performance Computing facilities to the results of this research. TeideHPC facilities are provided by the Instituto Tecnol\'ogico y de Energ\'ias Renovables (ITER, SA). URL: http://teidehpc.iter.es
\end{acknowledgements}

\bibliographystyle{aa}
\bibliography{biblio}

\begin{thebibliography}{44}
\expandafter\ifx\csname natexlab\endcsname\relax\def\natexlab#1{#1}\fi

\bibitem[{{Anan} {et~al.}(2017){Anan}, {Ichimoto}, \& {Hillier}}]{2017Anan}
{Anan}, T., {Ichimoto}, K., \& {Hillier}, A. 2017, \aap, 601, A103

\bibitem[{Anuchina {et~al.}(2004)Anuchina, Volkov, Gordeychuk, Es'kov,
  Ilyutina, \& Kozyrev}]{ANUCHINA}
Anuchina, N., Volkov, V., Gordeychuk, V., {et~al.} 2004, Journal of
  Computational and Applied Mathematics, 168, 11 , selected Papers from the
  Second International Conference on Advanced Computational Methods in
  Engineering (ACOMEN 2002)

\bibitem[{Arber {et~al.}(2007)Arber, Haynes, \& Leake}]{Arber_2007}
Arber, T.~D., Haynes, M., \& Leake, J.~E. 2007, The Astrophysical Journal, 666,
  541

\bibitem[{Berger {et~al.}(2017)Berger, Hillier, \& Liu}]{Berger_2017}
Berger, T., Hillier, A., \& Liu, W. 2017, The Astrophysical Journal, 850, 60

\bibitem[{{Berger} {et~al.}(2008){Berger}, {Shine}, {Slater}, {Tarbell},
  {Title}, {Okamoto}, {Ichimoto}, {Katsukawa}, {Suematsu}, {Tsuneta}, {Lites},
  \& {Shimizu}}]{Berger2008}
{Berger}, T.~E., {Shine}, R.~A., {Slater}, G.~L., {et~al.} 2008, \apjl, 676,
  L89

\bibitem[{{Berger} {et~al.}(2010){Berger}, {Slater}, {Hurlburt}, {Shine},
  {Tarbell}, {Title}, {Lites}, {Okamoto}, {Ichimoto}, {Katsukawa}, {Magara},
  {Suematsu}, \& {Shimizu}}]{2010Berger}
{Berger}, T.~E., {Slater}, G., {Hurlburt}, N., {et~al.} 2010, \apj, 716, 1288

\bibitem[{{Bhatia}(1974)}]{1974Bhatia}
{Bhatia}, P.~K. 1974, \apss, 26, 319

\bibitem[{{Carlyle} \& {Hillier}(2017)}]{Hillier2017}
{Carlyle}, J. \& {Hillier}, A. 2017, A\&A, 605, A101

\bibitem[{Chandrasekhar(1961)}]{Ch1961}
Chandrasekhar, S. 1961, Hydrodynamic and Hydromagnetic Stability, International
  series of monographs on physics (Clarendon Press)

\bibitem[{{D\'{\i}az} {et~al.}(2014){D\'{\i}az}, {Khomenko}, \&
  {Collados}}]{DiazKh2013}
{D\'{\i}az}, A.~J., {Khomenko}, E., \& {Collados}, M. 2014, A\&A, 564, A97

\bibitem[{D{\'{\i}}az {et~al.}(2012)D{\'{\i}}az, Soler, \&
  Ballester}]{Diaz2012}
D{\'{\i}}az, A.~J., Soler, R., \& Ballester, J.~L. 2012, The Astrophysical
  Journal, 754, 41

\bibitem[{Dolai {et~al.}(2016)Dolai, Prajapati, \& Chhajlani}]{rti-atwood1}
Dolai, B., Prajapati, R.~P., \& Chhajlani, R. 2016, Physics of Plasmas, 23,
  113704

\bibitem[{{Doludenko} {et~al.}(2019){Doludenko}, {Fortova}, {Shepelev}, \&
  {Son}}]{rti_visc3}
{Doludenko}, A.~N., {Fortova}, S.~V., {Shepelev}, V.~V., \& {Son}, E.~E. 2019,
  \physscr, 94, 094003

\bibitem[{{Felipe} {et~al.}(2010){Felipe}, {Khomenko}, \&
  {Collados}}]{Felipe2010}
{Felipe}, T., {Khomenko}, E., \& {Collados}, M. 2010, \apj, 719, 357

\bibitem[{Freed {et~al.}(2016)Freed, McKenzie, Longcope, \& Wilburn}]{freed}
Freed, M.~S., McKenzie, D.~E., Longcope, D.~W., \& Wilburn, M. 2016, The
  Astrophysical Journal, 818, 57

\bibitem[{Gardner {et~al.}(1988)Gardner, Glimm, McBryan, Menikoff, Sharp, \&
  Zhang}]{1988Gardner}
Gardner, C.~L., Glimm, J., McBryan, O., {et~al.} 1988, The Physics of Fluids,
  31, 447

\bibitem[{{Gerashchenko} \& {Livescu}(2016)}]{visc2}
{Gerashchenko}, S. \& {Livescu}, D. 2016, Physics of Plasmas, 23, 072121

\bibitem[{{Gilbert} {et~al.}(2007){Gilbert}, {Kilper}, \&
  {Alexander}}]{2007Gilbert}
{Gilbert}, H., {Kilper}, G., \& {Alexander}, D. 2007, \apj, 671, 978

\bibitem[{Gonz\'alez-Morales {et~al.}(2018)Gonz\'alez-Morales, Khomenko,
  Downes, \& de~Vicente}]{Pedro2018}
Gonz\'alez-Morales, P.~A., Khomenko, E., Downes, T., \& de~Vicente, A. 2018,
  \aap, submitted

\bibitem[{Hillier(2016)}]{Hillier2016}
Hillier, A. 2016, Monthly Notices of the Royal Astronomical Society, 462,
  stw1805

\bibitem[{Hillier(2018)}]{Hillier2018}
Hillier, A. 2018, Reviews of Modern Plasma Physics, 2

\bibitem[{Hillier(2019)}]{hillier2019}
Hillier, A. 2019, Physics of Plasmas, 26, 082902

\bibitem[{{Hillier} {et~al.}(2017){Hillier}, {Matsumoto, T.}, \& {Ichimoto,
  K.}}]{hillier}
{Hillier}, A., {Matsumoto, T.}, \& {Ichimoto, K.} 2017, A\&A, 597, A111

\bibitem[{{Jun} {et~al.}(1995){Jun}, {Norman}, \& {Stone}}]{1995Stone}
{Jun}, B.-I., {Norman}, M.~L., \& {Stone}, J.~M. 1995, \apj, 453, 332

\bibitem[{{Khomenko} {et~al.}(2016){Khomenko}, {Collados}, \&
  {D{\'{\i}}az}}]{2016Khomenko}
{Khomenko}, E., {Collados}, M., \& {D{\'{\i}}az}, A.~J. 2016, \apj, 823, 132

\bibitem[{{Khomenko} {et~al.}(2014){Khomenko}, {D{\'{\i}}az}, {de Vicente},
  {Collados}, \& {Luna}}]{2014bKh}
{Khomenko}, E., {D{\'{\i}}az}, A., {de Vicente}, A., {Collados}, M., \& {Luna},
  M. 2014, \aap, 565, A45

\bibitem[{{Leake} {et~al.}(2014){Leake}, {DeVore}, {Thayer}, {Burns},
  {Crowley}, {Gilbert}, {Huba}, {Krall}, {Linton}, {Lukin}, \&
  {Wang}}]{Leake2014}
{Leake}, J.~E., {DeVore}, C.~R., {Thayer}, J.~P., {et~al.} 2014, \ssr, 184, 107

\bibitem[{Leonardis {et~al.}(2012)Leonardis, Chapman, \& Foullon}]{leonardis}
Leonardis, E., Chapman, S.~C., \& Foullon, C. 2012, The Astrophysical Journal,
  745, 185

\bibitem[{Livescu(2004)}]{visc_incomp}
Livescu, D. 2004, Physics of Fluids, 16, 118

\bibitem[{Mitra {et~al.}(2016)Mitra, Roychoudhury, \& Khan}]{visc1}
Mitra, A., Roychoudhury, R., \& Khan, M. 2016, Physics of Plasmas, 23, 024503

\bibitem[{{Newcomb}(1983)}]{1983Newcomb}
{Newcomb}, W.~A. 1983, Physics of Fluids, 26, 3246

\bibitem[{{Orozco Su{\'a}rez} {et~al.}(2014){Orozco Su{\'a}rez}, {D{\'\i}az},
  {Asensio Ramos}, \& {Trujillo Bueno}}]{2014Ozorco}
{Orozco Su{\'a}rez}, D., {D{\'\i}az}, A.~J., {Asensio Ramos}, A., \& {Trujillo
  Bueno}, J. 2014, \apjl, 785, L10

\bibitem[{{Parchevsky} \& {Kosovichev}(2007)}]{2007ManchaPa}
{Parchevsky}, K.~V. \& {Kosovichev}, A.~G. 2007, \apjl, 666, L53

\bibitem[{{Popescu Braileanu} {et~al.}(2019){Popescu Braileanu}, {Lukin},
  {Khomenko}, \& {de Vicente}}]{Popescu+etal2018}
{Popescu Braileanu}, B., {Lukin}, V.~S., {Khomenko}, E., \& {de Vicente},
  {\'A}. 2019, \aap, 627, A25

\bibitem[{{Ribeyre} {et~al.}(2005){Ribeyre}, {Tikhonchuk}, \&
  {Bouquet}}]{2005Ribeyre}
{Ribeyre}, X., {Tikhonchuk}, V.~T., \& {Bouquet}, S. 2005, \apss, 298, 75

\bibitem[{{Ruderman} {et~al.}(2018){Ruderman}, {Ballai}, {Khomenko}, \&
  {Collados}}]{2018Rude}
{Ruderman}, M.~S., {Ballai}, I., {Khomenko}, E., \& {Collados}, M. 2018, \aap,
  609, A23

\bibitem[{{Ruderman} {et~al.}(2014){Ruderman}, {Terradas}, \&
  {Ballester}}]{2014Rude}
{Ruderman}, M.~S., {Terradas}, J., \& {Ballester}, J.~L. 2014, \apj, 785, 110

\bibitem[{{Terradas} {et~al.}(2015){Terradas}, {Soler}, {Luna}, {Oliver}, \&
  {Ballester}}]{Soler-rti1}
{Terradas}, J., {Soler}, R., {Luna}, M., {Oliver}, R., \& {Ballester}, J.~L.
  2015, \apj, 799, 94

\bibitem[{{Vandervoort}(1961)}]{1961Vander}
{Vandervoort}, P.~O. 1961, \apj, 134, 699

\bibitem[{Wiehr {et~al.}(2019)Wiehr, Stellmacher, \& Bianda}]{Wiehr2019}
Wiehr, E., Stellmacher, G., \& Bianda, M. 2019, The Astrophysical Journal, 873,
  125

\bibitem[{{Xue} \& {Ye}(2010)}]{2010Xue}
{Xue}, C. \& {Ye}, W. 2010, Physics of Plasmas, 17, 042705

\bibitem[{Youngs(1984)}]{YOUNGS1984}
Youngs, D.~L. 1984, Physica D: Nonlinear Phenomena, 12, 32

\bibitem[{Zhou(2017{\natexlab{a}})}]{ZHOU1}
Zhou, y. 2017{\natexlab{a}}, Physics Reports, 720-722

\bibitem[{Zhou(2017{\natexlab{b}})}]{ZHOU2}
Zhou, Y. 2017{\natexlab{b}}, Physics Reports, 723-725, 1 , rayleigh–Taylor
  and Richtmyer–Meshkov instability induced flow, turbulence, and mixing. II

\end{thebibliography}

\end{document}